\documentclass[final]{siamltex}
\usepackage{lscape}
\usepackage[center]{subfigure}
\usepackage{multicol}
\usepackage{multirow}

\usepackage{amsfonts}
\usepackage{graphicx,color}
\usepackage{amssymb}
\usepackage[matrix,arrow]{xy}

\newcommand{\vectwo}[2]{ \begin{pmatrix} #1 \\ #2 \end{pmatrix} }

\title{The large core limit of spiral waves in excitable media: A numerical approach}
\author{
Sebastian Hermann \thanks{School of Mathematics and Statistics, University of Sydney, NSW 2006, Australia
({\tt shermann@maths.usyd.edu.au}), supported by an Endeavour Australia-Europe Award, an EIPRS award and the Deutscher Akademischer Austauschdienst.}
\and
Georg A. Gottwald \thanks{ School of Mathematics and Statistics, University of Sydney, NSW 2006, Australia
({\tt georg.gottwald@sydney.edu.au}), supported by the Australian Research Council.}
}

\begin{document}

\maketitle


\begin{abstract}
We modify the freezing method introduced by {\emph {Beyn \& Th\"ummler, 2004,}} for analyzing rigidly rotating spiral waves in excitable media. The proposed method is designed to stably determine the rotation frequency and the core radius of rotating spirals, as well as the approximate shape of spiral waves in unbounded domains. In particular, we introduce spiral wave boundary conditions based on geometric approximations of spiral wave solutions by Archimedean spirals and by involutes of circles. We further propose a simple implementation of boundary conditions for the case when the inhibitor is non-diffusive, a case which had previously caused spurious oscillations. 

We then utilize the method to numerically analyze the large core limit. The proposed method allows us to investigate the case close to criticality where spiral waves acquire infinite core radius $r_c$ and zero rotation frequency $\omega$, before they begin to develop into retracting fingers. We confirm the linear scaling regime of a drift bifurcation for the rotation frequency and the core radius of spiral wave solutions close to criticality. This regime is unattainable with conventional numerical methods.

\end{abstract}




\section{Introduction}

Excitable media are abundant in nature, and appear in physical, chemical and biological systems. Prominent examples include cAMP waves in slime mold aggregation \cite{Siegert91} and intracellular calcium waves \cite{Berridge00}, electrical waves in cardiac and nerve tissue \cite{Winfree87,Davidenko92}, and the auto-catalytic Belousov-Zhabotinsky reaction \cite{Winfree72}. Excitable media involve the interwoven dynamics of so called {\it activators} and {\it inhibitors}. An important class of excitable media contains non-diffusive inhibitors.  This is the classic situation in cardiac dynamics where the inhibitor consists of (relatively) immobile ion channels. Excitable media support localized pulses and periodic wave trains. In two dimensions rotating vortices (or spirals) are possible \cite{Winfree72}, and in three dimensions scroll waves occur  \cite{Winfree72,Winfree90,Winfree94,Margerit01,Margerit02}.\\

\noindent
In this work we focus on rigidly rotating spiral waves of single-diffusive excitable media. These spiral waves perform a time-periodic motion where the spiral tip moves on a perfect circle around an unexcited region, the spiral core. Such spirals are characterized by their rotation frequency $\omega$ and their core radius $r_c$. Varying the excitability of the medium leads to qualitative changes in the type of motion the spiral is performing. For certain parameters rigidly rotating spirals bifurcate via a Hopf bifurcation into meandering spirals \cite{Winfree91,Barkley94,Hakim97}. A different type of bifurcation occurs at low excitabilities where rigidly rotating spirals develop into travelling fingers with an overall zero curvature. Approaching this bifurcation, the rotation frequency of a rigidly rotating spiral wave $\omega$ decreases down to zero and its core radius $r_c$ diverges. Past this bifurcation rigidly rotating spirals do not exist. Whereas the bifurcation to meandering spirals is well-understood numerically and theoretically \cite{Barkley94,Wulff96,Fiedler96,Golubitsky97,Sandstede97b,Sandstede99}, the latter one is not. Theoretically, several attempts have been made to study this large core limit, using kinematic theory \cite{Zykov87,Tyson88,Elkin98,Mikhailov94,Hakim97,Hakim99,Zykov09,Elkin02}, dynamical systems theory \cite{Ashwin99} and non-perturbative asymptotics \cite{Gottwald04}. Computational results, on the other hand, are rare. In \cite{Gottwald04} the large core limit was investigated in terms of the growing velocity of the spiral wave tip close to criticality. Using a non-perturbative approach it was shown that the growing velocity behaves linearly as a function of the excitability close to criticality. This was verified numerically.\\

\noindent
Here we set out to tackle the numerically much more difficult problem of determining the rotation frequency $\omega$ and the radius $r_c$ of the core of a spiral wave in this limit. This presents a huge computational challenge. Given current computer power, determining the rotation frequency and the core radius from direct simulations of the underlying reaction-diffusion system is restricted to small core spirals. In the large core limit, where the rotation frequency becomes zero and the core radius becomes infinite, no reliable results have been obtained so far.\\
 
\noindent
Our aim here is twofold. First we present a numerical method to study rigidly rotating spirals in excitable media which is able to capture the spiral wave even in the large core limit without having to solve for computationally expensive large domains. We then apply this method to obtain the rotation frequency and the core radius of spiral wave solutions in the large core limit, and establish their scaling behaviour close to criticality to study the type of bifurcation.\\

\noindent
The numerical method we present is based on the freezing method \cite{Beyn04} in which the dynamics of an equivariant  partial differential equation is split orthogonally into the shape dynamics and the dynamics of the associated symmetry group. This method has been successfully applied to many types of equivariant systems \cite{Beyn08,Thummler05,Thummler08a}. However, when applying the freezing method to single-diffusive excitable media several problems were encountered \cite{Beyn04,Thummler06,Beyn09}. Firstly, the inhibitor exhibits spurious oscillations, the amplitude of which increases when approaching the large core limit. It was suggested that these oscillations are caused by numerical instabilities linked to the mixed hyperbolic-parabolic nature of the problem. The oscillations could be partially controlled by an upwind-downwind scheme for small core spirals. Secondly, the application of Neumann boundary conditions causes the shape of the spiral wave to deviate near the boundary from the form expected in an unbounded domain. However, it turns out that although the shape is not properly resolved near the boundary, the group parameters, i.e. rotation frequency and core radius, are accurately reproduced for small core spirals. This can be understood on a phenomenological basis by considering that the spiral wave coils shield the solution near the core, and information only flows outwards for rotating spiral wave solutions.\\
\indent
We will present an implementation of the freezing method which will overcome these difficulties. At first, we will identify two separate types of spatial oscillations, which are caused by different mechanisms: oscillations in the interior of the computational domain linked to numerical instabilities of the non-diffusive inhibitor, and oscillations close to the boundary caused by the application of Neumann boundary conditions. We eliminate the former by employing a semi-implicit Crank-Nicolson scheme and the latter by imposing a subtle implementation of a different type of boundary condition. We implement so-called \emph{spiral wave boundary conditions}, approximating the spiral wave by an Archimedean spiral or by an involute of a circle, which approximately respect the symmetry of the solution, and produce the correct shape of a spiral wave in an effective unbounded domain at the boundary. We find that the spatial oscillations can be controlled if (a) boundary conditions which respect the symmetry of the solution are employed, and (b) the implementation of these boundary conditions is such that the boundary is coupled to the interior.\\

\noindent
The paper is organized as follows. In Section~\ref{sec:model} we present the excitable media model under consideration. In Section~\ref{sec:NumericalMethod} we describe the original freezing method. This method is then modified to suit excitable media with a non-diffusive inhibitor in Section~\ref{sec:BoundaryConditions}, where we introduce spiral wave boundary conditions and their discrete implementations. In Section~\ref{sec:Results} we apply our method to study the large core limit and determine the scaling behaviour of the rotation frequency and the core radius of spiral wave solutions close to criticality for varying excitability $\epsilon$. We conclude with a discussion in Section~\ref{sec:discussion}.


\section{Model}
\label{sec:model}
We consider here the Barkley model \cite{Barkley91} for an activator $u$ and an inhibitor $v$ described by
\begin{eqnarray}\label{eq:Barkley}
    \partial_t u &=&  \Delta  u  +{\cal{F}}(u,v),\quad {\cal{F}}(u,v) = \frac{1}{\epsilon}u(1-u)(u-\frac{v+b}{a}) \, , \\
    \partial_t v &=&  D_v \Delta v +  (u- v) \, .
\end{eqnarray}
Although the numerical method we will describe in Sections \ref{sec:NumericalMethod} and \ref{sec:BoundaryConditions} is independent of the particular model used, we illustrate some basic properties of excitable media with the Barkley model (\ref{eq:Barkley}). Our choice of model is motivated by the fact that it incorporates the ingredients of an excitable system in a compact and lucid way. Thus, for $u_s=b/a>0$ the rest state $u_0=v_0=0$ is linearly stable with decay rates $\sigma_1=u_s/\epsilon$ along the activator direction and $\sigma_2=1$ along the inhibitor direction. Perturbing $u$ above the threshold $u_s$ (in 0D) will lead to growth of $u$. In the absence of the inhibitor $v$ the activator will saturate at $u=1$ leading to a bistable system.  The positive inhibitor growth factor forces the activator to decay back to $u=0$. Finally also the inhibitor with the refractory time constant ${1}$ will decay back to $v=0$. For $a>b+1$ with $b>0$ the system is in 0D no longer excitable but instead bistable. 

We have included a diffusion term for the inhibitor $v$ in (\ref{eq:Barkley}). However, we will be concentrating on the case of vanishing diffusivity for $v$ with $D_v=0$. This case is more relevant for cardiac dynamics where the inhibitor models the (relatively) immobile potassium and sodium ion channels. Moreover, we will see in Section~\ref{sec:Results} that problems of applying the freezing method as proposed in \cite{Beyn04} to excitable media are caused by the lack of coupling of the boundary and the interior when $D_v=0$.

We shall fix in our numerical simulations $D_v=0$, $a=0.75$ and $b=0.01$ and vary the excitability parameter $\epsilon$ if not stated otherwise.\\

\begin{figure}[htbp]
\centering
\subfigure[Finger developing into a spiral] 
{
    \label{fig:CurlingFinger}
    \includegraphics[width=5.9cm]{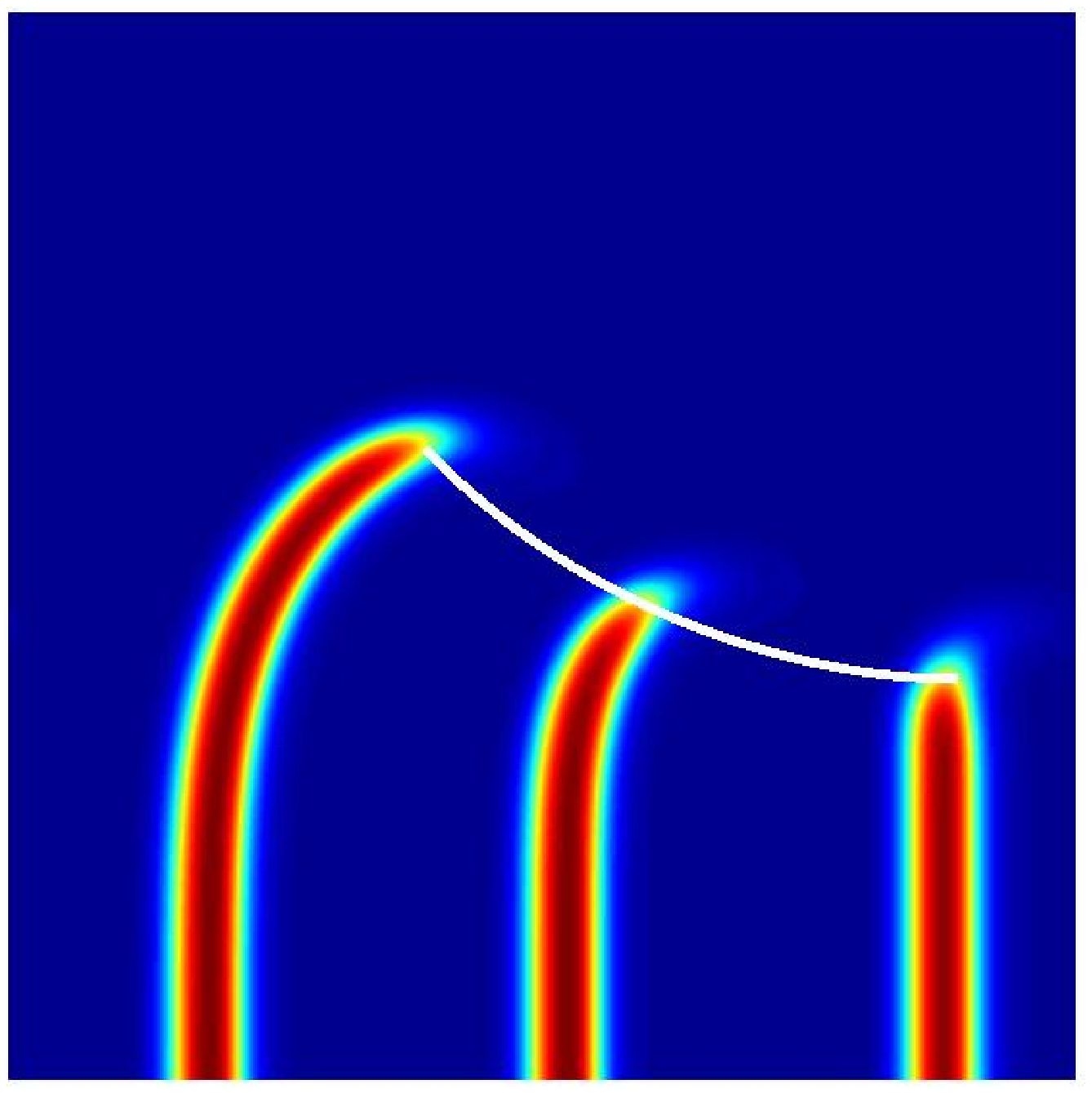}
}
\hspace{0.5cm}
\subfigure[Retracting finger] 
{
    \label{fig:RetractingFinger}
    \includegraphics[width=5.9cm]{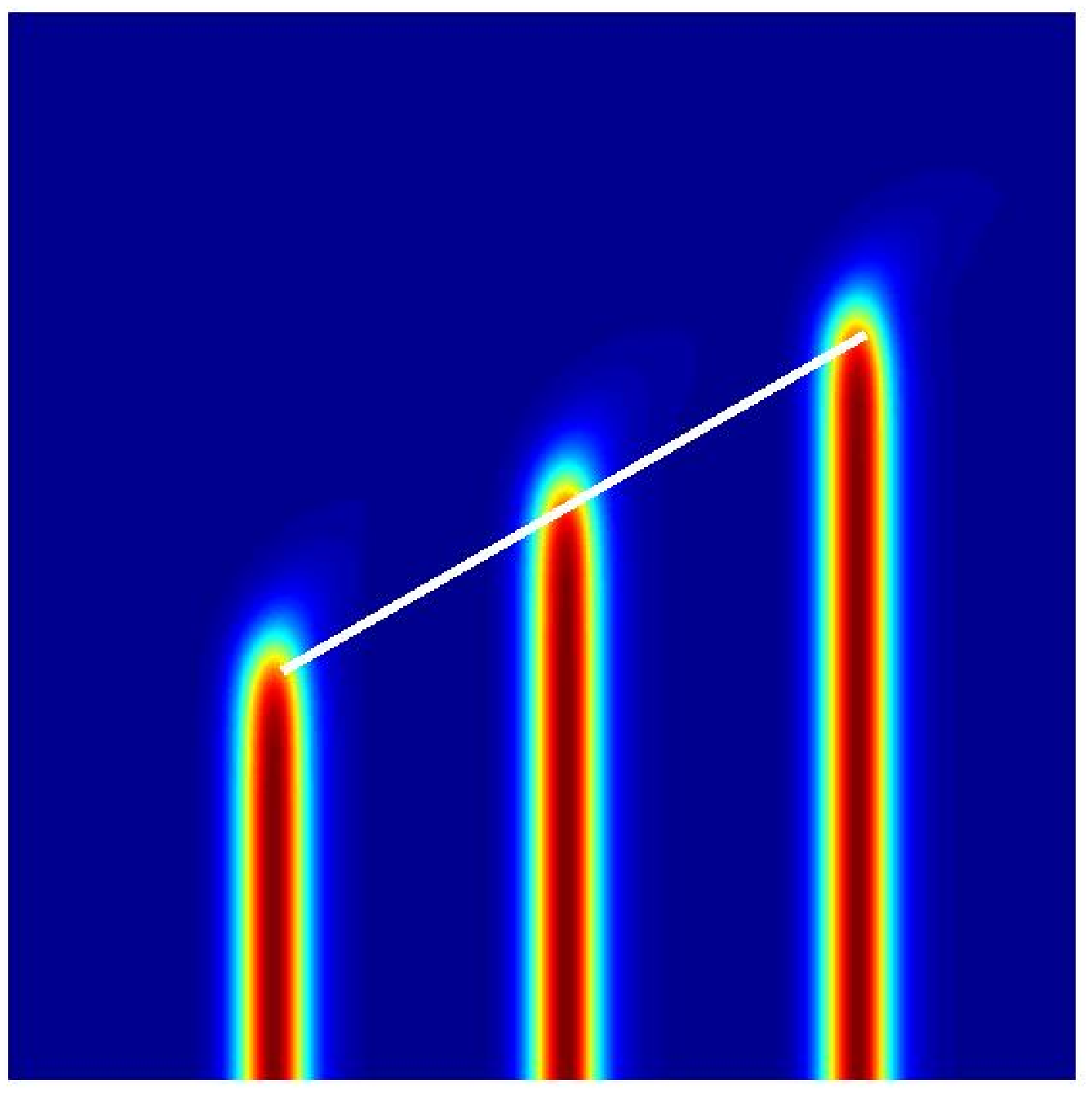}
}
\caption[Temporal evolution of a spiraling and a retracting finger.]{Temporal evolution of a retracting and spiraling finger solution $u$ of the Barkley model (\ref{eq:Barkley}) at 3 different time steps. The motion is clockwise for the spiraling finger and from right to left for the retracting finger. The white curve represents the trace of the tip as defined by (\ref{eq:DefinitionTip}).}
\label{fig:CurlingRetracting} 
\end{figure}

\noindent
The Barkley model supports, in a well-defined parameter region \cite{Barkley91}, rigidly rotating spirals. These spiral wave solutions are characterized by their rotation frequency $\omega$ and their core radius $r_c$. In the large core limit the rotation frequency approaches zero and the core radius becomes infinite at a critical value $\epsilon_c$ of the excitability parameter $\epsilon$. At criticality a rigidly rotating spiral wave becomes a finger propagating in the transverse direction only, with the speed of the corresponding travelling wave. Finger like initial conditions will curl up and eventually develop into rigidly rotating spirals below criticality, or into retracting fingers above criticality (see Figure~\ref{fig:CurlingRetracting}).
Whereas the rotation frequency is well defined for rigidly rotating spirals, the definition of the core radius is less clear (see \cite{Gray09} for a recent discussion). We define the core of a rigidly rotating spiral to be the maximal region which has never been excited during the course of one revolution of the spiral. To estimate this region we describe the boundary of the core as the trace of the tip during one revolution where the tip is defined 
as the intersection of the two contour lines
\begin{equation}\label{eq:DefinitionTip}
u^*=0.5 \quad \mbox{and} \quad v^*=\frac{a}{2}-b\, , 
\end{equation}
of the activator and the inhibitor, respectively \cite{Barkley91}. The value $v^\star$ solves ${\cal{F}}(u^*,v^*)=0$ with $u^*=0.5$. This definition ensures that the distance of the tip to the centre of the spiral is close to minimal. Note that this definition is arbitrary. In the large core limit, however, the value of the core radius $r_c$ calculated as the radius of the circle traced by the spiral tip does not vary much proportionally for different choices of the contour lines used to define the location of the tip.


\section{Freezing method}
\label{sec:NumericalMethod}
In this Section we briefly outline the so called \emph{freezing method} as introduced in \cite{Beyn04} before we propose our modifications to this method in Section~\ref{sec:BoundaryConditions}.

We consider reaction-diffusion systems on the plane,
\begin{equation}
\label{eq:ReactionDiffusionSystem}
  U_t=D\triangle U+f(U)=:F(U) \, ,
\end{equation}
with $U(z,t)=(U_1,\cdots,U_d)^T\in\mathbb{R}^d$, $d\geq 1$, $z=(x,y)^T\in\mathbb{R}^2$, $f:\mathbb{R}^d\mapsto\mathbb{R}^d$ and a diffusion matrix $D\in \mathbb{R}^{d\times d}$ with constant coefficients. The system (\ref{eq:ReactionDiffusionSystem}) is equivariant under the action of the special Euclidean group $SE(2)=S^1\ltimes\mathbb{R}^2$ consisting of rotations and translations. The Barkley model (\ref{eq:Barkley}) described in the previous section belongs to this class of equations with $d=2$. 

Equivariant systems of the form (\ref{eq:ReactionDiffusionSystem}) can be solved by the freezing method introduced in \cite{Beyn04} for equivariant systems. This beautiful numerical method utilizes the fact that the dynamics of equivariant systems can be decomposed into two parts: the dynamics on the symmetry group and the dynamics orthogonal to it. Equivariant systems can be cast into a skew product form whereby the dynamics on the symmetry group is driven by the so called shape dynamics which is orthogonal to the group dynamics. This idea was developed in the seminal paper \cite{Barkley94} and then put on a more rigorous footing in \cite{Krupa90,Fiedler96,Fiedler98,Golubitsky97,Sandstede97b,Sandstede97a,Sandstede99,Golubitsky00,Sandstede06}. The method can be applied for any symmetry group. Here we restrict the exposition of the method to equivariance with respect to the Euclidean group, and follow closely \cite{Beyn04,Thummler06}. For a more general framework of the method we refer to \cite{Beyn08,Beyn07,Thummler05,Thummler08a}.


\subsection{General Setup}
\label{subsec:GeneralSetup}
Let $\gamma=(\Theta,\beta)^T\in S^1\times\mathbb{R}^2$ be an element of the special Euclidean group $SE(2)$ consisting of the angle of rotation $\Theta$ and translation $\beta=(\beta_1, \beta_2)^T$. Two group elements are linked by the operation $\gamma^{(1)} \circ \gamma^{(2)}=(\Theta^{(1)}+\Theta^{(2)},\beta^{(1)}+\varrho_{\Theta^{(1)}}\beta^{(2)})$ with 
\begin{eqnarray*}
  \varrho_{\Theta}=\left(
  \begin{array}{cc}
    \cos{\Theta} & -\sin{\Theta}\\
    \sin{\Theta} & \cos{\Theta}\\
  \end{array}
  \right) \, .
\end{eqnarray*}
The action $a$ of the Euclidean group on functions $U\in \mathbb{R}^d$ can be defined by
\begin{equation}
\nonumber
  [a(\gamma)U](z)=U(\varrho_{-\Theta}(z-\beta)) \, ,\quad z\in\mathbb{R}^2 \, .
\end{equation}
The group action satisfies the properties
\begin{equation}
\nonumber
  a(e)=I\, ,\quad a(\gamma^{(1)}\circ\gamma^{(2)})=a(\gamma^{(1)})a(\gamma^{(2)})\, ,
\end{equation}
where $e$ is the unit element of $SE(2)$ and $I$ the identity matrix. Note that system (\ref{eq:ReactionDiffusionSystem}) is equivariant under the action of $SE(2)$, i.e. 
\begin{equation}
\label{eq:Equivariance}
  F(a(\gamma)U)=a(\gamma)F(U)\, .
\end{equation}
The invariance of equation (\ref{eq:ReactionDiffusionSystem}) with respect to the Euclidean group implies that it is possible to construct new solutions from a given solution by applying symmetry operations. In particular,  we can rewrite a solution $U(t)$ as 
\begin{equation}
\label{eq:SplitGroupMotion}
  a(\gamma(t))W(t)=U(t) \, ,
\end{equation}
with $W(t)\in \mathbb{R}^d$. By formally differentiating (\ref{eq:SplitGroupMotion}) with respect to $t$ we obtain, upon using the equivariance condition (\ref{eq:Equivariance}),
\begin{equation}
\nonumber
  U_t = [a_{\gamma}(\gamma)W]\gamma_t+a(\gamma)W_t = a(\gamma)F(W) \; ,
\end{equation}
which can be rearranged to yield
\begin{eqnarray}
\label{eq:FreezingFormal} 
  \nonumber W_t &=& F(W)-a(\gamma^{-1})[a_{\gamma}(\gamma)W]\gamma_t\\
         &=& F(W)-S(W,\gamma)\nu \; ,
\end{eqnarray}
where we used $a(\gamma)^{-1}=a(\gamma^{-1})$ and set $\nu=\gamma_t$. 
The derivative $[a_{\gamma}(\gamma)W]$ of the group action with respect to $\gamma=(\Theta,\beta)\in S^1\times\mathbb{R}^2$ can be calculated as
\begin{eqnarray*}
  [a_{\gamma}(\gamma)W] \nu 
&=&
\sum_i \partial_{\gamma_i}[W(\rho_{-\Theta}(z-\beta))]\nu_i\\
&=&
-\nabla W \left(\varrho_{-\Theta}(z-\beta)\right) \varrho_{-\Theta}\varrho_{\frac{\pi}{2}}(z-\beta)\nu_1
-\nabla W \left(\varrho_{-\Theta}(z-\beta)\right) \varrho_{-\Theta}(\nu_2,\nu_3)^T\, ,
\end{eqnarray*}
where $\nu_1=\Theta_t=\omega$ and $(\nu_2,\nu_3)^T=\beta_t$. The gradient acts on vector-valued functions as $(\nabla W)_{ij}=\partial W_i/\partial z_j$. This yields the expression
\begin{eqnarray}\label{FreezingTerms}
  \nonumber 
  S(W,\gamma)\nu \mathrel{\mathop:}&=&
  -\nabla W  \left[\nu_1\varrho_{\frac{\pi}{2}}\, z+
    \varrho_{-\Theta}\left(\nu_2,\nu_3\right)^T\right]\\
  &=& (yW_x-xW_y)\nu_1-\left(W_x,W_y\right)\varrho_{-\Theta}\left(\nu_2,\nu_3\right)^T  \, .
\end{eqnarray}
Introducing new group variables $(\Theta,\alpha) \in  S^1\times\mathbb{R}^2$ by setting $\alpha=\varrho_{-\Theta}\beta$ and defining parameters $\mu_1=\nu_1$ and $(\mu_2,\mu_3)^T=\varrho_{-\Theta}(\nu_2,\nu_3)^T$ allows for an elimination of the group variable $\Theta$ in (\ref{FreezingTerms}). Using this transformation we write (\ref{FreezingTerms}) as
\begin{equation}\label{FreezingTermsNew}
  \widehat{S}(W)\mu 
  \mathrel{\mathop:}= (yW_x-xW_y)\mu_1-W_x\mu_2-W_y\mu_3 \, .
\end{equation}
The equations for the group variables $(\Theta,\alpha)$ are given by
\begin{eqnarray}
  \label{eq:FrozenBarkleyGroupVariablesRotation} \Theta_t&=&\mu_1 \quad \mathrm{with} \quad \Theta(0)=0 \, ,\\
  \label{eq:FrozenBarkleyGroupVariablesTranslation} \alpha_t&=&\mu_1\varrho_{\frac{\pi}{2}}\alpha +
  \left(\mu_2,\mu_3\right)^T \quad \mathrm{with} \quad \alpha(0)=0 \, .
\end{eqnarray}
For constant $\mu_i$ equation (\ref{eq:FrozenBarkleyGroupVariablesTranslation}) describes a rotation on a circle with centre at
\begin{equation}
\label{eq:Centre}
    (x_M,y_M)=\left(-\frac{\mu_3}{\mu_1},\frac{\mu_2}{\mu_1}\right),
\end{equation}
and radius of rotation
\begin{equation}
\label{eq:Radius}
    r_p=\sqrt{(x_p-x_M)^2+(y_p-y_M)^2}
\end{equation}
for some point $(x_p,y_p)$ on the circle. The core radius $r_c$ can be determined by applying equation (\ref{eq:Radius}) to the tip of the spiral (as defined by (\ref{eq:DefinitionTip})), using the centre of the core $(x_M,y_M)$ given by (\ref{eq:Centre}). The core radius can thus be calculated from the group parameters, which are calculated in the freezing procedure.\\

\noindent
So far, the path $\gamma(t)$ on the group in (\ref{eq:FreezingFormal}) is arbitrary. Therefore three additional degrees of freedom, equaling the dimension of the group $SE(2)$, exist. To close the system we fix the location on the group orbit, and augment the equations by a phase condition
\begin{equation}\label{Phasecondition}
  \Psi(W,\gamma)=0 \,,
\end{equation}
where the functional $\Psi$ maps into $\mathbb{R}^3$. The phase condition $\Psi$ can be chosen in several ways. For the time--dependent problem (\ref{eq:FreezingFormal}) we will use the condition 
\begin{equation}\label{MinimizingPhasecondition}
  \Psi_{\mathtt{min}}(W,\gamma)=\int_{\mathbb{R}^2}\widehat{S}(W)^TW_t\, dx \, dy \, ,
\end{equation}
which assigns the location on the group orbit by minimizing the temporal change of $\|W_t\|_2$ at each time step. Note that this condition is equivalent to requiring that $W_t$ is orthogonal to the group orbit at $W$. This condition allows us to determine the freezing parameters $\mu_i$, which are implicitly contained in (\ref{MinimizingPhasecondition}) through $W_t$. These values are then subsequently fed into the shape dynamics to update $W$ (see Section~\ref{subsubsec:NumericsFreezing} for details on the implementation).

When considering the stationary problem of (\ref{eq:FreezingFormal}), we will use the following slightly simpler phase condition
\begin{equation}\label{FixedPhasecondition}
  \Psi_{\mathtt{fix}}(W) =\int_{\mathbb{R}^2}\widehat{S}(W_0 )^T(W-W_0)\, dx \, dy \, ,
\end{equation}
which determines the location on the group orbit by minimizing the distance of $W$ to some template function $W_0$. The phase condition (\ref{FixedPhasecondition}) is independent of $\mu$, and can therefore not be used directly in the time-dependent setting with a semi-implicit Crank-Nicolson scheme. Note that in order to use $\Psi_{\mathtt{fix}}$ the template function $W_0$ must be sufficiently close to a solution of (\ref{eq:ReactionDiffusionSystem}). For further details on phase conditions see \cite{Beyn07,Foulkes10}.\\

\noindent
We summarize the closed system of partial differential algebraic equations (\ref{eq:FreezingFormal}), (\ref{FreezingTermsNew}) -- (\ref{eq:FrozenBarkleyGroupVariablesTranslation}) and (\ref{Phasecondition}). The group dynamics is given by
\begin{eqnarray}\label{eq:GroupDynamics}
    \Theta_t&=&\mu_1 \, ,\quad \Theta(0)=0 \, ,\\
    \alpha_t&=&\mu_1\varrho_{\frac{\pi}{2}}\alpha + \left(\mu_2,\mu_3\right)^T \, ,\quad \alpha(0)=0 \, ,
\end{eqnarray}
which is driven through the parameters $\mu_i$ by the shape dynamics
\begin{eqnarray}\label{eq:FreezingSystem}
    W_t&=&D\triangle W+f(W)-\widehat{S}(W)\mu \, ,\quad W(0)=W_0 \, ,\\
    0&=&\Psi(W,\gamma) \, ,
\end{eqnarray}
with $\widehat{S}(W)\mu$ given by (\ref{FreezingTermsNew}). This system is called the \textit{frozen system} since for rigidly rotating spirals its solution $W$ evolves into a stationary solution for $t\to \infty$.\\

\noindent
Our aim is to numerically determine the shape of rigidly rotating spiral wave solutions of the Barkley model (\ref{eq:Barkley}) and the corresponding core radius and rotation frequency. Note that the shape dynamics  (\ref{eq:FreezingSystem}) entirely determines the solution $W$ and the parameters $\mu_i$ which can then be used to determine the rotation frequency $\omega=\mu_1$ and the centre of the core via (\ref{eq:Centre}). To this end, it would be sufficient to solve the \textit{stationary} problem corresponding to (\ref{eq:FreezingSystem}), using for example a Newton solver (as done in \cite{Thummler08a}). However, the high-dimensionality of the stationary problem, needed for an accurate resolution of the solution, requires a sufficiently good guess for the initialization of the Newton-Raphson method which otherwise would not converge. The initial guess for the Newton solver will be generated by the application of the freezing method for the time-dependent problem (\ref{eq:FreezingSystem}).


\subsection{Discretization}
\label{subsec:DynamicOrStationaryProblem}
We implement the freezing method for the Barkley model (\ref{eq:Barkley}) with $W=(u,v)^T$ for Cartesian coordinates and for polar coordinates. The choice of the coordinate system depends on the parameter range and the specific situation. Cartesian coordinates are better suited for the investigation of the large core limit where a finger-like solution can intersect the boundary almost perpendicularly and Neumann boundary conditions are a good approximation for the unbounded domain. Polar coordinates are better suited for small core radii where several revolutions of the spiral are usually inside the computational domain.\\

\noindent
Simulations in Cartesian coordinates $(x,y)$ are performed on a rectangular domain $[0,L_x]\times [0,L_y]$ with $M\times N$ points and gridsizes $\Delta x=L_x/(M-1)$ and $\Delta y=L_y/(N-1)$. The shape dynamics reads for Cartesian coordinates as
\begin{eqnarray}\label{eq:FrozenBarkley}
    \partial_t u &=&  \Delta  u  + {\cal{F}}(u,v) 
                         + \nabla u \left[\mu_1\varrho_{\frac{\pi}{2}}\left(x,y\right)^T+
    \left(\mu_2,\mu_3\right)^T\right] \, ,\\
    \partial_t v &=& u- v + \nabla v \left[\mu_1\varrho_{\frac{\pi}{2}}\left(x,y\right)^T+
    \left(\mu_2,\mu_3\right)^T\right] \, ,\\
    0&=&\Psi(u,v,\mu) \, ,
\end{eqnarray}
with ${\cal{F}}(u,v) = u(1-u)(u-(v+b)/a)/\epsilon$. The group variables $(\Theta,\alpha)$ can be determined by equations (\ref{eq:GroupDynamics}). 

Simulations in polar coordinates $(r,\varphi)$ with $(x,y)=(r\cos(\varphi),r\sin(\varphi))$ are performed on a rectangular domain $[0,R]\times [0,2\pi-\Delta \varphi]$ consisting of $M\times N$ grid points with gridsizes $\Delta r=R/(M-1)$ and $\Delta \varphi=2\pi/N$. The analogue to (\ref{eq:FrozenBarkley}) reads as
\begin{eqnarray}\label{eq:FrozenBarkleyPolar}
    \partial_t u &=&  u_{rr}+\frac{1}{r}u_r+\frac{1}{r^2}u_{\varphi\varphi} + {\cal{F}}(u,v) 
    + \mu_1u_\varphi+(u_r,\frac{1}{r} u_\varphi) \varrho_{-\varphi}
    \left(\mu_2,\mu_3\right)^T \, ,\\
    \partial_t v &=& u- v+ \mu_1v_\varphi+(v_r,\frac{1}{r} v_\varphi) \varrho_{-\varphi}
    \left(\mu_2,\mu_3\right)^T \, ,\\
    0&=&\Psi(u,v,\mu) \, ,
\end{eqnarray}
and the dynamics of the group variables is again given by (\ref{eq:GroupDynamics}). \\

\noindent
The temporal integration for both (\ref{eq:FrozenBarkley}) and (\ref{eq:FrozenBarkleyPolar}), is done in discrete time steps $\Delta t$. We denote the values of the fields $u$ and $v$ at the $n$-th time-step $t=n\Delta t$ and spatial location $((i-1)\Delta x,(j-1)\Delta y)$ (or $((i-1)\Delta r,(j-1)\Delta \varphi)$) with $i=1,\ldots,M$, $j=1,\ldots,N$ by $u_{i, j}^n$ and $v_{i, j}^n$, respectively. Spatial derivatives are evaluated using second-order central differences.


\subsubsection{Time-dependent freezing}
\label{subsubsec:NumericsFreezing}
We use a second order semi-implicit Crank-Nicolson scheme to solve the time-dependent problem (\ref{eq:FrozenBarkley}) (or (\ref{eq:FrozenBarkleyPolar})) whereby the linear terms are treated implicitly and the nonlinear term $\cal{F}$$(u,v)$ is treated explicitly with an Adams-Bashforth scheme \cite{Press92,Durran99}. The nonlinear freezing term can be rendered as an effectively linear term $\widehat{S}(\frac{W^{n+1}+W^n}{2})\mu^{n+1}$ to be included into the Crank-Nicolson part of the temporal discretization, by first obtaining $\mu^{n+1}$ through the phase condition 
\begin{equation}
\nonumber
\Psi(W^n,\mu^{n+1}) 
= 
\Psi_{\mathtt{min}}(W^n,\mu^{n+1})
=0 \; .
\end{equation}
The additional computational costs of solving the semi-implicit equations is far outweighed by the less restrictive time-step required for the semi-implicit method when compared to an explicit Euler method.\\

\noindent
In \cite{Beyn04} an explicit Euler scheme with Neumann boundary conditions was used which exhibited spurious spatial oscillations in both, the interior and at the boundary of the domain. An upwind-downwind scheme was introduced to control the oscillations with partial success for large excitabilities at small values of $\epsilon$ only. The semi-implicit scheme does not exhibit oscillations in the interior of the domain. However, spatial oscillations at the boundary persist. In Sections~\ref{subsec:SpiralBoundaryCondition} and \ref{subsec:OscillationFreeBC} we will present an explanation for the oscillations at the boundary, and provide a simple method to eliminate them. 


\subsubsection{Stationary freezing}
\label{subsubsec:RelativeEquilibria}
Travelling waves and spiral waves are both examples of relative equilibria \cite{Hoyle06,Golubitsky03}. Relative equilibria have the representation $U(t)=a(\gamma(t))\widetilde{W}$ with a time-independent $\widetilde{W}$. Hence the temporal evolution of a solution can be described entirely by the dynamics on the group. We may therefore find the spiral wave solutions of (\ref{eq:FrozenBarkley}) by solving its associated stationary problem
\begin{equation}\label{eq:FrozenBarkleyStationary}
  0=\mathbf{\hat{F}}(u,v,\mu)=\left(
  \begin{array}{c}
    \Delta  u  + {\cal{F}}(u,v) + \nabla u \left[\mu_1\varrho_{\frac{\pi}{2}}\left(x,y\right)^T+
    \left(\mu_2,\mu_3\right)^T\right]\\
    u- v + \nabla v  \left[\mu_1\varrho_{\frac{\pi}{2}}\left(x,y\right)^T+
    \left(\mu_2,\mu_3\right)^T\right]\\
    \Psi_{\mathtt{fix}}(u,v)
  \end{array}
  \right) 
\end{equation}
or, analogously for polar coordinates and (\ref{eq:FrozenBarkleyPolar})
\begin{equation}\label{eq:FrozenBarkleyStationaryPolar}
  0=\mathbf{\hat{F}}(u,v,\mu)=\left(
  \begin{array}{c}
    \Delta_{r,\varphi}u + {\cal{F}}(u,v) 
    + \mu_1u_\varphi+(u_r,\frac{1}{r} u_\varphi) \varrho_{-\varphi}
    \left(\mu_2,\mu_3\right)^T\\
    u- v+ \mu_1v_\varphi+(v_r,\frac{1}{r} v_\varphi) \varrho_{-\varphi}
    \left(\mu_2,\mu_3\right)^T\\
    \Psi_{\mathtt{fix}}(u,v)
  \end{array}
  \right)
\end{equation}
with $\Delta_{r,\varphi}u = u_{rr}+\frac{1}{r}u_r+\frac{1}{r^2}u_{\varphi\varphi}$. We solve the system for the $2MN+3$ unknowns $\left(u_{1,1},...,u_{M,N},v_{1,1},...,v_{M,N},\mu_1,\mu_2,\mu_3\right)$. Typically we use $M>200$ and $N>200$ for Cartesian coordinates and $M>150$ and $N>500$ for polar coordinates. To solve this high-dimensional nonlinear system we use the Newton-Raphson method with line searches and backtracking (see for example \cite{Press92}) with a terminating condition $0.5\,  \mathbf{\hat{F}}^T\, \mathbf{\hat{F}}\le10^{-10}$.


\section{Boundary Conditions}
\label{sec:BoundaryConditions}
In the introduction, two issues were described which, so far, prevented a successful stable application of the freezing method to excitable media. Firstly, numerical instabilities in the form of spatial oscillations spoiled results in the case of a non-diffusive inhibitor. In the previous Section we have eliminated spatial oscillations in the interior of the domain by using a semi-implicit Crank-Nicolson scheme in the time-dependent freezing problem (\ref{eq:FrozenBarkley}) (or (\ref{eq:FrozenBarkleyPolar})). Spatial oscillations near the boundaries, however, persist. Similarly, solving the stationary freezing problem (\ref{eq:FrozenBarkleyStationary}) (or (\ref{eq:FrozenBarkleyStationaryPolar})) leads to spatial oscillations near the boundary. Secondly, the usual Neumann boundary conditions do not reproduce the correct shape of a spiral in an unbounded domain. These problems are associated with the type of boundary condition used, and how they are formulated in the discretization.\\

\noindent
Naturally, computations are performed on a bounded domain and appropriate boundary conditions have to be chosen. The boundary conditions should preferably reflect the nature of the investigated system and its solutions. We are interested here in approximating spiral wave solutions in unbounded domains. By unbounded domains we mean either the infinite limit, where spiral wave solutions do not decay at infinity, or finite domains with boundary conditions but where the computational domain is much smaller than the actual physical domain and the physical boundaries can be ignored. In these cases the usual Neumann boundary conditions are, in general, not well suited. We discuss their impact on the freezing method in Section \ref{subsec:NeumannBoundaryCondition}. In Section \ref{subsec:SpiralBoundaryCondition} we present spiral boundary conditions and show how they can be implemented for freezing methods. The standard implementation of both these boundary conditions leads to oscillations of the inhibitor near the boundary. In Section \ref{subsec:OscillationFreeBC} we will therefore introduce an implementation for both Neumann and spiral wave boundary conditions, which does not exhibit spurious spatial oscillations.


\subsection{Neumann boundary conditions}
\label{subsec:NeumannBoundaryCondition}
Most work on spiral waves in excitable media uses either Dirichlet or Neumann boundary conditions (NBC) (e.g. \cite{Barkley91}). These boundary conditions are physically meaningful for simulations of excitable media on bounded domains, e.g. in chemical experiments. However, if simulations are performed with the intention to understand the behaviour of spirals in unbounded domains, they have the disadvantage of not respecting the underlying symmetry at the boundary. Nevertheless, Neumann boundary conditions have been used extensively. For simplicity we restrict our discussion to polar coordinates, for which Neumann boundary conditions are formulated as $u_r=0$ and $v_r=0$ at $r=R$. These are discretized according to
\begin{equation}
\label{eq:NBC}
  u_{M+1, j}=u_{M-1, j} \quad \mathrm{and} \quad v_{M+1, j}=v_{M-1, j}\; , 
\end{equation}
for $j=1,\cdots,N$. The values of $u_{M+1, j},\, v_{M+1, j}$ can then be used in the evaluation of the diffusion and advection terms on the boundary.
\begin{figure}[htpb]
\centering
\subfigure[NBC] 
{
    \label{fig:NegativeBoundaryEffectsContourCircular}
    \includegraphics[width=3.8cm]{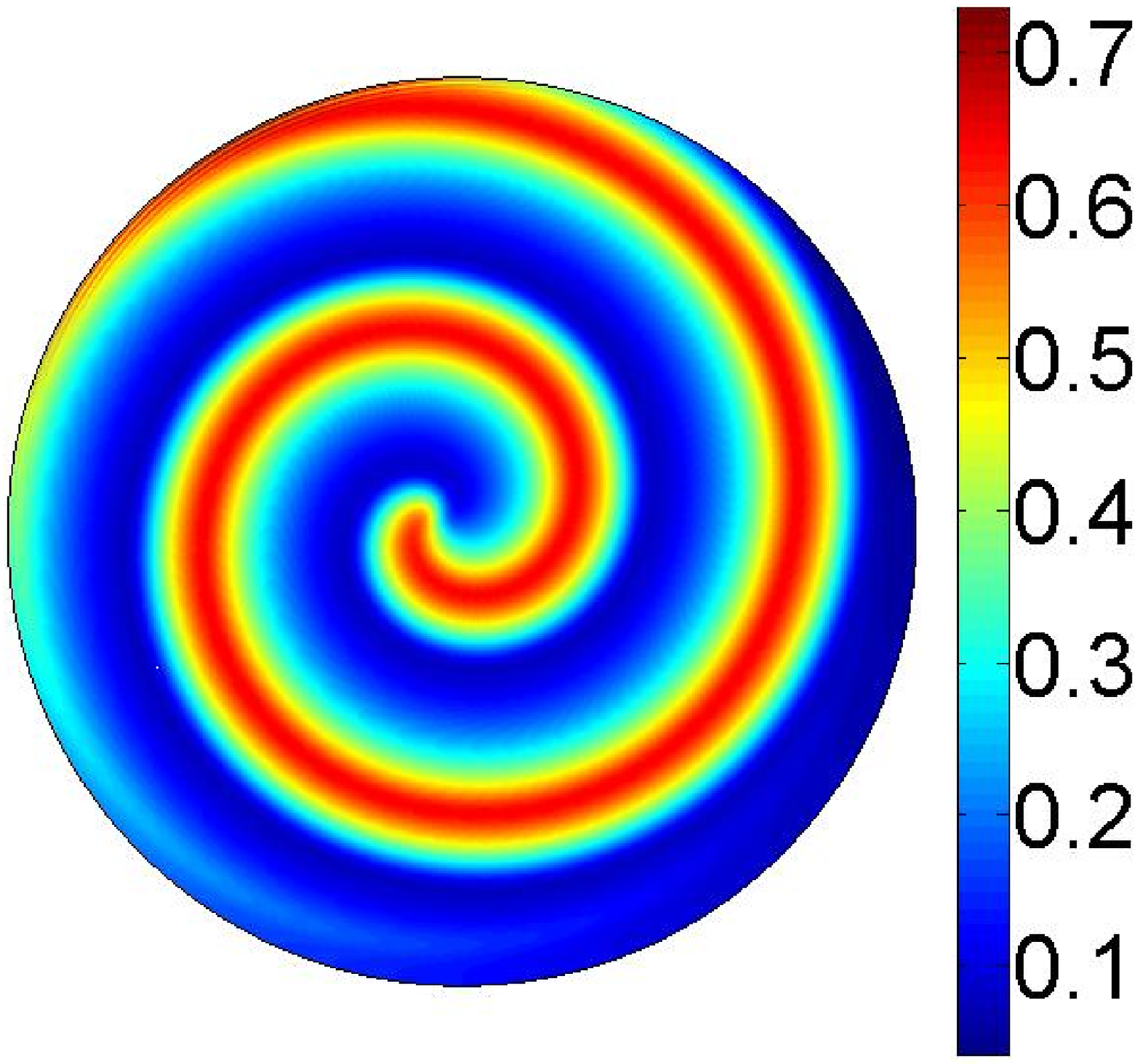}
}
\hspace{0.2cm}
\subfigure[] 
{
    \label{fig:NegativeBoundaryEffectsContourRectangular}
    \includegraphics[width=3.8cm]{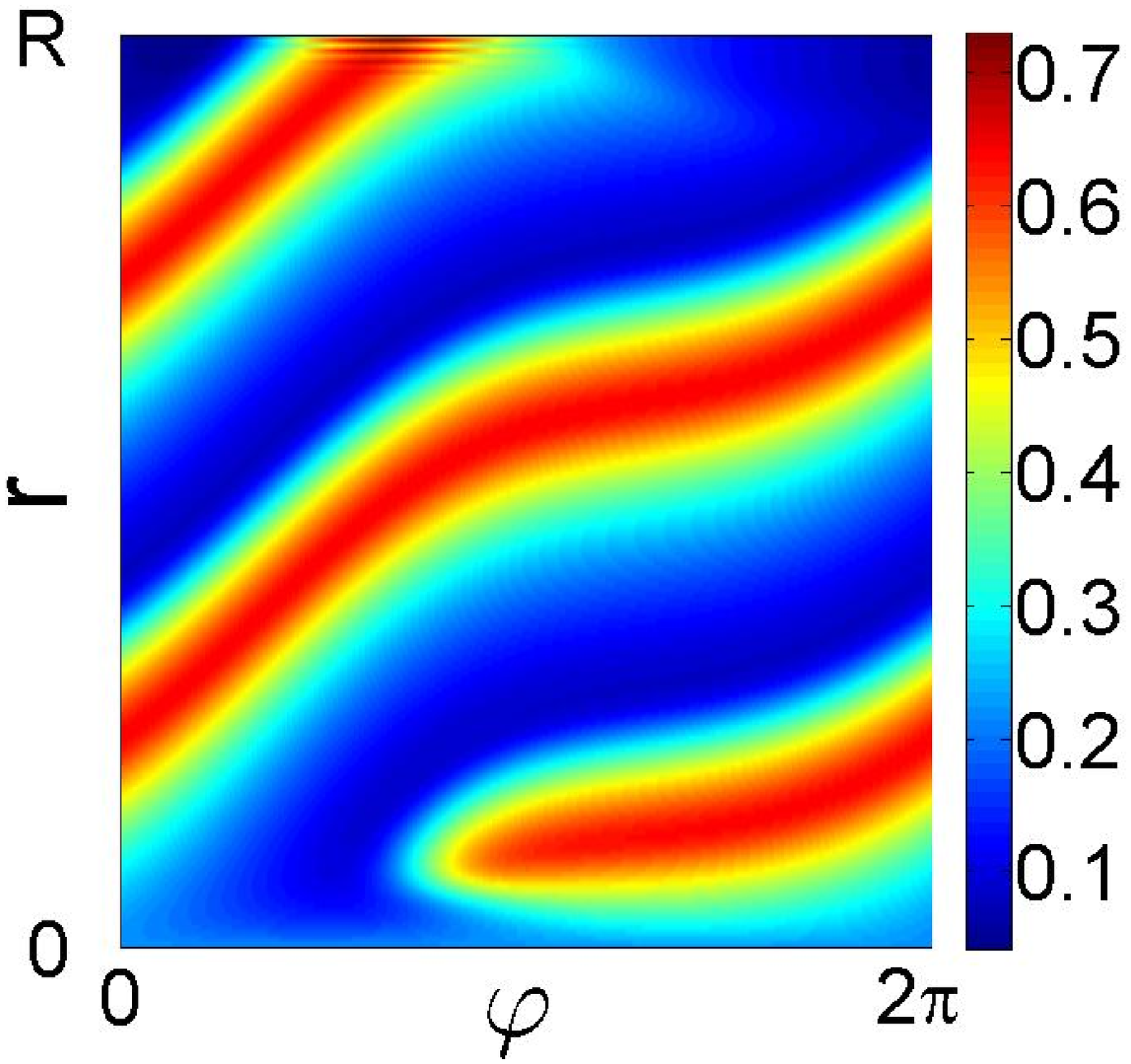}
}
\hspace{0.2cm}
\subfigure[] 
{
    \label{fig:NegativeBoundaryEffectsOscillationsContour}
    \includegraphics[width=3.8cm]{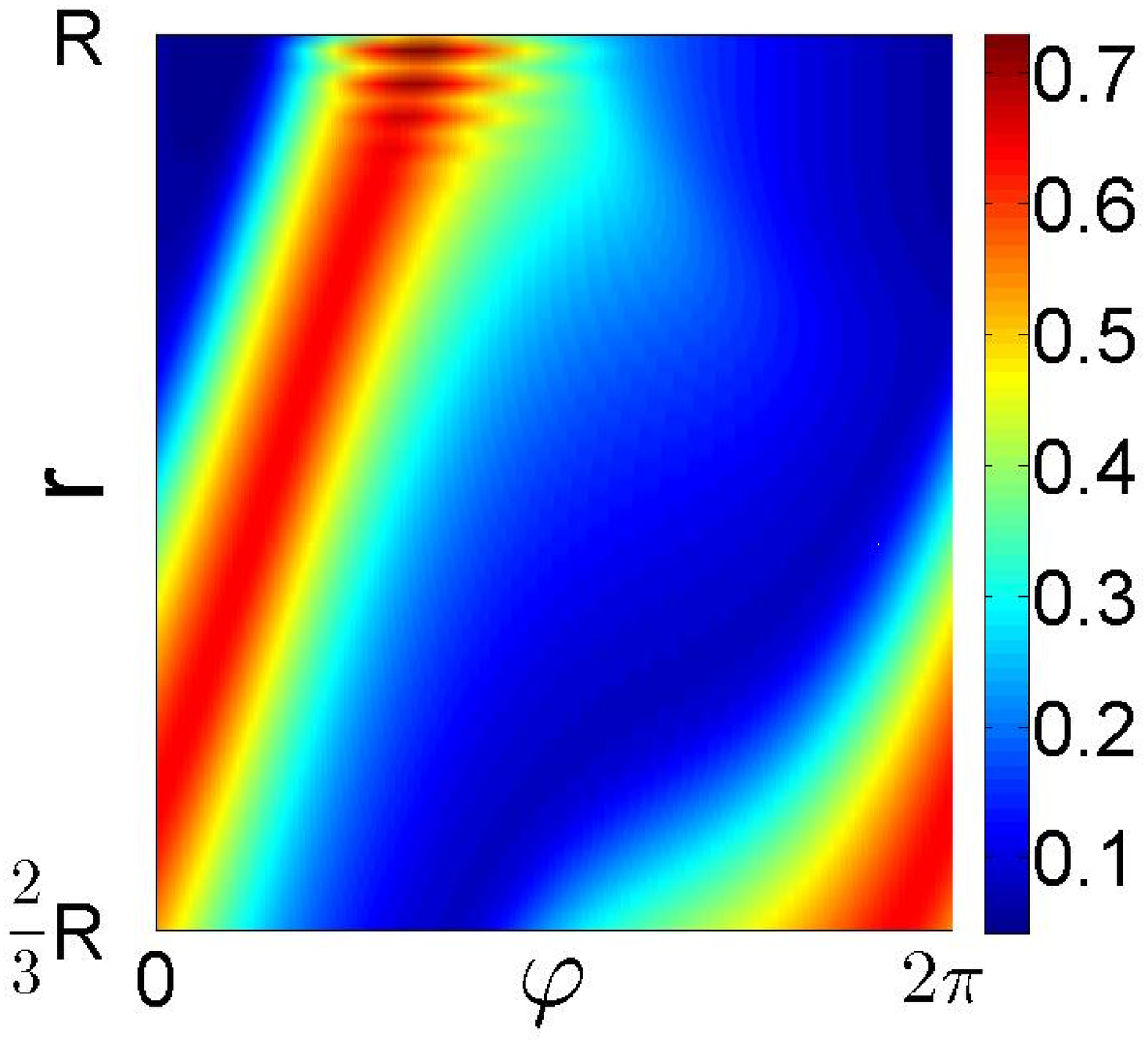}
}
\vspace{0.1cm}
\subfigure[SBC (\ref{eq:SBCCoefficientInvolute}) for $u$ and $v$] 
{
    \label{fig:NegativeBoundaryEffectsContourSBC_oldCircular}
    \includegraphics[width=3.8cm]{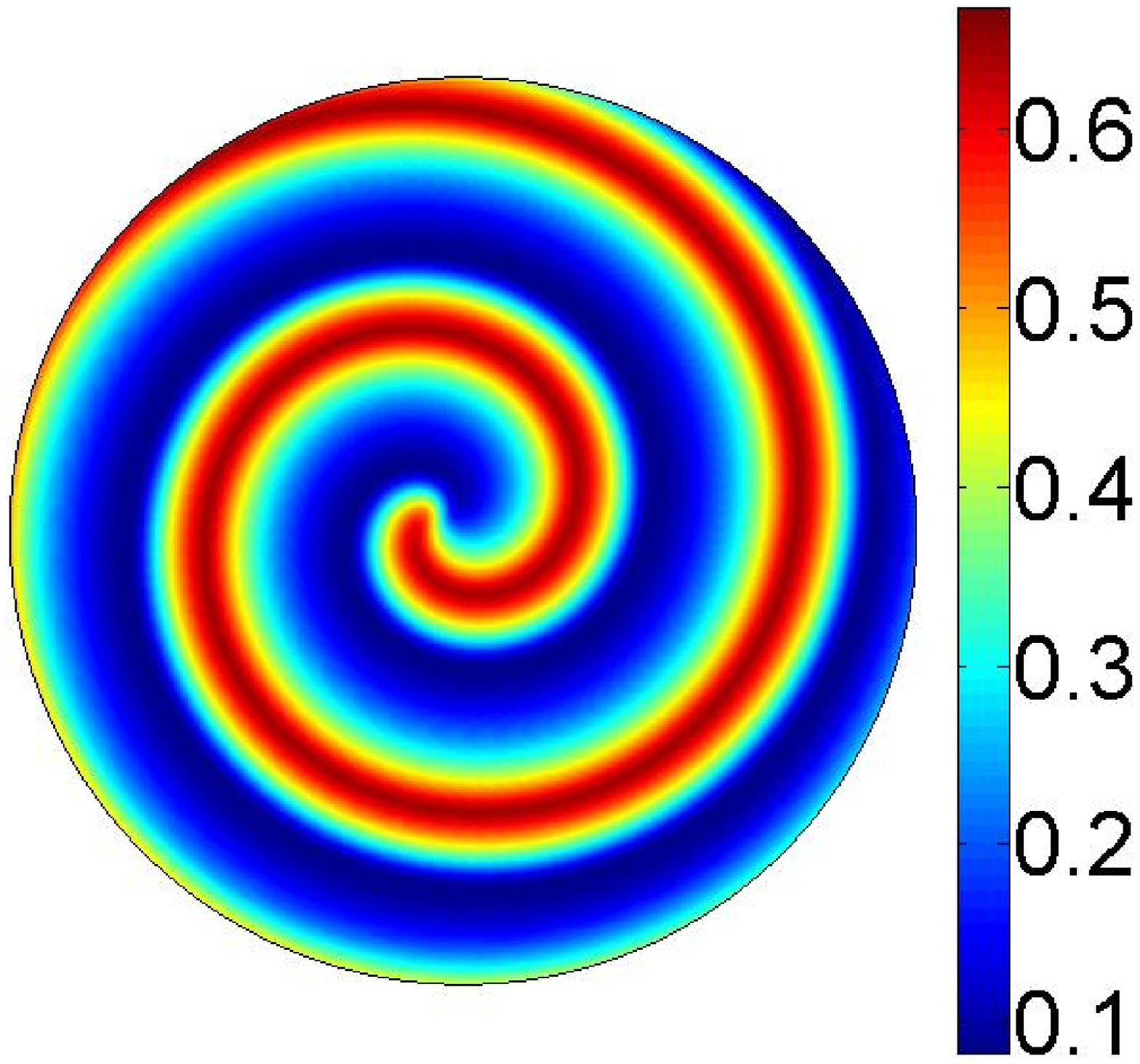}
}
\hspace{0.2cm}
\subfigure[] 
{
    \label{fig:NegativeBoundaryEffectsContourSBC_oldRectangular}
    \includegraphics[width=3.8cm]{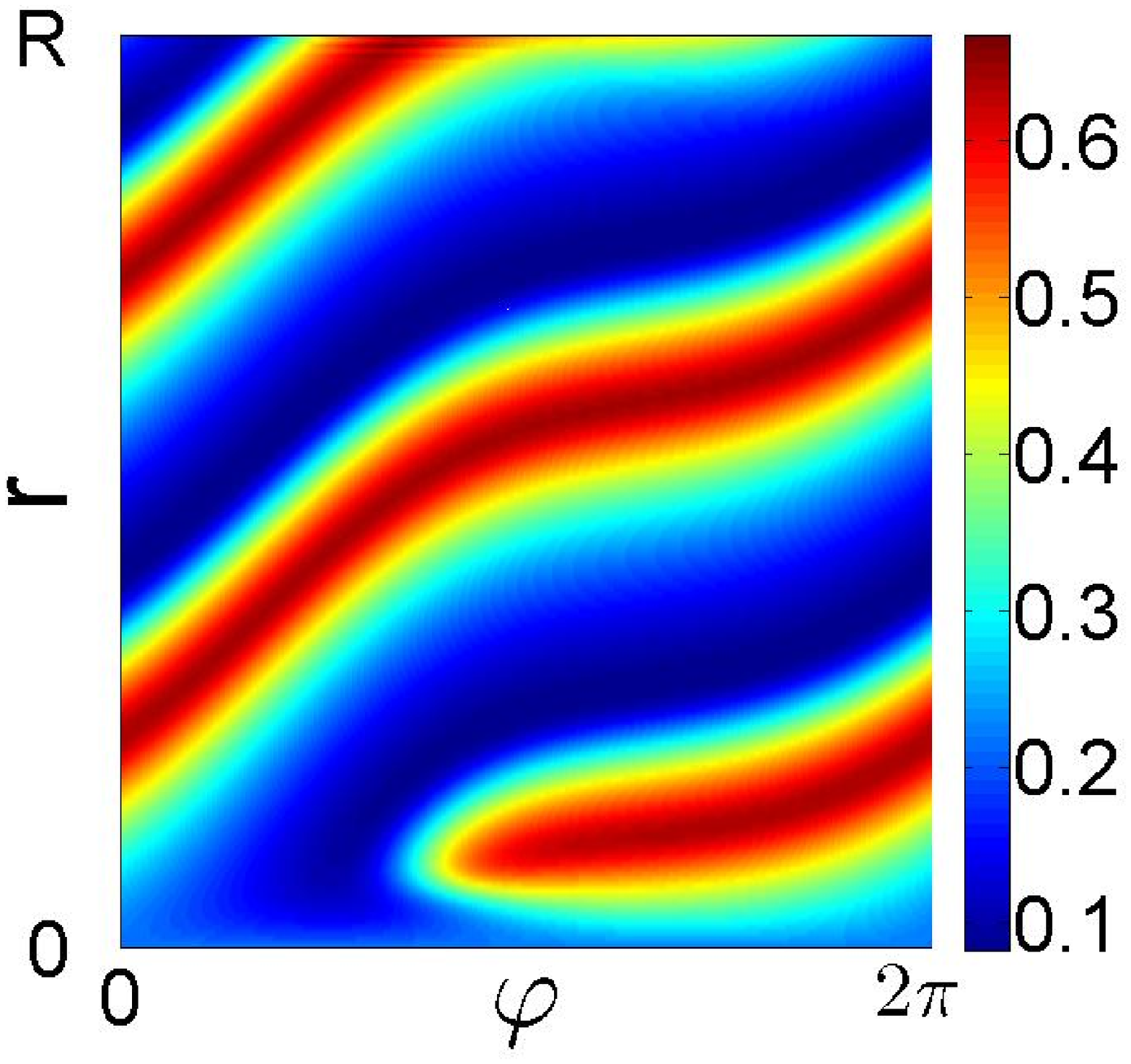}
}
\hspace{0.2cm}
\subfigure[] 
{
    \label{fig:NegativeBoundaryEffectsOscillationsSBC_oldContour}
    \includegraphics[width=3.8cm]{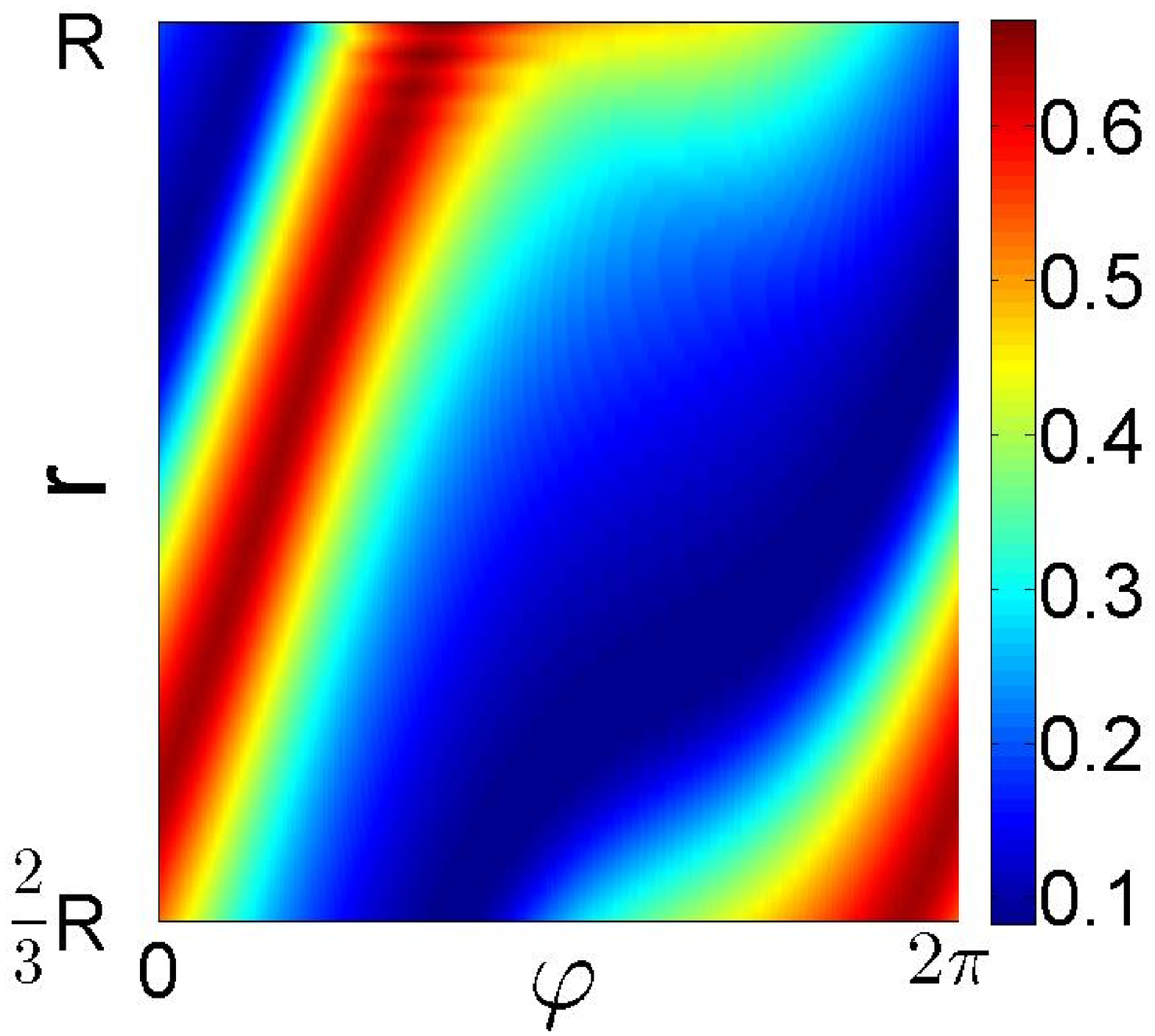}
}
\vspace{0.1cm}
\subfigure[SBC (\ref{eq:SBCCoefficientInvolute}) for $u$, and (\ref{eq:OneSidedBC}) for $v$] 
{
    \label{fig:NegativeBoundaryEffectsContourSBCCircular}
    \includegraphics[width=3.8cm]{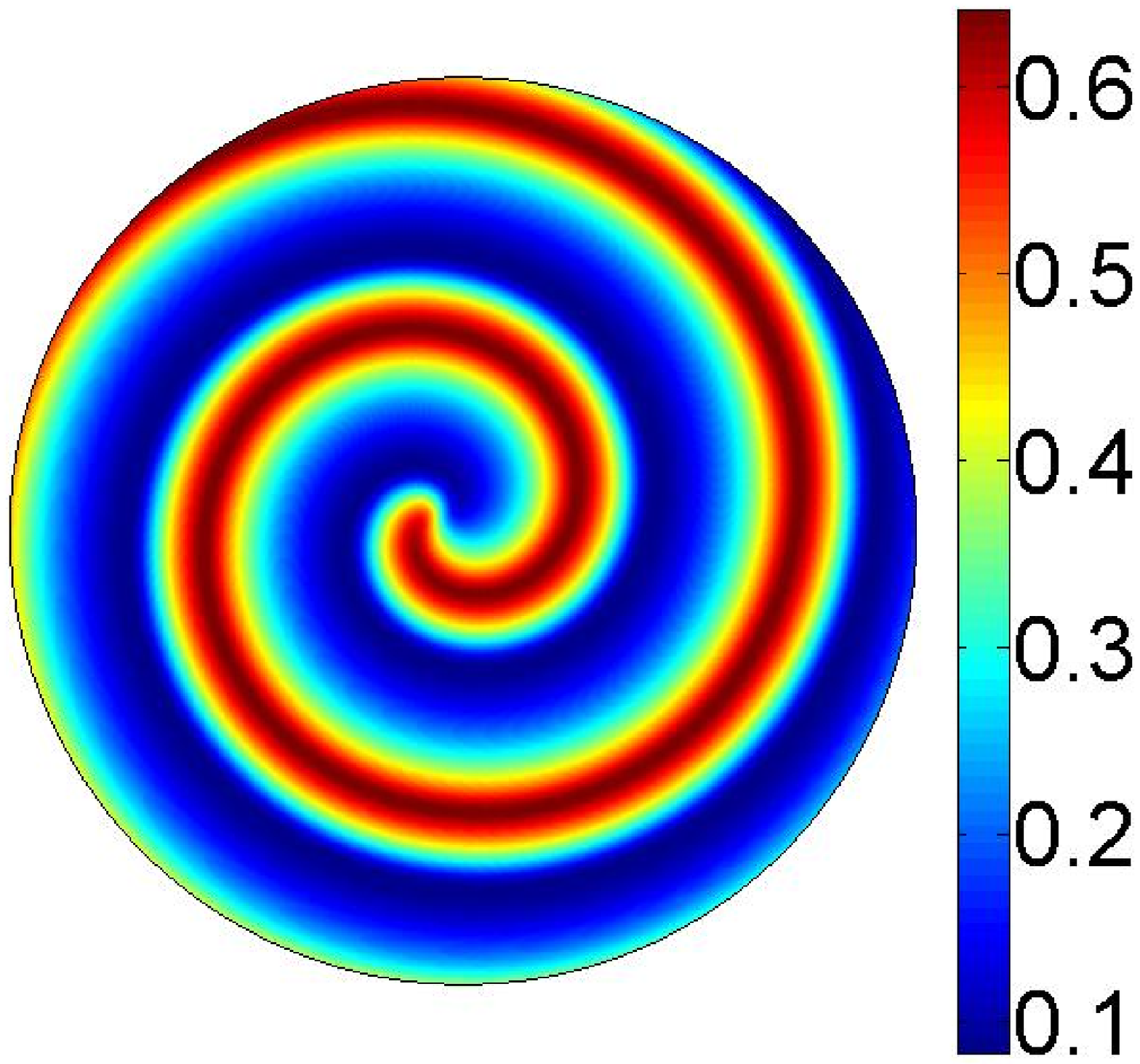}
}
\hspace{0.2cm}
\subfigure[] 
{
    \label{fig:NegativeBoundaryEffectsContourSBCRectangular}
    \includegraphics[width=3.8cm]{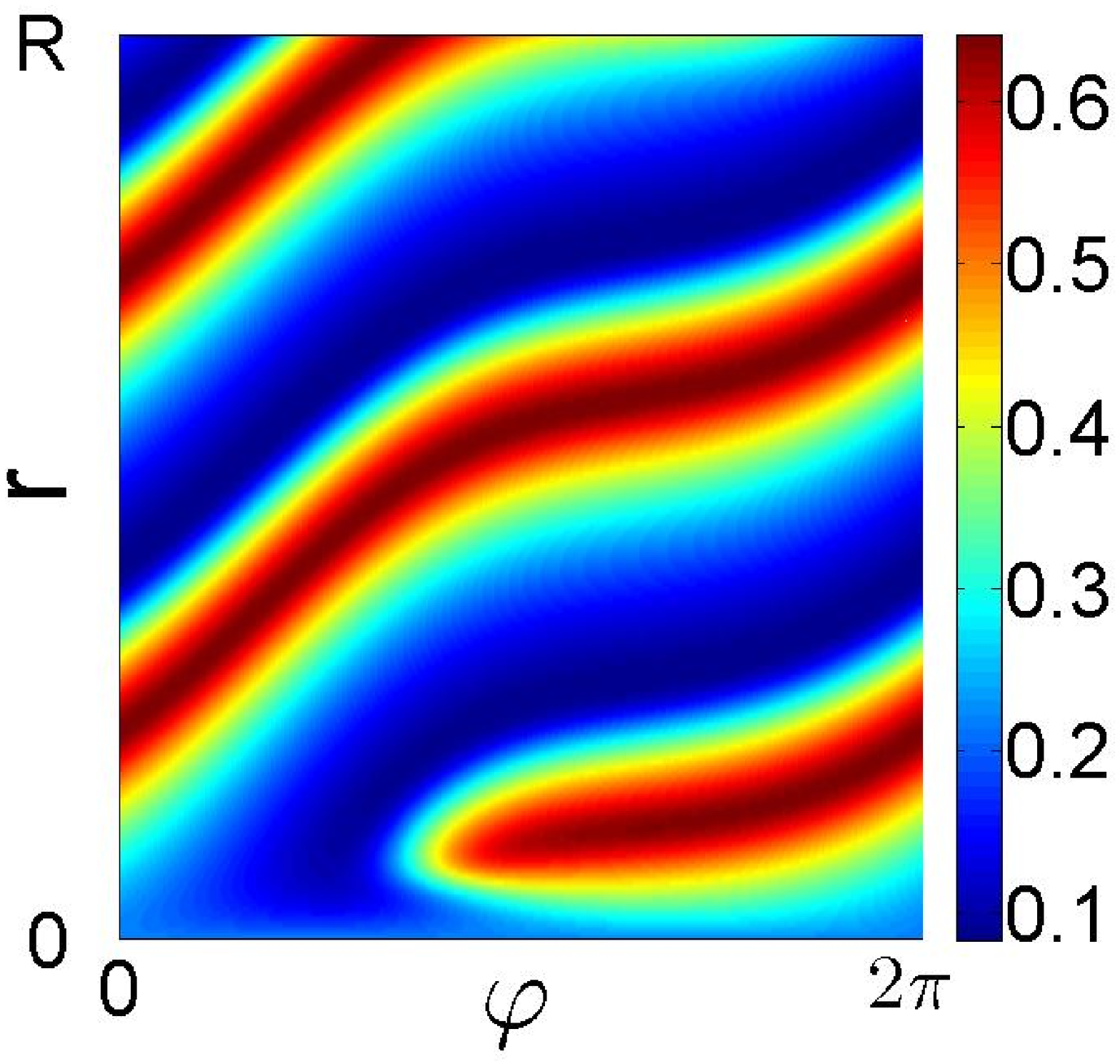}
}
\hspace{0.2cm}
\subfigure[] 
{
    \label{fig:NegativeBoundaryEffectsContourSBCCloseUp}
    \includegraphics[width=3.8cm]{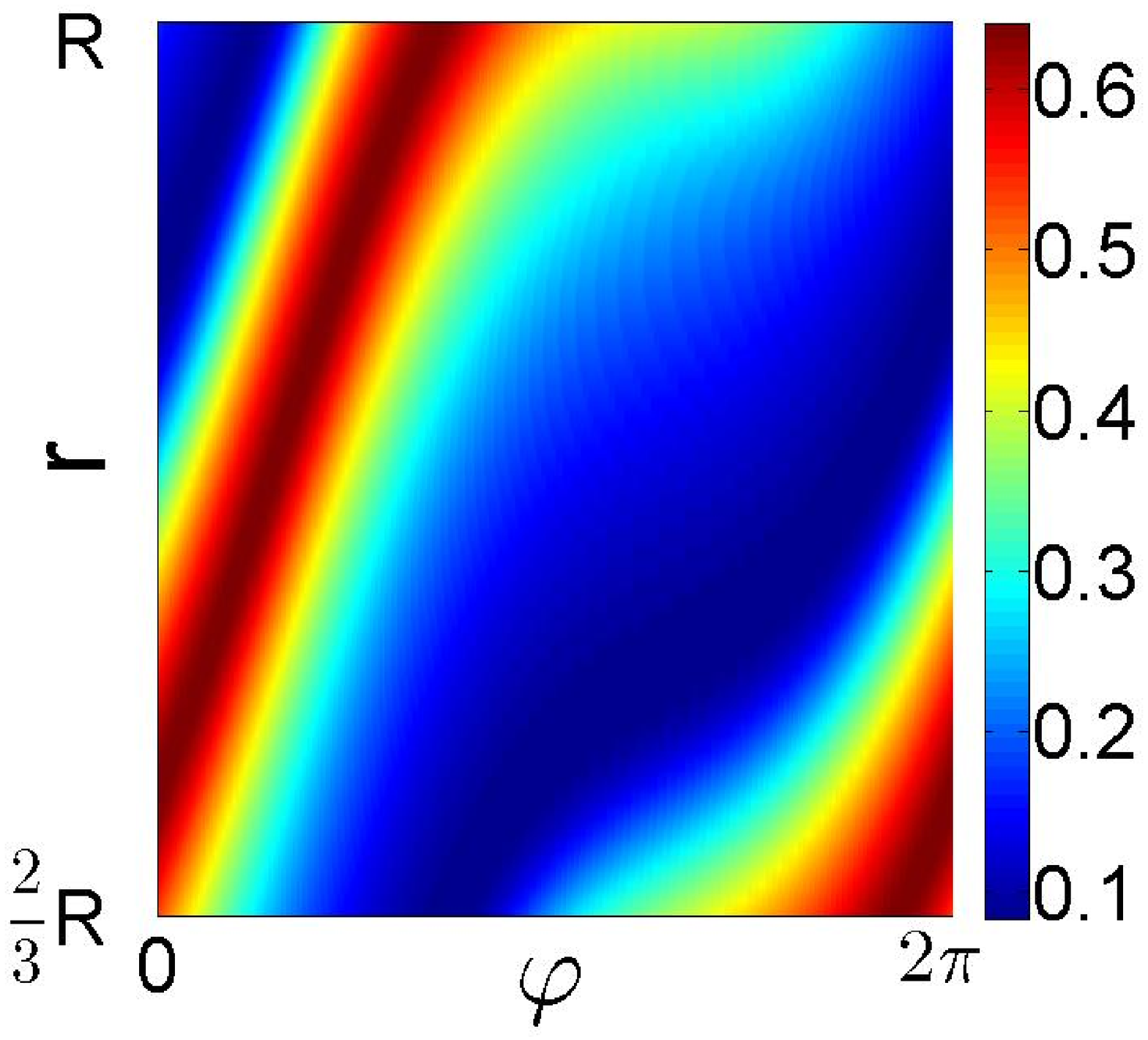}
}
\caption{Inhibitor $v$ of a frozen spiral wave solution calculated via the stationary problem (\ref{eq:FrozenBarkleyStationaryPolar}). From left to right: representation in the Cartesian plane $(x,y)=(r\cos(\varphi),r\sin(\varphi))$, in the polar $(r,\varphi)$-plane, and a close-up of the radial boundary. (a)--(c): Neumann boundary condition (NBC) (\ref{eq:NBC}), (d)--(f): spiral boundary condition (\ref{eq:SBCCoefficientInvolute}) for both, activator and inhibitor, (g)--(i): spiral boundary condition (SBC) (\ref{eq:SBCCoefficientInvolute}) for activator only and free boundary condition (\ref{eq:OneSidedBC}) for the inhibitor. We chose $\epsilon=0.025$ in the Barkley model (\ref{eq:Barkley}), and $R=21.74$, $\Delta r=0.1257$ and $\Delta \varphi=0.01$ for the spatial discretization.}
\label{fig:NegativeBoundaryEffectsContour} 
\end{figure}

In the two top left panels of Figure \ref{fig:NegativeBoundaryEffectsContour} we show contour plots of the inhibitor $v$ calculated by solving the stationary problem (\ref{eq:FrozenBarkleyStationaryPolar}) in polar coordinates. Figure \ref{fig:NegativeBoundaryEffectsContourCircular} shows the spiral solution on the circular domain with radius $R$ in the Cartesian plane $(x,y)=(r\cos(\varphi),r\sin(\varphi))$. In Figure \ref{fig:NegativeBoundaryEffectsContourRectangular} we show the same solution in the $(r,\varphi)$-plane. One sees clearly how the contours bend towards the boundary to satisfy the Neumann boundary condition $u_r=0$, $v_r=0$ at $r=R$. This kink is localized near the boundary and does not extend far into the domain. However, the bending of the spiral wave solution leads to an effective wider transversal cross-section of the solution near the boundary. A wider activator profile allows the inhibitor to adopt larger amplitudes (cf. the darker color (online red) of the inhibitor in the close--up in Figure \ref{fig:NegativeBoundaryEffectsOscillationsContour}). The boundaries, however, do not affect the dynamics near the core -- provided they are located sufficiently far away from the spiral wave tip -- and hence the presence of a kink in the solution near the boundary does not affect the values of the rotation frequency $\omega$ and the core radius $r_c$.

\begin{figure}[htpb]
\center{
\subfigure[]  
  {\label{fig:NegativeBoundaryEffectsOscillationsSliceNBC}
  \includegraphics[width=5.5cm]{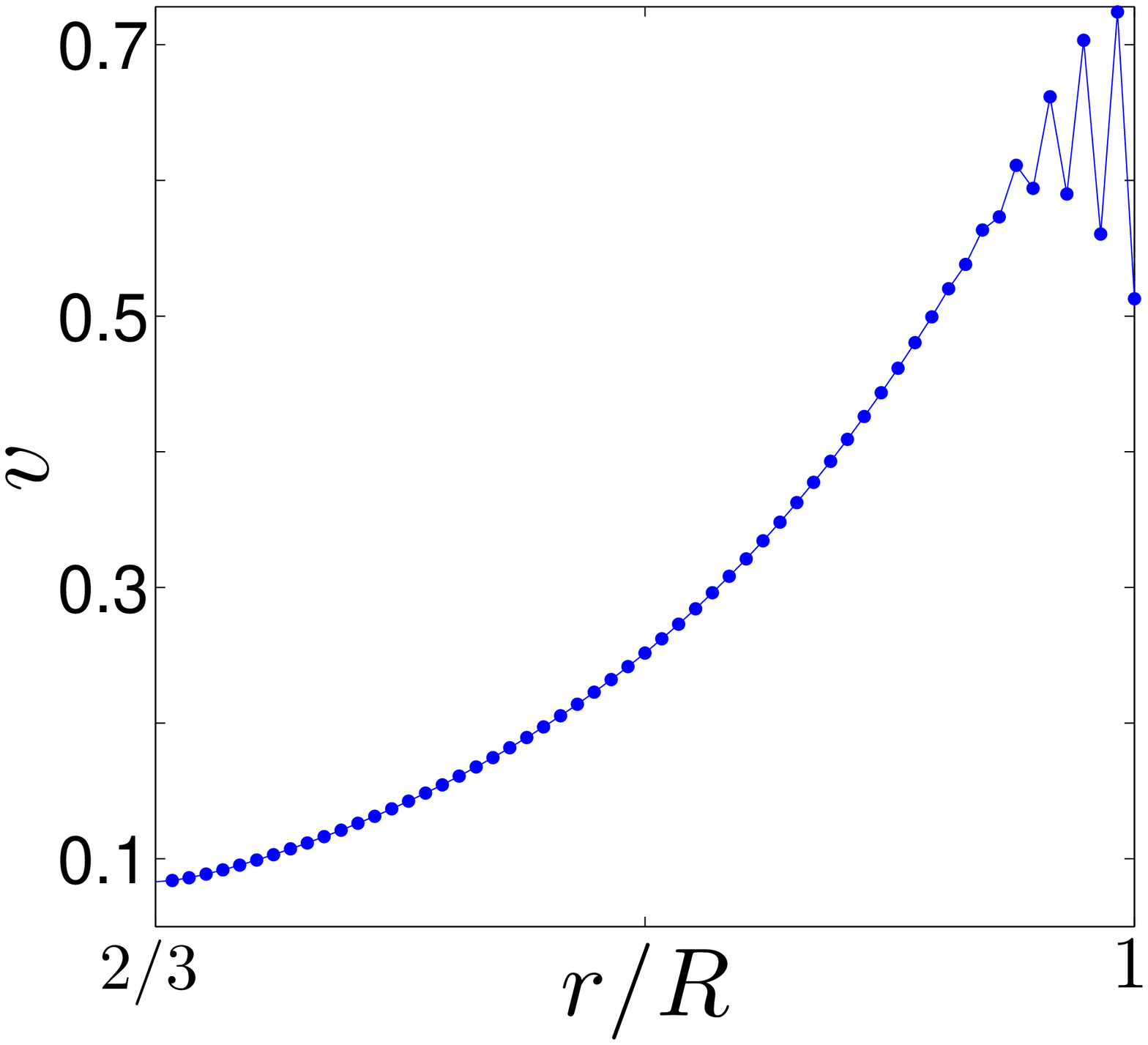}}
\subfigure[]  
  {\label{fig:NegativeBoundaryEffectsOscillationsSliceFREE}
  \includegraphics[width=5.5cm]{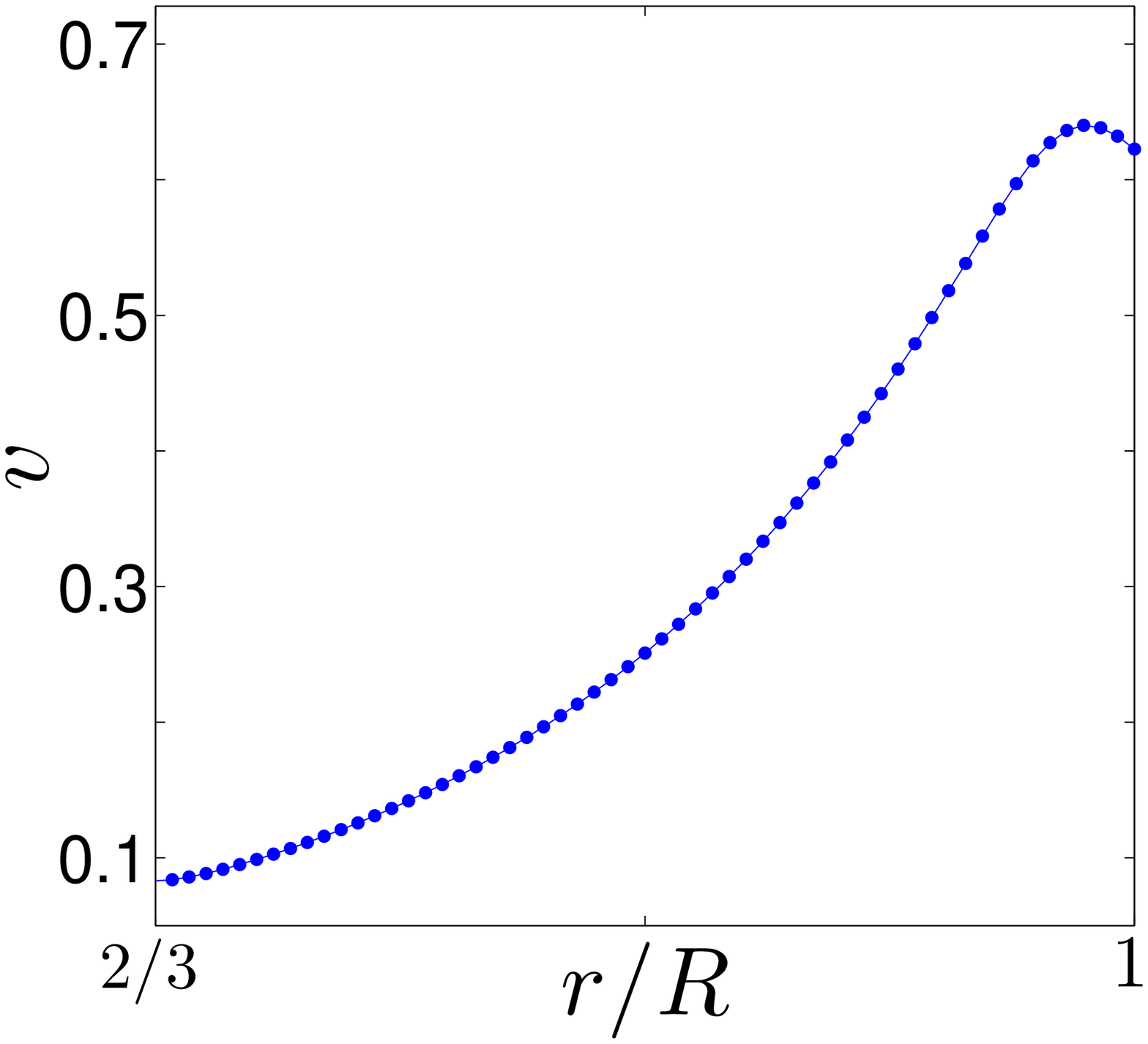}}
  \caption{Left: Slice of the close--up in Figure~\ref{fig:NegativeBoundaryEffectsOscillationsContour} along a radial grid line for the inhibitor solution $v$ where NBCs are used for activator $u$ and inhibitor $v$, demonstrating spurious spatial oscillations near the boundary. Right: The same slice but now with SBC for the activator $u$ and the one-sided derivative (\ref{eq:OneSidedBC}) as boundary condition for the inhibitor $v$, corresponding to Figure~\ref{fig:NegativeBoundaryEffectsContourSBCCloseUp}.}
  \label{fig:NegativeBoundaryEffectsOscillationsSlice}}
\end{figure}

More important, from a numerical stability point of view, is the following issue. In the close--up of the inhibitor near the boundary in \ref{fig:NegativeBoundaryEffectsOscillationsContour}, strong spatial oscillations are clearly visible. For a more detailed view we show a slice of the inhibitor near the boundary along the grid line $\varphi=3\pi/4$ in Figure \ref{fig:NegativeBoundaryEffectsOscillationsSliceNBC}. These oscillations are a well known problem of the freezing method for excitable media, and are associated with the non-diffusive nature of the inhibitor \cite{Beyn04,Thummler06,Beyn09}. These spatial oscillations occur only in the non-diffusive inhibitor $v$, and are absent for the diffusive activator $u$. We have checked that there are no oscillations for sufficiently large diffusion coefficient $D_v\neq 0$ of the inhibitor. 

The amplitude and the extent of the oscillations increase strongly for larger values of $\epsilon$, which prohibits an accurate numerical analysis of the large core limit. 

%


\subsection{Spiral wave boundary conditions}
\label{subsec:SpiralBoundaryCondition}
In this Section we introduce two boundary conditions based on simple geometrical approximations of spiral waves in order to mimic the behaviour of spirals in unbounded domains, or in finite domains when the size of the computational domain is much smaller than the physical domain. A natural version of this type of boundary conditions using Archimedean spirals was introduced in \cite{Dellnitz95}. Therein,  spiral wave solutions were ``grown" with a predefined wavelength $\lambda$. We will expand and generalize this idea with the aim to apply it to the freezing method. Here the wavelength is not known \textit{a priori} and, moreover, the centre of the spiral wave generally moves during the freezing procedure. In addition to the approximation by an Archimedean spiral, we will present boundary conditions where the spiral wave is approximated by an involute of a circle \cite{Winfree72}. The two approximations coincide in the far field but differ near the spiral wave tip. We will formulate spiral boundary conditions for polar and Cartesian coordinates. It has to be noted that both, the Archimedean spiral and the involute of a circle, are just geometrical approximations of contour lines of the actual spiral wave solution in unbounded domains. There exists, to our knowledge, no rigorous theoretical justification for these approximations. However, we will see that these approximations may serve as convenient constructs to formulate boundary conditions. 

\begin{figure}[t]
\centering
\subfigure[$\epsilon=0.025$] 
{
    \label{fig:ArchimedeanSpiral}
    \includegraphics[width=5.8cm]{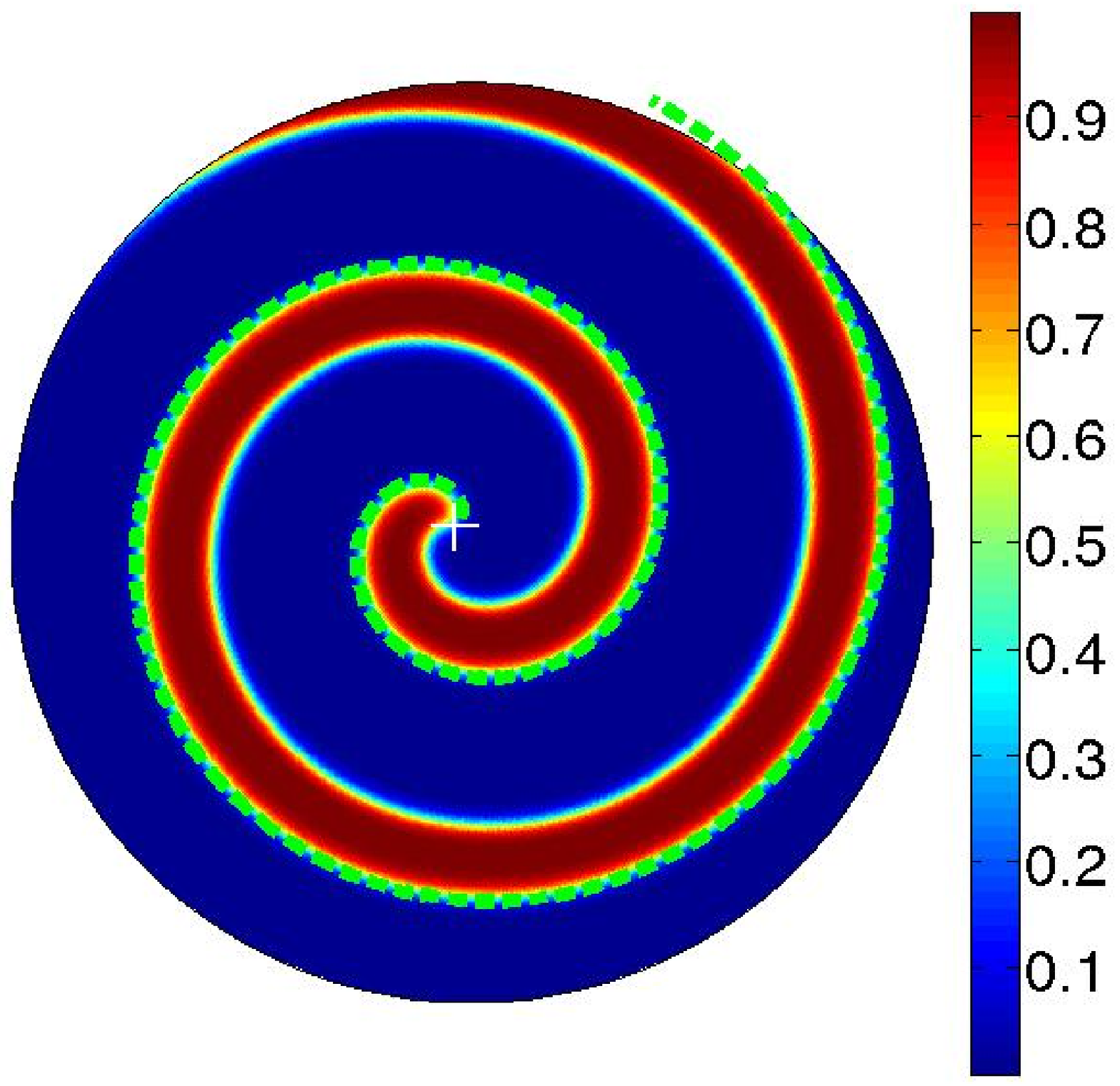}
}
\hspace{0.2cm}
\subfigure[$\epsilon=0.065$] 
{
    \label{fig:ArchimedeanSpiralAndInvolute}
    \includegraphics[width=6.0cm]{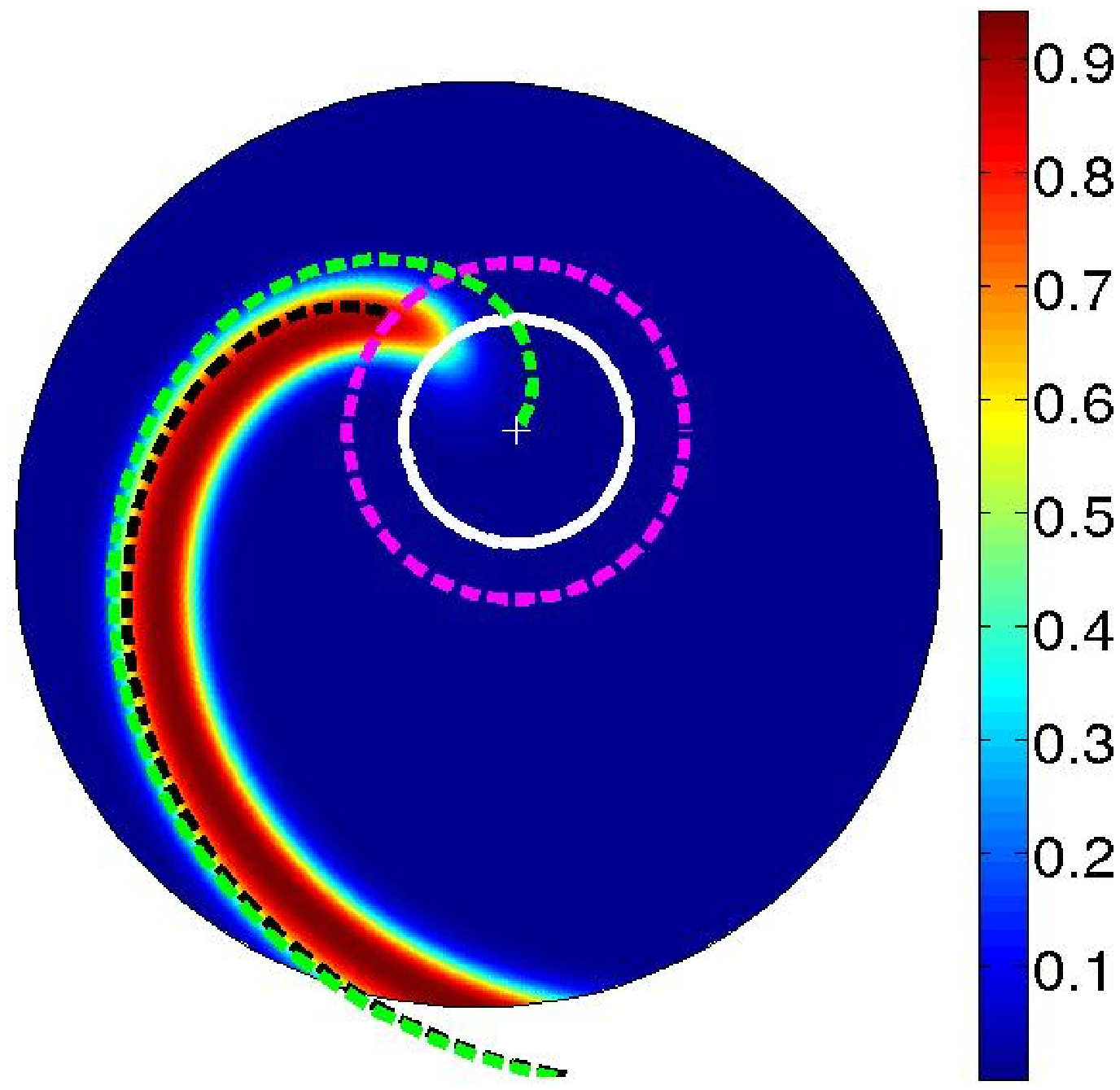}
}
\caption{Spiral wave solution $u$ calculated by solving the stationary frozen system (\ref{eq:FrozenBarkleyStationaryPolar}) on a circular domain of radius $R=21.74$. (a) Spiral solution for $\epsilon=0.025$ with a superimposed Archimedean spiral (dashed line; online: green) with $\widetilde{m}=2\pi/\lambda=0.59$. (b) Spiral solution for $\epsilon=0.065$ with superimposed Archimedean spiral (light dashed line; online: green) with $\widetilde{m}=2\pi/\lambda=0.13$ and superimposed involute of a circle (dashed black line) with $r_I=7.9$ (dashed circle; online: magenta). The smaller circle (solid line; online: white) with radius $r_c=5.3$ is the trace of the tip of the spiral wave $u$ defined in (\ref{eq:DefinitionTip}).}
\label{fig:FittedSpirals} 
\end{figure}

We assume a spiral wave solution $U(\tilde{r},\tilde{\varphi})$ with a constant far field wavelength $\lambda$, centred at the origin $(x_0,y_0)$ of the polar coordinate system $(\tilde{r},\tilde{\varphi})$. A common choice to describe such a spiral wave geometrically is by means of an Archimedean spiral. In this approximation contour lines of the spiral wave are given by  
\begin{equation}\label{eq:ArchimedeanSpiral}
  \tilde{\Phi}_s=\widetilde{m}\tilde{r}+\tilde{\varphi} = {\it const} \; ,
\end{equation}
allowing us to simplify $U(\tilde{r},\tilde{\varphi})=V_s(\widetilde{m}\tilde{r}+\tilde{\varphi})$. The parameter $\widetilde{m}$ is given by $\widetilde{m}=2\pi/\lambda$, which can easily be seen by noting that a spiral wave profile is $2\pi$-periodic with $V_s({\tilde \Phi}_s)=V_s({\tilde \Phi}_s+2\pi)$, and therefore $\widetilde{m}\tilde{r}+\tilde{\varphi}=\widetilde{m}(\tilde{r}+\frac{2\pi}{\widetilde{m}})+\tilde{\varphi}$. The approximation of a spiral wave solution of an excitable medium by an Archimedean spiral is fairly accurate in the far field, away from the spiral wave core, but fails close to it. This is illustrated in Figure~\ref{fig:FittedSpirals}, where two examples of rigidly rotating spiral wave solutions of the Barkley model (\ref{eq:Barkley}) are shown for different excitabilities $\epsilon$. The spiral wave in Figure \ref{fig:ArchimedeanSpiral} rotates around a circular core which is small compared to the computational domain. The contour lines of this spiral wave coincide well with a fitted Archimedean spiral (light dashed line; online: green). In comparison, for higher values of $\epsilon$, when the core of the spiral wave solution is larger, as depicted in Figure~\ref{fig:ArchimedeanSpiralAndInvolute} (smaller dashed circle; online: red), the contour lines of a spiral wave solution are approximated by an Archimedean spiral  (light dashed line; online: green) only sufficiently far away from the core. This effect worsens for larger core radii. The inability of Archimedean spirals to approximate spiral wave solutions near the core can be understood by considering that spiral waves possess locally an approximately constant velocity which is normal to the wave front. However, for uniformly rotating Archimedean spirals the radial velocity is constant. This is particularly problematic close to the origin where the normal direction of the wave is considerably different to the radial direction. Especially for numerical investigations in the large core limit, where numerical domains are not able to include several spiral wave coils, Archimedean spirals will not serve as good approximations within the computational domain.\\ \indent Note that logarithmic corrections to (\ref{eq:ArchimedeanSpiral}) have been considered (see for example \cite{Sandstede06}). However, these corrections become negligible in the far field near the boundary, and more importantly for our purpose here as boundary conditions, their derivatives with respect to the $\tilde r$ will be small at the boundary.\\

\noindent
Based on the assumption of a locally normal velocity for spiral waves, it was suggested in \cite{Wiener46,Winfree72,Lazar95} to approximate spiral waves by involutes of a circle. In this case contour lines of the spiral wave are given by  
\begin{equation}\label{eq:Involute}
\tilde{\Phi}_I =   \tilde{\varphi}+\arccos(r_I/\tilde{r})- \sqrt{(\tilde{r}/{r_I})^2-1} -\frac{\pi}{2} = {\it const}\; ,
\end{equation}
allowing us to simplify $U(\tilde{r},\tilde{\varphi})=V_I( \tilde{\varphi}+\arccos(r_I/\tilde{r})-\sqrt{(\tilde{r}/{r_I})^2-1}-\pi/2)$. The parameter $r_I$ denotes the radius of the circle that is revolved by the involute. The involute is only defined for $\tilde{r}\geq r_I$. The radius $r_I$ is somewhat arbitrary and we have to choose a proper definition that suits our application. We will later explain how to choose an appropriate $r_I$. Note that for $\tilde r\to \infty$ we have $\tilde{\Phi}_s-\tilde{\Phi}_I \to 0$, i.e. the two approximations of Archimedean spiral and involute of a circle coincide in the far field. In Figure~\ref{fig:ArchimedeanSpiralAndInvolute} we show how the approximation of an involute of a circle compares to the one by an Archimedean spiral.
In \cite{Mueller87} the two geometric approximations were compared in their ability to approximate experimental data of the Belousov-Zhabotinsky reaction. As the experimentally obtained spiral waves were rotating around small cores, both approximations showed equally good agreement. It is clear that both approximations fail close to the spiral tip \cite{Winfree01}, however, for our purpose, as illustrated in Figure~\ref{fig:ArchimedeanSpiralAndInvolute}, involutes of circles are better suited, especially for the large core limit.

We can use the approximations of an Archimedean spiral or of an involute of a circle, to formulate boundary conditions. Under these approximations we can use contour lines to express spiral wave solutions as 
\begin{equation}
\label{eq:uSpiralContours}
  U(\tilde{r},\tilde{\varphi})=V_{s,I}(\tilde{\Phi}_{s,I}(\tilde{r},\tilde{\varphi})) \, ,
\end{equation}
where $\tilde{\Phi}_{s,I}(\tilde{r},\tilde{\varphi})$ is given by equation (\ref{eq:ArchimedeanSpiral}) or (\ref{eq:Involute}), respectively. Differentiating (\ref{eq:uSpiralContours}) leads to
\begin{equation}\label{eq:SBCGeneral}
  U_{\tilde{r}}=\tilde{\alpha}(\tilde{r},\tilde{\varphi}) \, U_{\tilde{\varphi}} \, ,
\end{equation}
where $\tilde{\alpha}(\tilde{r},\tilde{\varphi})=\partial \tilde \Phi _{s,I}(\tilde r,\tilde \varphi)/\partial \tilde r$ is a coefficient depending on which spiral approximation has been chosen. For the Archimedean spiral we find
\begin{equation}\label{eq:SBCCoefficientArchimedean}
  \tilde{\alpha}(\tilde{r},\tilde{\varphi})=\widetilde{m} \, .
\end{equation}
Note that $\tilde{\alpha}$ is constant for Archimedean spirals. 
For the involute of a circle we find 
\begin{equation}\label{eq:SBCCoefficientInvolute}
  \tilde{\alpha}(\tilde{r},\tilde{\varphi})=-\frac{1}{r_I}\sqrt{1-\left(\frac{r_I}{\tilde{r}}\right)^2}\, .
\end{equation}
When evaluated at the boundary of the computational domain, we coin this type of boundary condition {\emph {spiral boundary condition}} (SBC). SBCs have the advantage that derivatives of a spiral wave solution $U$ on the boundary can be expressed entirely by known values of $U$ from inside the domain and from the boundary. Note that the geometric approximations can still be used to formulate SBCs, even, if close to the tip, they are not accurate.\\ 

\noindent
The spiral boundary conditions (\ref{eq:SBCGeneral}) are formulated within the coordinate system $(\tilde{r},\tilde{\varphi})$ which is the coordinate system with origin at the centre of the spiral wave $(x_0,y_0)$. This does, in general, not coincide with the polar grid of the computational domain. Assume computations are performed on a circular domain of radius $r=R$ in a coordinate system $(r,\varphi)$ with centre $(0,0)$. 

\begin{figure}[htpb]
  \center{
  \includegraphics[scale=0.35]{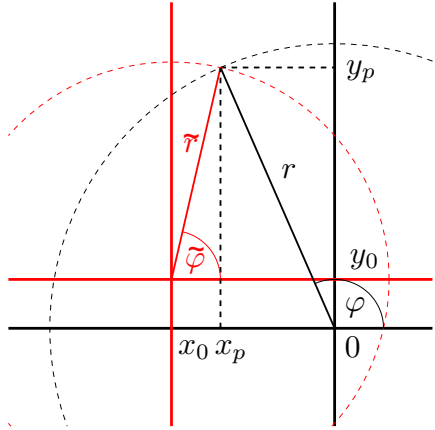}\\
  \caption{Coordinates of a point $(x_p,y_p)$ in two different polar coordinate systems $(\tilde{r},\tilde{\varphi})$ and    $(r,\varphi)$.}
  \label{fig:SpiralPolarGrid}}
\end{figure}

To complicate things, the centre of the spiral $(x_0,y_0)$ and the associated coordinate system $(\tilde{r},\tilde{\varphi})$ typically shift during the process of freezing. We therefore need to perform a coordinate transformation relating the two coordinate systems, $(\tilde{r},\tilde{\varphi})$ and $(r,\varphi)$, in order to express the spiral boundary conditions (\ref{eq:SBCGeneral}) in terms of the coordinate system of the computational domain $(r,\varphi)$. Using elementary trigonometric relations (see Figure \ref{fig:SpiralPolarGrid}) we can write
\begin{eqnarray}
\label{eq:tilde_r} 
  \tilde{r} &=& \sqrt{\left(r\cos\varphi-x_0\right)^2+\left(r\sin\varphi-y_0\right)^2} \, ,\\
\label{eq:tilde_phi}  
  \tilde{\varphi} &=& \arctan\left(\frac{r\sin\varphi-y_0}{r\cos\varphi-x_0}\right) \, .
\end{eqnarray}
We can now formulate SBCs on the actual computational domain by inserting the transformation into (\ref{eq:SBCGeneral}), to obtain
\begin{equation}
\label{eq:SBCTransformed}
  U_r = \alpha(r,\varphi)U_{\varphi} \, ,
\end{equation}
with
\begin{equation}\label{eq:AlphaTransformed}
  \alpha(r,\varphi) = \left(\frac{1}{r}\,\frac{\tilde{\alpha}(\tilde{r},\tilde{\varphi})\,\tilde{r}\cos(\varphi-\tilde{\varphi})+\sin(\varphi-\tilde{\varphi})}
  {-\tilde{\alpha}(\tilde{r},\tilde{\varphi})\,\tilde{r}\sin(\varphi-\tilde{\varphi})+\cos(\varphi-\tilde{\varphi})}\right)
   \, ,
\end{equation}
where $\tilde{\alpha}(\tilde{r},\tilde{\varphi})$ is given by (\ref{eq:SBCCoefficientArchimedean}) for Archimedean spirals and by (\ref{eq:SBCCoefficientInvolute}) for involutes of circles.\\
%


\subsubsection{Determination of the parameters of the spiral boundary condition}
\label{sec:detpara}
\noindent
The SBC requires knowledge of the location of the centre of the spiral wave $(x_0,y_0)$ and its wavelength in the case of the Archimedean spiral (\ref{eq:SBCCoefficientArchimedean}), and knowledge of the parameter $r_I$ for the case of the involute of a circle (\ref{eq:SBCCoefficientInvolute}). In the following we will explain how these parameters can be determined.

We start with the Archimedean spiral. The freezing method automatically determines the instantaneous centre of rotation of the spiral wave solution via (\ref{eq:Centre}). We may therefore determine $(x_0,y_0)$ on the fly during the freezing procedure as
\begin{equation}
\label{eq:Centre0M}
  (x_0,y_0)\approx (x_M,y_M) = \left(-\frac{\mu_3}{\mu_1},\frac{\mu_2}{\mu_1}\right)\, ,
\end{equation}
which becomes exact once the group parameters $\mu_i$ have become constant. Note that for rigidly rotating spirals $\mu_1=\omega\neq 0$ unless we are at the bifurcation from rigidly rotating spirals to retracting fingers. Approaching this limiting case, $(x_M,y_M)$ becomes larger, reflecting the increase of the core radius which diverges at criticality. If the tip were fixed at the origin, the core radius could be determined exactly by $r_c=\|(x_M,y_M)\|$ using only the group parameters. In this case the rotation frequency and the core radius are exactly inversely proportional to each other. In the large core limit, an offset of the spiral tip from the origin of the computational domain, however, becomes negligible as demonstrated later in Figure~\ref{fig:log(r)-log(2pi_cInfty_o_omega)}. See \cite{Foulkes10} where a phase condition is used which pins the tip at the centre of the domain.

The parameter $\widetilde{m}=2\pi/\lambda$ can be estimated on the fly as well during the freezing procedure.
For small core spirals we can do so by employing the dispersion relation of travelling wave trains \cite{Winfree91}. Before the application of the full $2$D-freezing procedure, the dispersion relation $c_w(\lambda)$ for the velocity of a travelling wave train is determined for each fixed $\epsilon$. This is done by applying the freezing method to the Barkley model (\ref{eq:Barkley}) in a $1$D periodic domain with length $\lambda$. Assuming that the 1D velocity $c_w(\lambda)$ is a good approximation for the normal velocity of the far field spiral wave coils, this velocity should equal the asymptotic far field velocity of a spiral wave $c_s=(\omega/2\pi)\lambda$, which is consistent with our geometric approximation of Archimedean spirals. The rotation frequency $\omega$ is determined during the freezing procedure and given simply by $\omega=\mu_1$. The wavelength $\lambda$ of the spiral wave is then obtained as the solution of $c_s(\lambda)=c_w(\lambda)$. For sufficiently large core radii, when the spiral wave coils do not interact with each other, we may replace $c_w(\lambda)$ by the 1D velocity of an isolated pulse $c_\infty$. The wavelength can then be estimated by $\lambda=2  \pi c_\infty /\omega$. The velocity $c_\infty$ can also be determined using a $1$D freezing procedure, or via direct simulations of the $1$D version of the Barkley model (\ref{eq:Barkley}), where the box length is chosen large compared to the decay length of the inhibitor.

For large core spirals, a simpler method can be used to determine $\lambda$. In this limit we can employ the approximation $\lambda=2\pi r_c$ which is independent of $c_\infty$ or the dispersion relation $c_w(\lambda)$,  which we would need to determine in advance. In Section~\ref{sec:lambda} we will present numerical results corroborating these approximations.\\

\noindent
The SBC involving the approximation of involutes of a circle requires that we specify the radius $r_I$ of the circle around which the tip revolves. We found that the involute of a circle with radius $r_I=c_{\infty}/\omega$ is a better approximation of contour lines of the actual spiral wave solution than the involute of a circle with the somewhat arbitrary radius $r_c$, in particular in weakly excitable media with large core radii.\\


\subsubsection{Applicability of the spiral boundary conditions}
\label{sec:SBCappl}
\noindent
In the following we investigate the applicability of the spiral boundary conditions (\ref{eq:SBCTransformed}) and (\ref{eq:AlphaTransformed}). There are two possible problems that can occur: Firstly, the case when there are grid points on the boundary at which $U_\varphi=0$ but $U_r\neq 0$, violating (\ref{eq:SBCTransformed}). Secondly, the expression for $\alpha(r,\varphi)$ in (\ref{eq:AlphaTransformed}) can possess singularities. We will derive conditions on the location of the spiral centre $(x_0,y_0)$ to ensure that the spiral boundary condition is applicable and both possible problems are avoided. For simplicity, we will use the example of the involute of a circle. 

At first we investigate, when there are points on the boundary such that $U_\varphi=0$ but $U_r\neq 0$. This is equivalent to asking whether there exist points on the boundary at which the involute is tangent to the circular boundary of the computational domain. The involute of a circle with radius $r_I$ can be described by
\begin{equation}
\nonumber 
  \nonumber \mathcal{I}(s)=r_I\vectwo{\cos s+s\sin s}{\sin s-s\cos s}+\vectwo{x_0}{y_0} \, , \quad s\in [0,\infty) \, .
\end{equation}
A parametric equation for the circular boundary reads as
\begin{equation}
\nonumber
  \nonumber \mathcal{C}(\sigma)=R\vectwo{\cos \sigma}{\sin \sigma} \, , \quad \sigma\in [0,2\pi) \, .
\end{equation}
Without loss of generality we may set $y_0=0$. A necessary and sufficient condition for a tangency at the boundary is given by the solution of the two equations $\mathcal{C}^T \mathcal{\dot{I}}=0$ and $\mathcal{C}=\mathcal{I}$, which is found to be $x_0\cos s=-r_I$. Hence, SBCs are violated whenever $|x_0|>r_I$.

In a next step we investigate the conditions for singularities of $\alpha(r,\varphi)$. Zeros of the denominator can be found for $\tan(\varphi-\tilde{\varphi}) = 1/(\tilde{r} \, \tilde{\alpha}(\tilde{r},\tilde{\varphi}))$. Using the trigonometric identity $\tan(\varphi-\tilde{\varphi})=(\tan\varphi-\tan\tilde{\varphi})/(1+\tan\varphi\tan\tilde{\varphi})$ and the coordinate transformation (\ref{eq:tilde_r}), (\ref{eq:tilde_phi}) we find $x_0\sin\varphi=r_I$, implying as before $|x_0|>r_I$. We note here without derivation, that for Archimedean spirals one finds $|x_0|>r_c$ \cite{HermannThesis}.\\

\noindent
In the small core limit, when the core radius is small compared to the computational domain, the centre of the core can be placed close to the origin of the computational domain, assuring $|x_0|<r_c$. If the core becomes larger upon increasing $\epsilon$, the centre of the spiral wave core will move outside the computational domain, if the spiral wave tip remains resolved within the domain. In this case the conditions $|x_0|<r_I$ can be satisfied, by placing the tip of the spiral wave solution close to the centre of the computational domain, which implies $|x_0| \approx r_c$. Recalling our definition $r_I=c_{\infty}/\omega$, we have
\begin{equation}\label{eq:r_I_geq_r_c}
  r_I=\frac{c_{\infty}}{\omega}>\frac{c_c}{\omega}=r_c \, ,
\end{equation}
where we define $c_c$ to be the velocity of the spiral tip tangential to the circle with radius $r_c$. Because of the positive curvature of the wavefront close to the core, this velocity is smaller than the velocity $c_\infty$ of the spiral coils in the far field with vanishing curvature. Hence, if the tip is placed close to the centre of the computational domain, $|x_0|<r_I$ is satisfied, assuring that the SBC is neither singular nor violated. 

In the large core limit, however, several problems arise, which cannot be resolved by an appropriate positioning of the spiral wave solutions within the computational domain. For the approximation with involutes of circles different problems may arise which need to be addressed. First, for large core spirals the involute circle (or the spiral core) intersects the boundary of the computational domain, and therefore there are parts of the boundary for which $\tilde{r}\le r_I$, and the SBC  (\ref{eq:SBCTransformed}) with (\ref{eq:AlphaTransformed}) and (\ref{eq:SBCCoefficientInvolute}) is not defined anymore. However, in this case we can formulate mixed boundary conditions. We keep SBCs on the part of the boundary where $\tilde{r}\geq r_I$, and we impose the NBC $U_r=0$ for the remainder of the boundary where $\tilde{r}< r_I$, i.e.
\begin{eqnarray}\label{eq:MixedBCPolar}
  U_r=\left\{\begin{array}{cc}
\alpha(r,\varphi)U_\varphi & {\rm{for}} \;\, \tilde{r}\geq r_I \\
\\
0 &  {\rm{for}} \;\, \tilde{r}<r_I
\end{array}
\right. \;,
\end{eqnarray}
where $\alpha$ is given by (\ref{eq:AlphaTransformed}) using (\ref{eq:SBCCoefficientArchimedean}) for the Archimedean spiral or (\ref{eq:SBCCoefficientInvolute}) for the involute of a circle. This is justified since, per definition, the core region of a rigidly rotating spiral is the part of the domain which is never excited by the spiral wave, with $U\approx 0$ and $U_r\approx0$. For the part of the domain which does not lie within the core region, but within the circle of radius $r_I>r_c$, we may also set $U=U_r= 0$ since in the large core limit, the fields decay fast enough. 

Approaching the critical point with $r_c \to \infty$, we inevitably reach the point when $r_I-r_c$ is larger than the domain size $R$. At this point none of the geometric approximations we discussed is valid anymore. In this case, NBCs, formulated in Cartesian coordinates, prove to be a good approximation, as the curvature of the spiral wave solution becomes negligible and the spiral wave appears to have the shape of a travelling finger. \\

\noindent
Analogously to (\ref{eq:SBCTransformed}) we can formulate spiral boundary conditions for Cartesian grids $(x,y)$. We perform a transformation between $(\tilde{r},\tilde{\varphi})$-coordinates of the spiral system and $(x,y)$-coordinates of the computational grid
\begin{eqnarray}
   \label{eq:cartpol}
    \tilde{r}&=&\sqrt{\left(x-x_0\right)^2+\left(y-y_0\right)^2}\, , \\
    \tilde{\varphi}&=&\arctan\left(\frac{y-y_0}{x-x_0}\right)\, .
 \end{eqnarray}
We find
\begin{equation}
\label{eq:SBCTransformed_x}
  U_x = \hat{\alpha}(x,y)U_y
\end{equation}
for boundaries parallel to the $y$-axis, and
\begin{equation}
\label{eq:SBCTransformed_y}
  U_y = \hat{\alpha}(x,y)^{-1}U_x
\end{equation}
for boundaries parallel to the $x$-axis. Here $\hat{\alpha}(x,y)$ is given by
\begin{equation}\label{eq:AlphaTransformed_xy}
  \hat{\alpha}(x,y) =
  \left(\frac{\tilde\alpha(\tilde{r},\tilde{\varphi})\,\tilde{r}\cos\tilde{\varphi}-\sin\tilde{\varphi}}
  {\tilde\alpha(\tilde{r},\tilde{\varphi})\,\tilde{r}\sin\tilde{\varphi}+\cos\tilde{\varphi}}\right)
   \, ,
\end{equation}
where ${\tilde \alpha}$ is given again by (\ref{eq:SBCCoefficientArchimedean}) for the Archimedean spiral and by (\ref{eq:SBCCoefficientInvolute}) for the involute of a circle. The variables $\tilde{r}$ and $\tilde{\varphi}$ are expressed as functions of $(x,y)$ using the coordinate transformation (\ref{eq:cartpol}). 

\begin{figure}[htb]
\centering
\subfigure[]
{
    \label{fig:ContourLargeCoreSBC}
    \includegraphics[width=5.9cm]{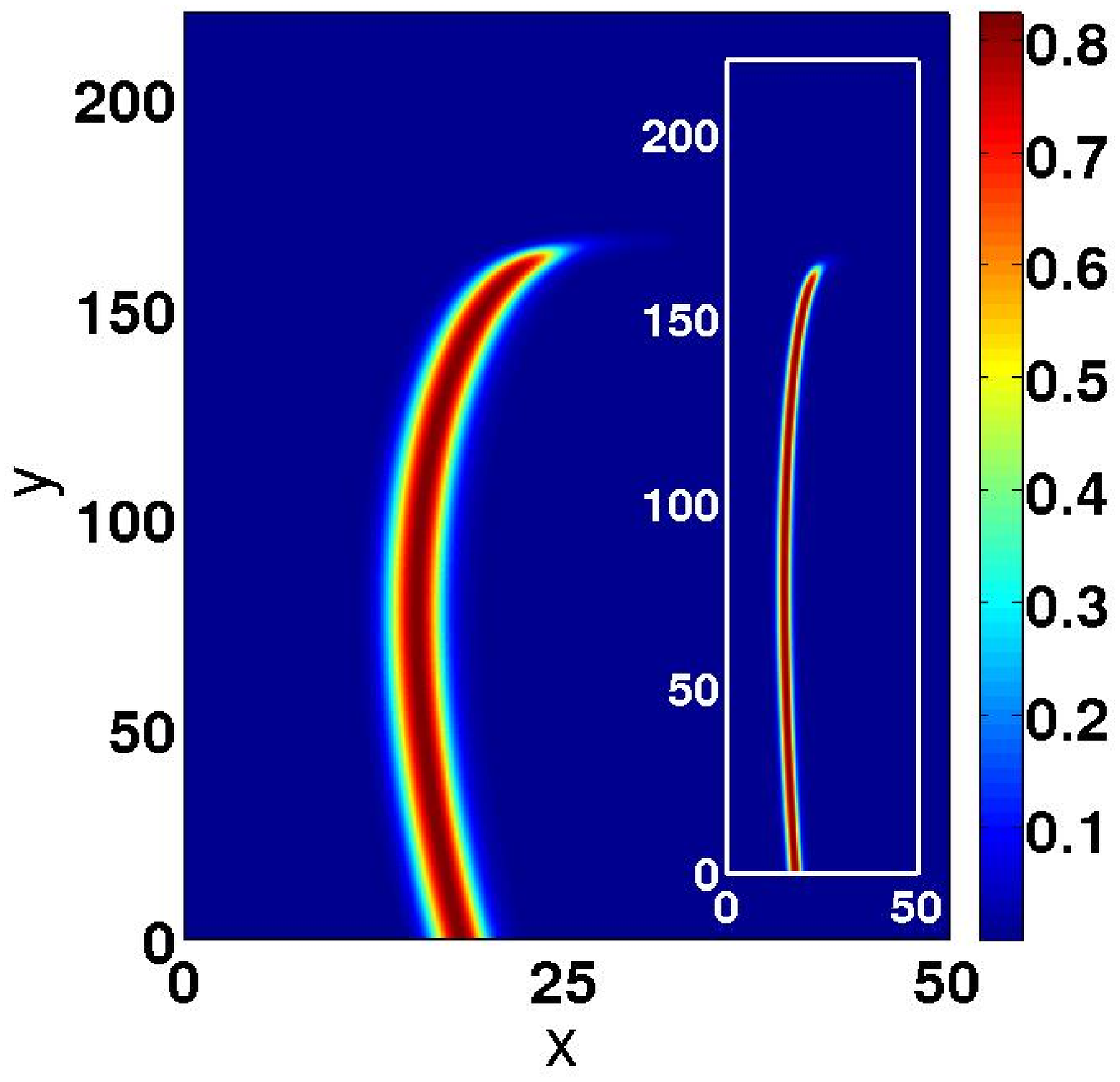}
}
\hspace{0.5cm}
\subfigure[]
{
    \label{fig:ContourLargeCoreSBC+NBC}
    \includegraphics[width=5.9cm]{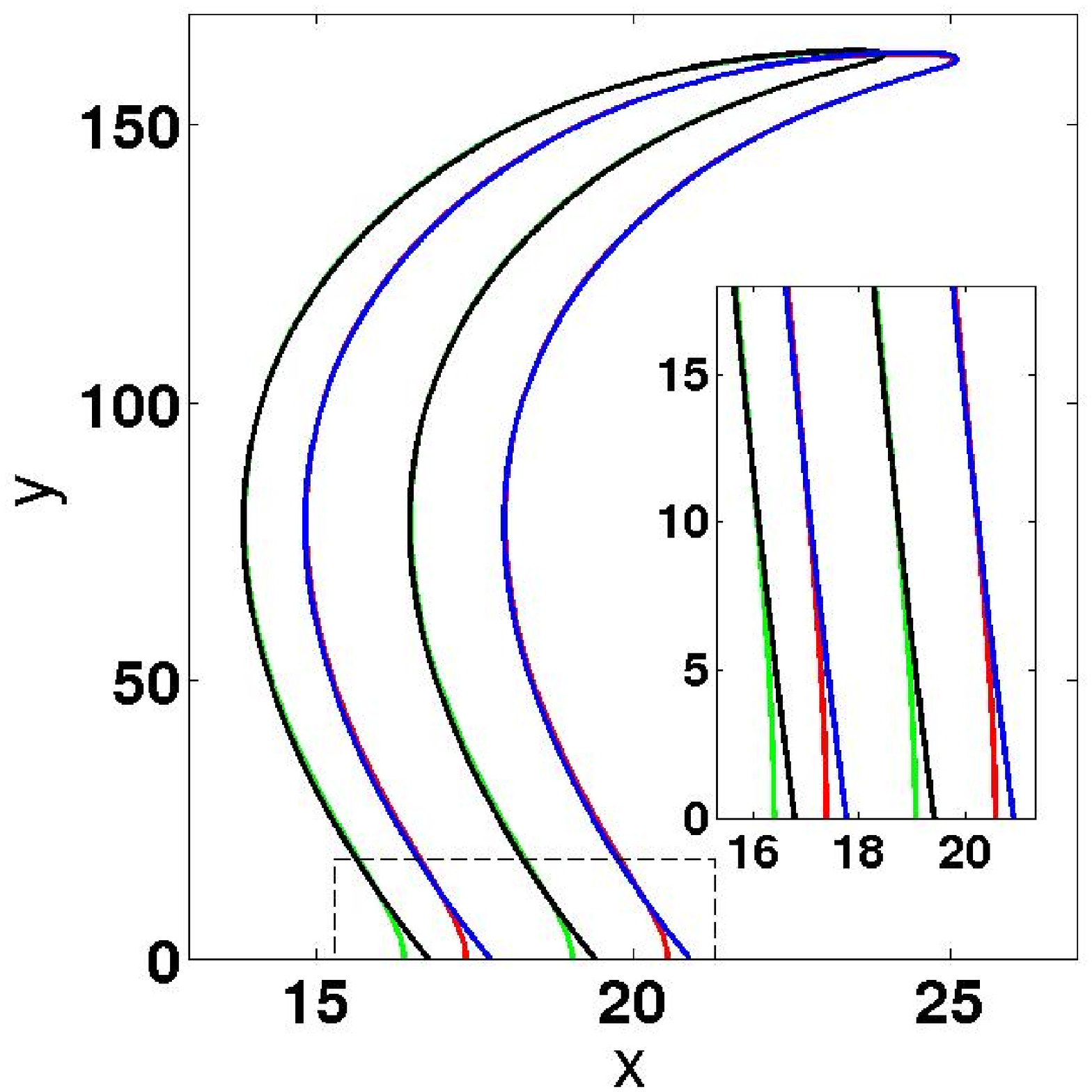}
}
\caption{Spiral wave solution $u$ in the large core limit obtained by solving the stationary frozen system (\ref{eq:FrozenBarkleyStationary}) in Cartesian coordinates. The excitability parameter is $\epsilon=0.0799$, and the spatial discretization  is $\Delta x=\Delta y=0.125$. (a) Contour plot of activator $u$. The inset is added to show the correct aspect ratio. (b) Contour lines $u=0.5$ (online: black, green) and $v=0.5\,a-b$ (online: blue, red) for SBC (involute of a circle) and NBC respectively (inset: zoom into dashed box close to the boundary at $y=0$).}
\label{fig:LargeCoreSpiral} 
\end{figure}

Spiral boundary conditions in Cartesian coordinates can only be applied for large core spirals when the core radius is considerably larger than the box length of the rectangular computational domain. This is due to the fact that on rectangularly shaped domains covering at least one revolution of the spiral arm, there are always points on the boundary at which the spiral arm is tangent to the boundary, violating the SBCs (\ref{eq:SBCTransformed_x}) or (\ref{eq:SBCTransformed_y}). Therefore we have to restrict the method to large core spirals where only a finger-like part of the spiral is resolved within the computational domain. The orientation of the domain can be chosen such that only one boundary intersects with the spiral wave arm. In this case we can set-up boundary conditions, by imposing an SBC on that boundary and Neumann boundary conditions on the remaining boundaries where activator and inhibitor are assumed to be approximately zero. Without loss of generality we choose $y=0$ to be the boundary which intersects with the spiral. The mixed SBC-NBC boundaries are then written as
\begin{equation}\label{eq:MixedBCCartesian}
  U_x=\left\{\begin{array}{cc}
0 &{ \rm{for}} \;\, x=0 \\
\\
0 & \;\; {\rm{for}} \;\, x=L_x
\end{array}
\right.
\quad \mbox{and} \quad
  U_y=\left\{\begin{array}{cc}
\hat{\alpha}(x,y)^{-1}U_x & {\rm{for}} \;\, y=0 \\
\\
0 & \;\; {\rm{for}} \;\, y=L_y
\end{array}
\right.
\, .
\end{equation}
In Figure \ref{fig:ContourLargeCoreSBC} a contour plot of the activator $u$ in Cartesian coordinates is shown, where we used boundary conditions (\ref{eq:MixedBCCartesian}). Figure \ref{fig:ContourLargeCoreSBC+NBC} shows an overlay of this solution with a similar solution where computations were performed with NBCs for all boundaries. We show the contour lines of $u=0.5$ and $v=0.5\,a-b$ respectively, and an inset zooming into the area close to the boundary. For the solution obtained by using SBCs, the boundary appears transparent, whereas the solution obtained by using NBCs for all boundaries exhibits a kink near the boundary. The difference between the two solutions is confined near the boundary, and is absent in the interior.

%


\subsection{Oscillation-free implementation of boundary conditions}
\label{subsec:OscillationFreeBC}

In the middle row of Figure \ref{fig:NegativeBoundaryEffectsContour} we show results of the stationary problem (\ref{eq:FrozenBarkleyStationaryPolar}) in polar coordinates using the spiral wave boundary condition (\ref{eq:SBCTransformed}) for the involute of a circle with (\ref{eq:AlphaTransformed}) and (\ref{eq:SBCCoefficientInvolute}). Comparing Figure \ref{fig:NegativeBoundaryEffectsContourRectangular} and Figure \ref{fig:NegativeBoundaryEffectsContourSBC_oldRectangular}, we see that the spiral boundary condition respects the shape of the solution at the boundary and does not include spurious kinks. However, as can be seen in Figure \ref{fig:NegativeBoundaryEffectsOscillationsSBC_oldContour}, the spiral boundary conditions as described in this section, are not able to control the oscillations near the boundary.

In this section we present an explanation for these oscillations, and suggest a simple way to eliminate them. To understand what causes spurious oscillations for NBC and SBC, we investigate their numerical implementation in more detail. Without loss of generality we restrict the discussion here to polar coordinates.

For both boundary conditions, NBCs and SBCs, the equation for the inhibitor involves only first derivatives. This implies, that at the radial  boundary $(R=(M-1)\Delta r,(j-1)\Delta\varphi)$, the inhibitor $v$ is computed only from values of $v$ and $u$ on the boundary from the previous time step. Whereas the discrete Laplacian, present in the equation for the activator $u$, couples values $u_{M, j}$ at the boundary to those in the interior, i.e. $u_{M-1, j}$, the boundary values of the inhibitor $v_{M, j}$ are decoupled from the interior, and do not receive the outward flowing information from the interior. It is this decoupling of the boundary from the interior for the non-diffusive inhibitor which causes the spatial oscillations near the boundary. (We recall that there are no oscillations for the diffusive activator $u$, and also no oscillations when diffusion is added to the equation for the inhibitor.)\\

\noindent
For both cases, NBCs and SBCs, we propose a simple method to overcome this decoupling. For the activator $u$ we use NBCs or SBCs as discussed. However, instead of invoking this boundary condition for the inhibitor as well, we evaluate the derivative of the inhibitor at the boundary by a one-sided second-order discretization according to
\begin{equation}\label{eq:OneSidedBC}
  v_{r\left|_{M, j}\right.}=\frac{3v_{M, j}-4v_{M-1,j}+v_{M-2, j}}{2\Delta r}\, .
\end{equation}
For Cartesian coordinates we use equivalent expressions.\\
\indent
Considering the original unfrozen Barkley system (\ref{eq:Barkley}), there are two reasons why this simple trick works. Firstly, the boundary condition is consistent with the fact that information flows outwards when studying spiral waves, and therefore outer grid points are influenced by inner grid points only. Secondly, the inhibitor is ``slaved" to the activator in the sense that if the activator is known, the linear equation for the inhibitor can be integrated explicitly. The boundary conditions of the activator are therefore (approximately) inherited by the inhibitor.\\ 

\begin{figure}[htp]
\centering
\subfigure[] 
{
    \label{fig:SurfaceStationaryNBCLargeCore}
    \includegraphics[width=5.9cm]{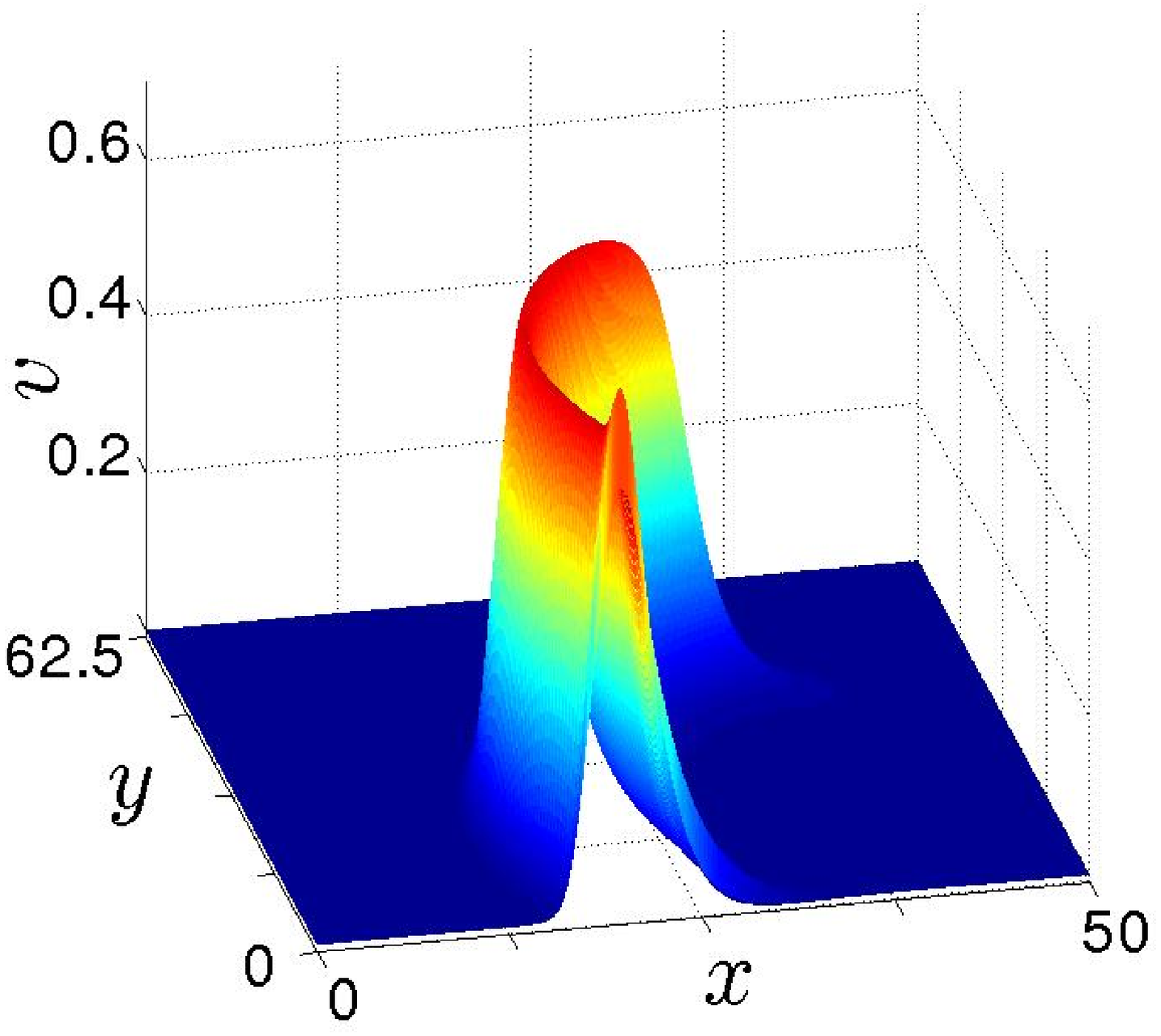}
}
\hspace{0.2cm}
\subfigure[] 
{
    \label{fig:SurfaceStationaryNBCLargeCoreCloseUp}
    \includegraphics[width=5.9cm]{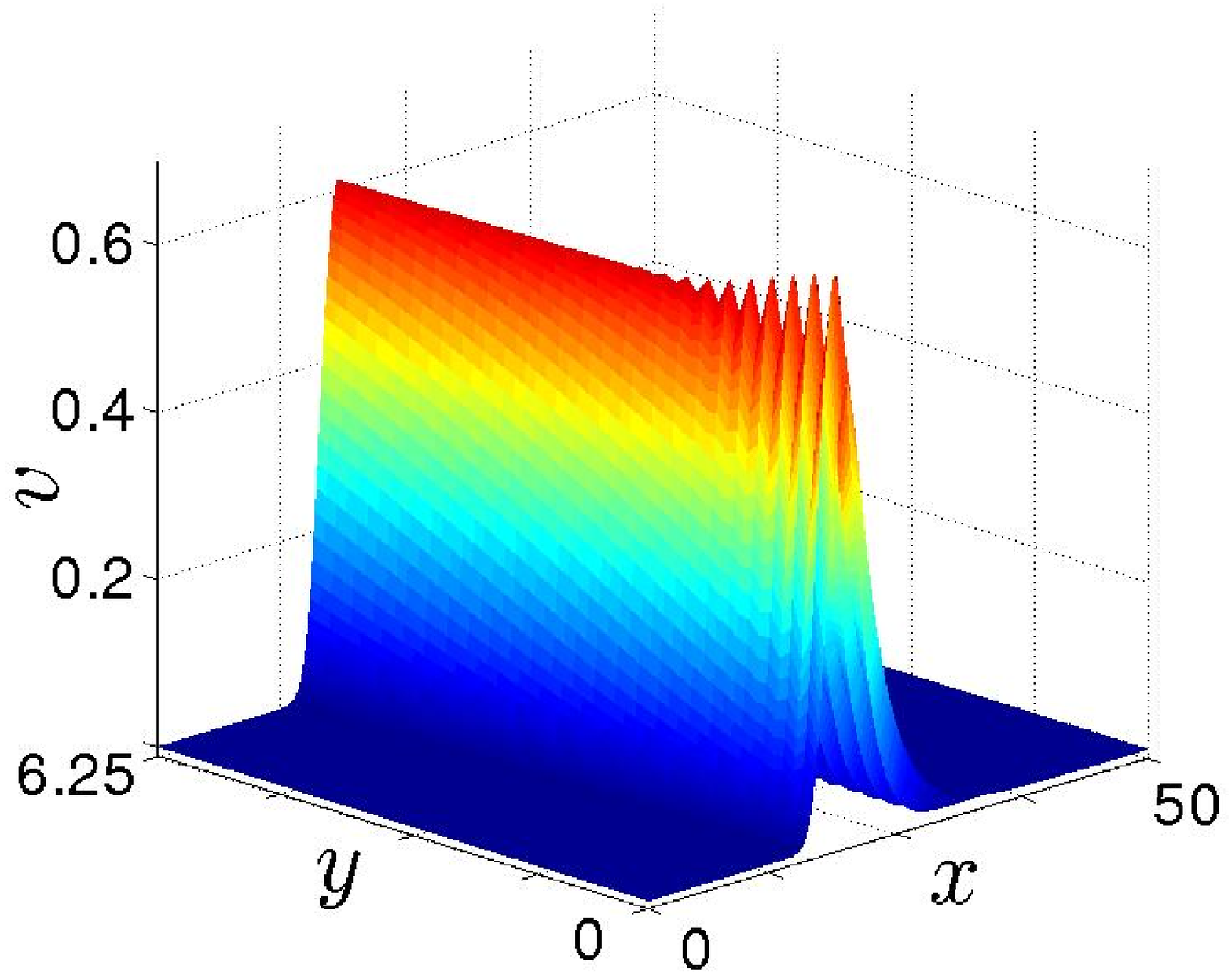}
}
\vspace{0.2cm}
\subfigure[] 
{
    \label{fig:SurfaceStationaryNBCOneSidedLargeCore}
    \includegraphics[width=5.9cm]{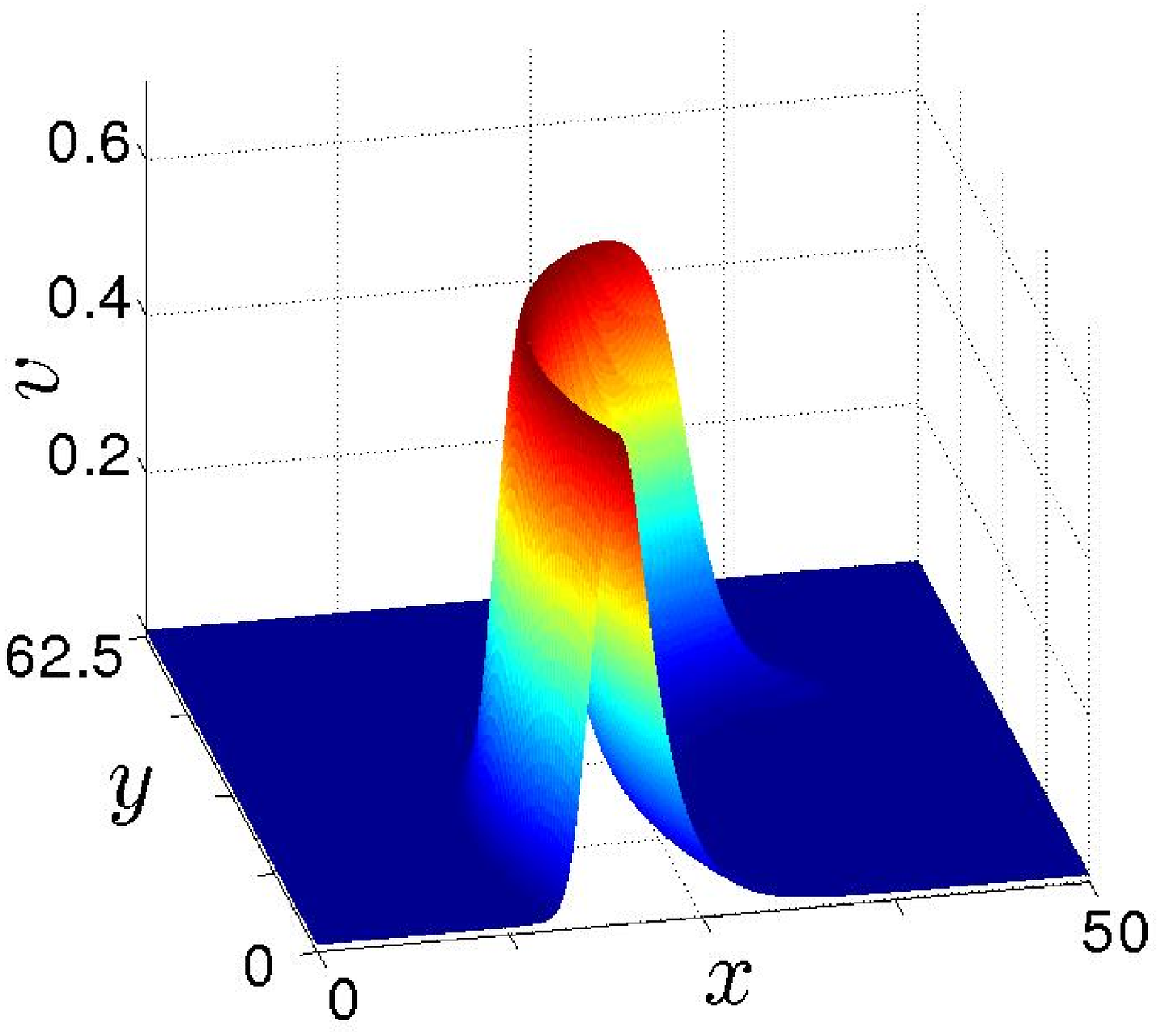}
}
\hspace{0.2cm}
\subfigure[] 
{
    \label{fig:SurfaceStationaryNBCOneSidedLargeCoreCloseUp}
    \includegraphics[width=5.9cm]{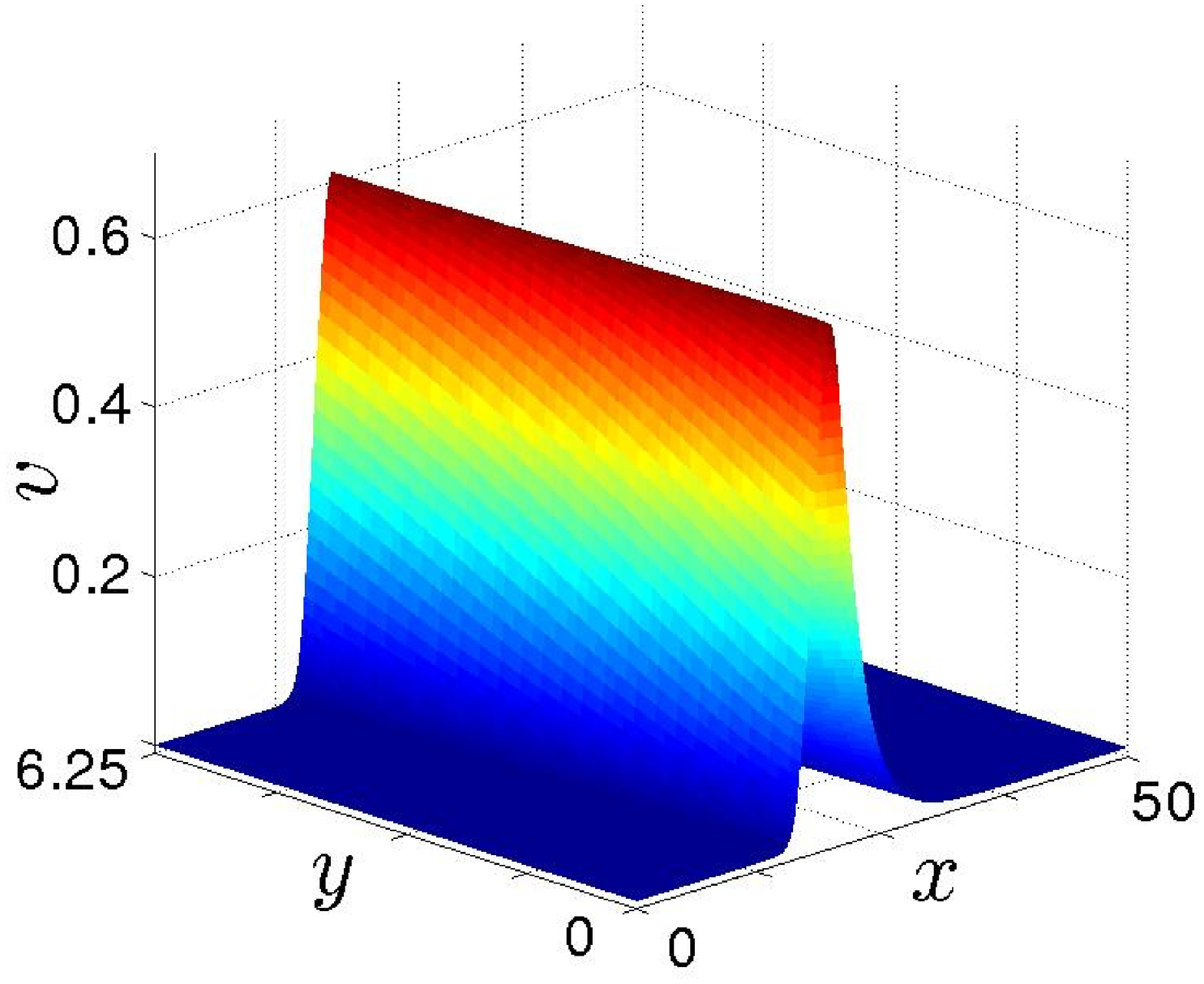}
}
\caption[Spiral solutions with NBCs in Cartesian coordinates.]{Inhibitor $v$ in the weakly excitable case with $\epsilon=0.075$, calculated from the stationary frozen system (\ref{eq:FrozenBarkleyStationary}) in Cartesian coordinates with NBCs for activator and inhibitor (top panels) and NBCs for activator only and one-sided derivative (\ref{eq:OneSidedBC}) for the inhibitor (bottom panels). (b) and (d) are close-ups of the respective solutions depicted in (a) and (c), zooming in on the boundary at $y=0$.}
\label{fig:StationaryNBCLargeCore} 
\end{figure}

\noindent
In Figure \ref{fig:NegativeBoundaryEffectsContour} (bottom row) we show how the implementation of the spiral wave boundary condition for the activator and the one-sided derivative (\ref{eq:OneSidedBC}) for the inhibitor in the stationary problem (\ref{eq:FrozenBarkleyStationaryPolar}) in polar coordinates suppresses the spatial oscillations and avoids changes in shape and amplitude (see also Figure~\ref{fig:NegativeBoundaryEffectsOscillationsSliceFREE}). In Figure~\ref{fig:StationaryNBCLargeCore} we show how the implementation of Neumann boundary conditions for the activator and the one-sided derivative (\ref{eq:OneSidedBC}) for the inhibitor controls the spatial oscillations and the spiral wave shape in the stationary problem (\ref{eq:FrozenBarkleyStationary}) in Cartesian coordinates in the weakly excitable regime.\\

\noindent
During extensive numerical tests \cite{HermannThesis} we found that in highly excitable media only spiral boundary conditions (\ref{eq:SBCTransformed}) in conjunction with the one-sided boundary condition (\ref{eq:OneSidedBC}) give accurate results for the stationary and the time-dependent freezing method without spurious oscillations and without unnatural deformations of the spiral shape. For the large core limit, in weakly excitable media, Neumann boundary conditions in Cartesian coordinates are the optimal choice to avoid numerical oscillations and to properly resolve the shape of the spiral wave solution. Note that in the large core limit NBCs resemble SBCs. This is due to the fact that the overall curvature of the spiral wave, captured in the computational domain, is close to zero. In a way NBCs can be viewed as a simpler and more convenient formulation of SBCs in the large core limit, which do not involve the core radius and the wavelength of the spiral wave.\\ 
\indent
In the following we will be using SBCs in polar coordinates for highly excitable media, and NBCs in Cartesian coordinates in the large core limit. We will also employ the one-sided boundary condition (\ref{eq:OneSidedBC}) in all numerical simulations to avoid oscillations.
%


\section{Numerical investigation of the large core limit of spiral waves}
\label{sec:Results} 
In this Section we provide numerical results on the wavelength $\lambda$, the core radius $r_c$ and the rotation frequency $\omega$ of spiral waves in the large core limit. The rotation frequency $\omega$ and the core radius $r_c$ are computed from the group parameters $\mu_i$. We recall that $\omega=\mu_1$ and $r_c$ can be calculated using (\ref{eq:Radius}) with the centre of the spiral given by (\ref{eq:Centre}) and the location of the tip defined as the intersection of the contour lines (\ref{eq:DefinitionTip}). We approach numerically criticality where the core radius develops a singularity and the frequency approaches zero. 
Before we present these results, we collect and discuss some practical aspects of the different methods we introduced to study spiral waves in the large core limit.\\

\noindent
We solve the stationary freezing problem (\ref{eq:FrozenBarkleyStationary}), increasing $\epsilon$. As initial guess for the Newton-Raphson method we use the frozen solutions of the previous value for $\epsilon$. We use a discretization of $\Delta x=\Delta y=0.125$ for Cartesian simulations, unless stated otherwise. We have checked the quadratic convergence of the error in the calculation of the core radius and the rotation frequency with respect to the spatial discretization, and found that at this resolution the values have sufficiently converged.

For polar coordinates we use a discretization $\Delta r=0.125$ and $\Delta \varphi=2\pi/640\approx0.01$. For a computational domain with $R=20$ this corresponds roughly to an equivalent discretization of $\Delta x=0.125$ and $\Delta y\approx 0.2$ near the boundary. This immediately alludes to practical limitations of polar coordinates for large computational domains; wave profiles away from the centre of the computational domain are not properly resolved, and $R$ has to be sufficiently small, unless one employs a computationally expensive fine angular discretization. At this point it is important to repeat, that in highly excitable media with small core radii at sufficiently small values of $\epsilon$, polar coordinates work very well. This is due to two effects: First, transverse wave profiles are wider for smaller values of $\epsilon$, and second, the smaller wavelength in this case causes a cross section of the spiral wave at the boundary which is wider than its transverse profile.\\

\begin{figure}[htb]
\centering
{
    \label{fig:r-eps_polar}
    \includegraphics[width=5.9cm]{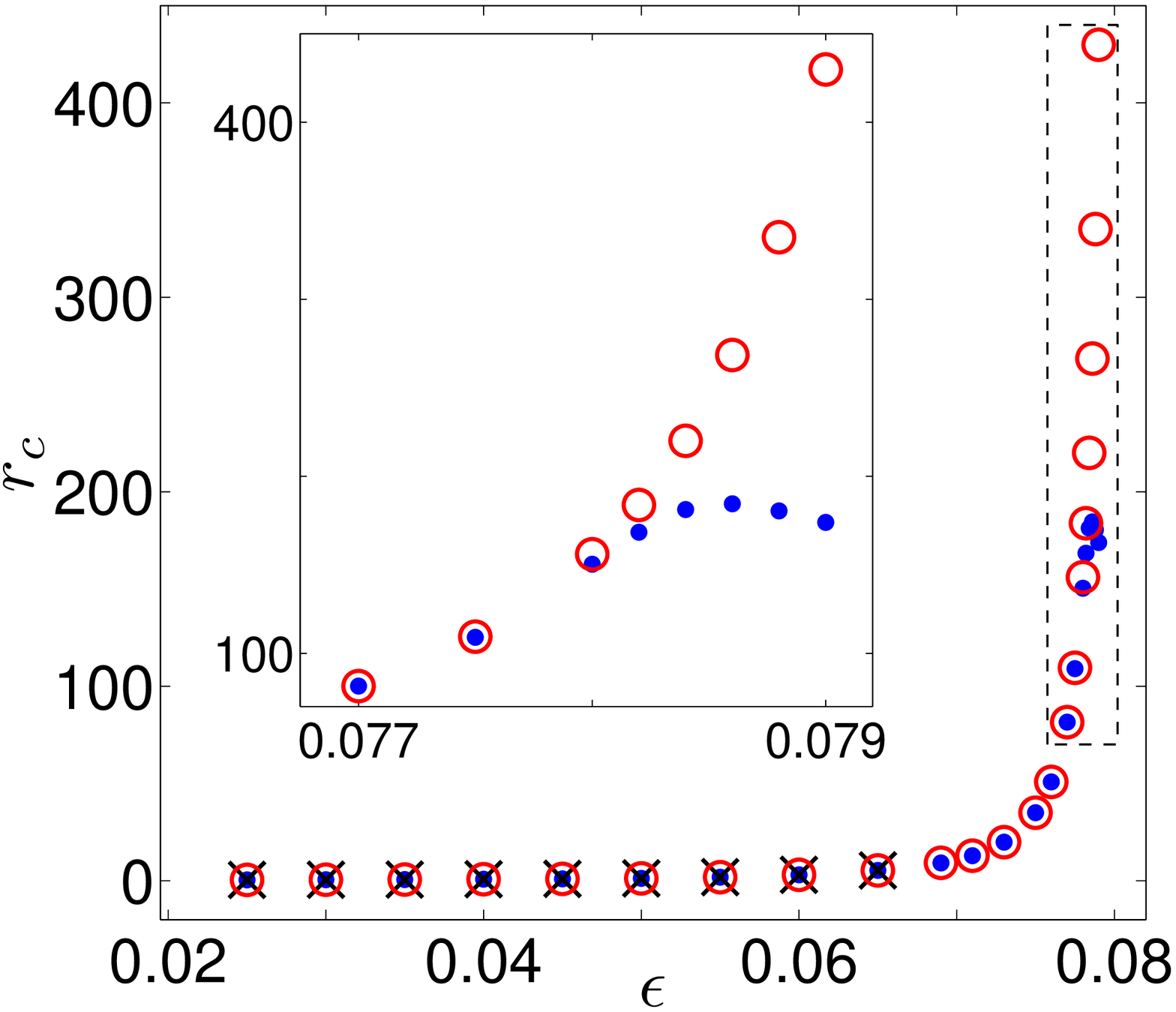}
}
\hspace{0.5cm}
{
    \label{fig:omega-eps_polar}
    \includegraphics[width=5.9cm]{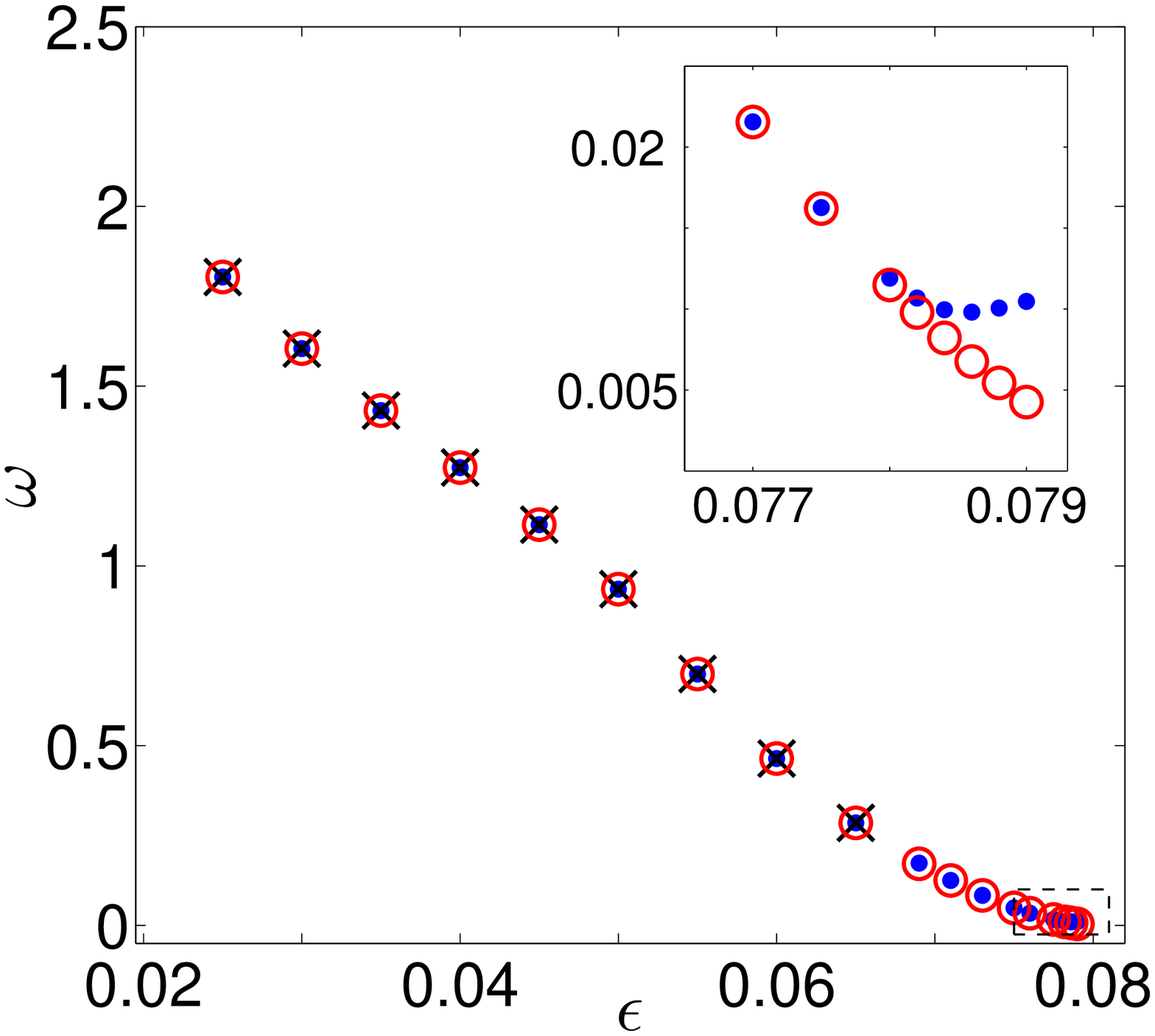}
}
\caption{Core radius $r_c$ and rotation frequency $\omega$ of spiral waves as functions of $\epsilon$ computed in polar coordinates. Here `$\times$' denotes values obtained by a direct simulation of the Barkley model (\ref{eq:Barkley}), and `$\bullet$' and `$\circ$' represent results from solving the stationary frozen system (\ref{eq:FrozenBarkleyStationaryPolar}) with NBCs and SBCs respectively. A computational domain with $R=21.74$ was used.}
\label{fig:r-omega_plots_polar} 
\end{figure}

\begin{figure}[htbp]
\hspace{1cm}
   \begin{minipage}{7cm}
   \subfigure[$\epsilon=0.025$] 
   {
       \label{fig:ContourEps_0_025}
       \includegraphics[width=5.9cm]{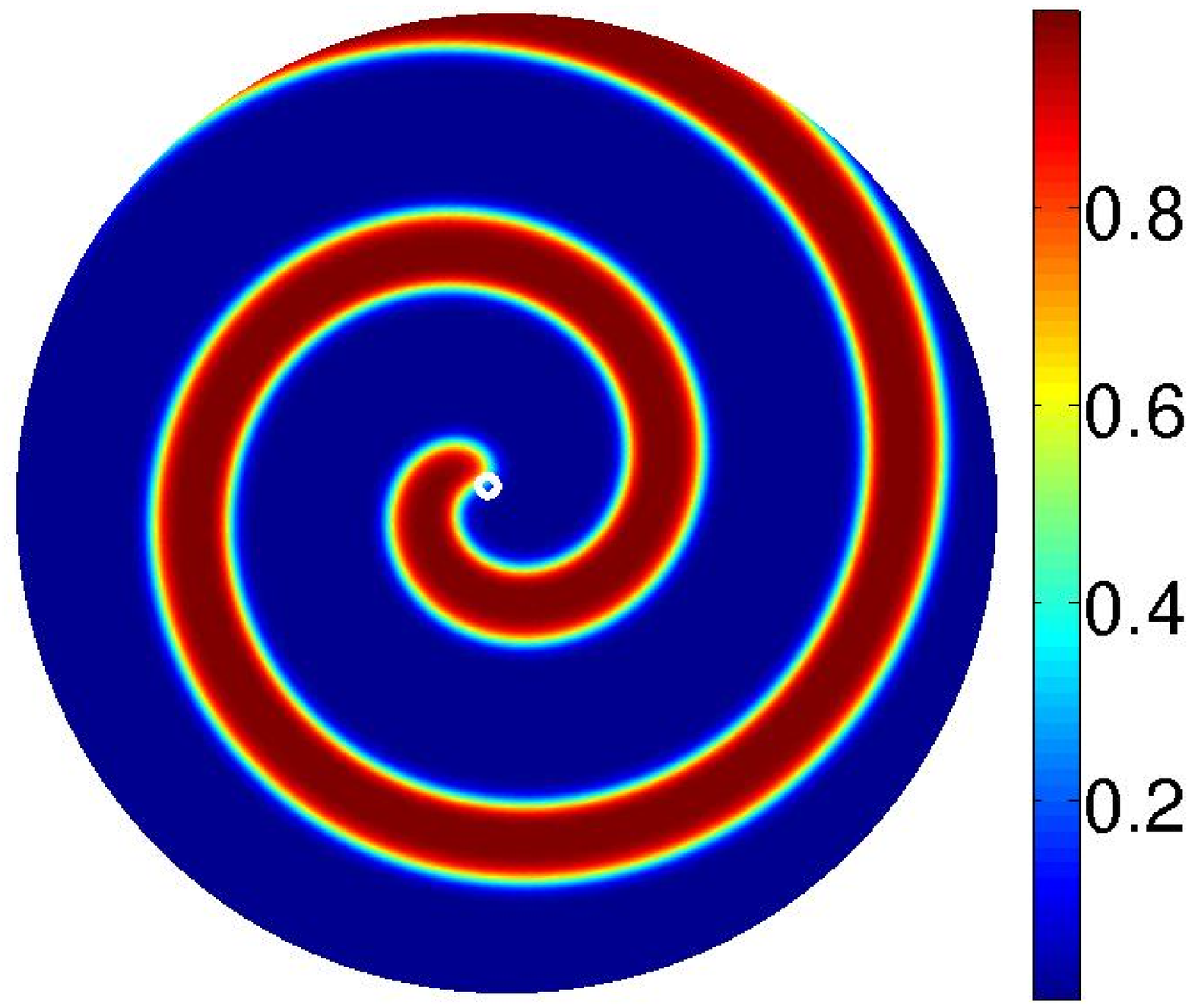}
   }
   \vspace{0.5cm}
   \subfigure[$\epsilon=0.055$] 
   {
       \label{fig:ContourEps_0_055}
       \includegraphics[width=5.9cm]{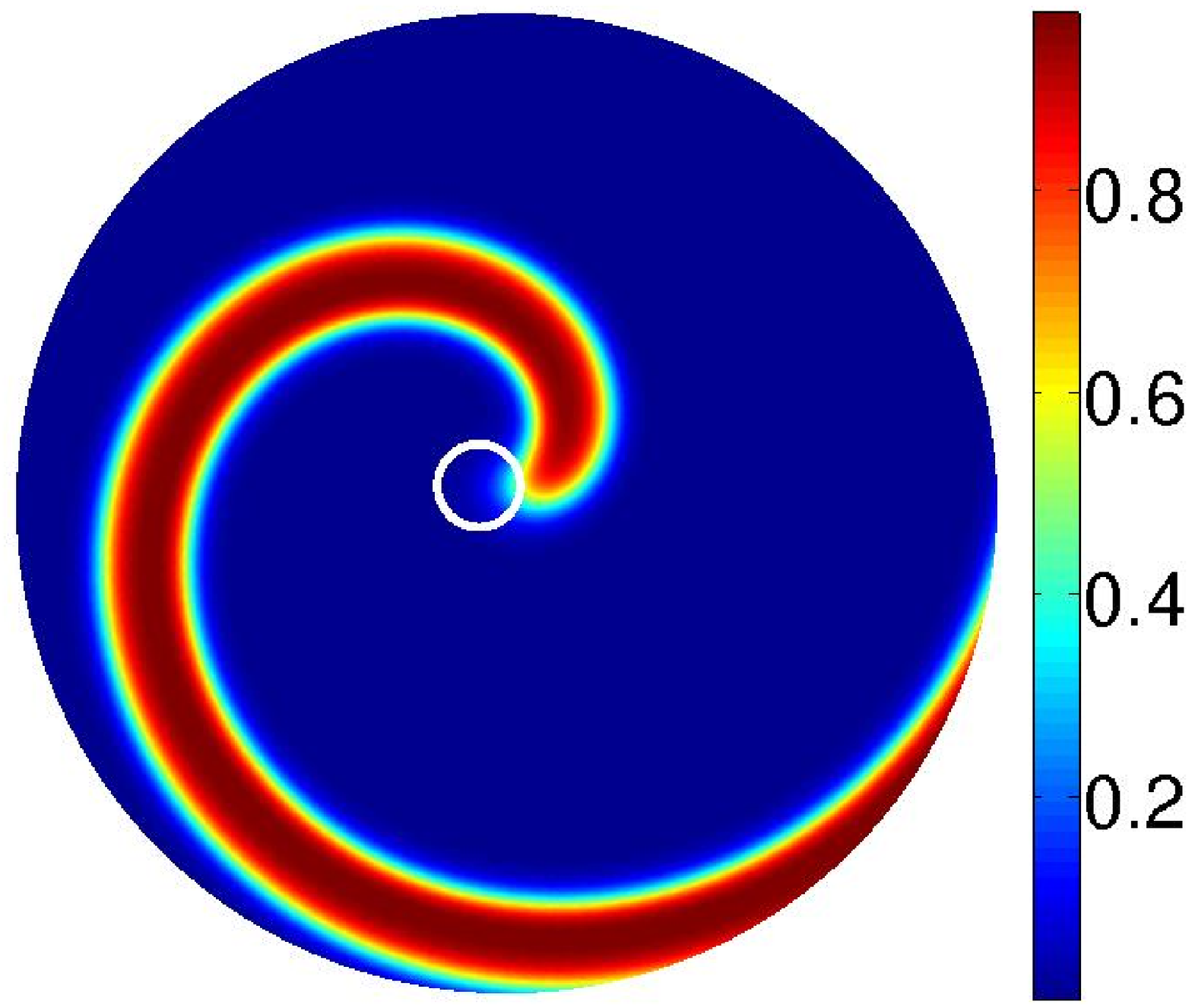}
   }
   \vspace{0.5cm}
   \subfigure[$\epsilon=0.075$] 
   {
       \label{fig:ContourEps_0_075}
       \includegraphics[width=5.9cm]{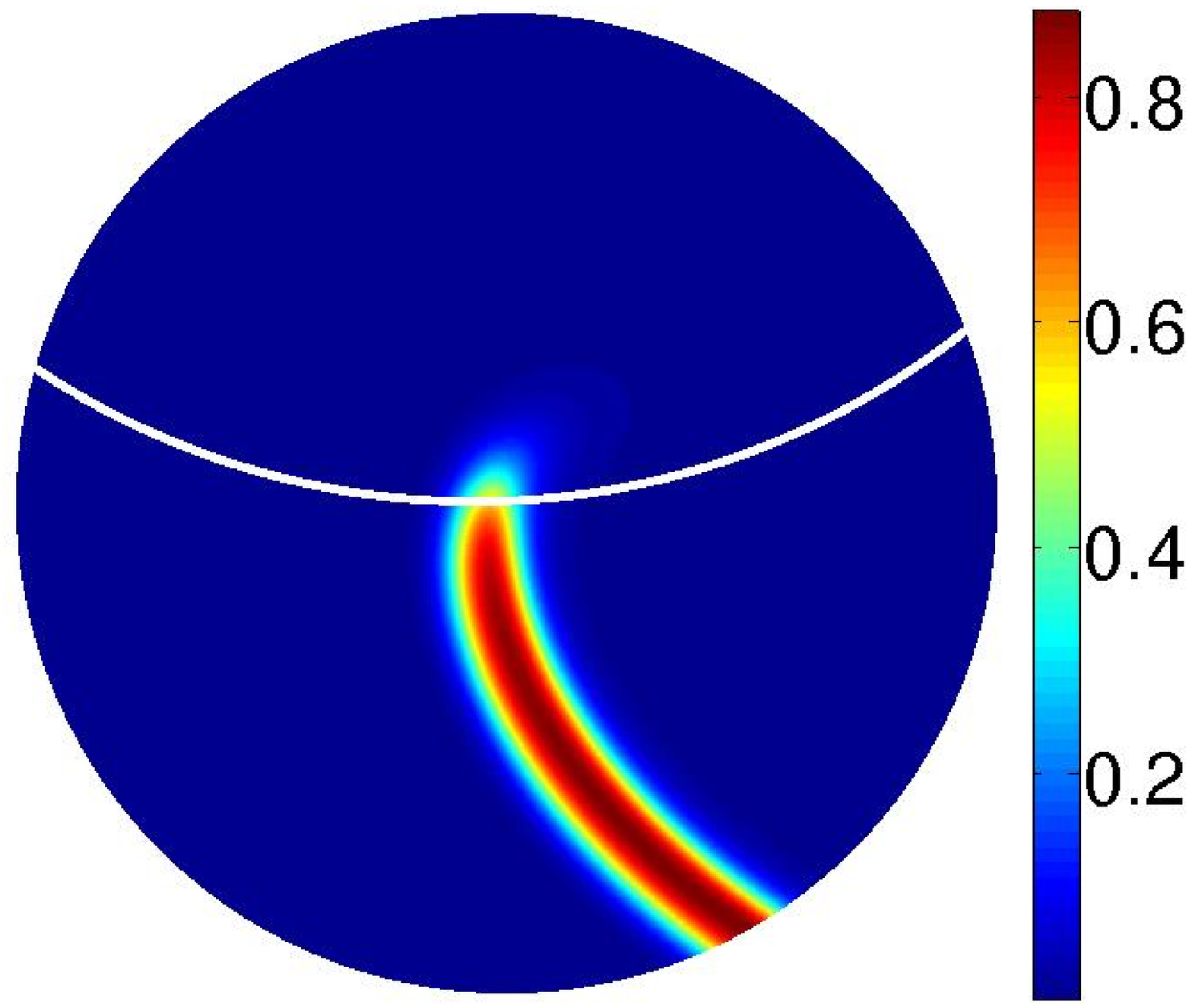}
   }
   \end{minipage}
   \begin{minipage}{7cm}
   \subfigure[$\epsilon=0.068$] 
   {
       \label{fig:ContourEps_0_068}
       \includegraphics[width=5.9cm]{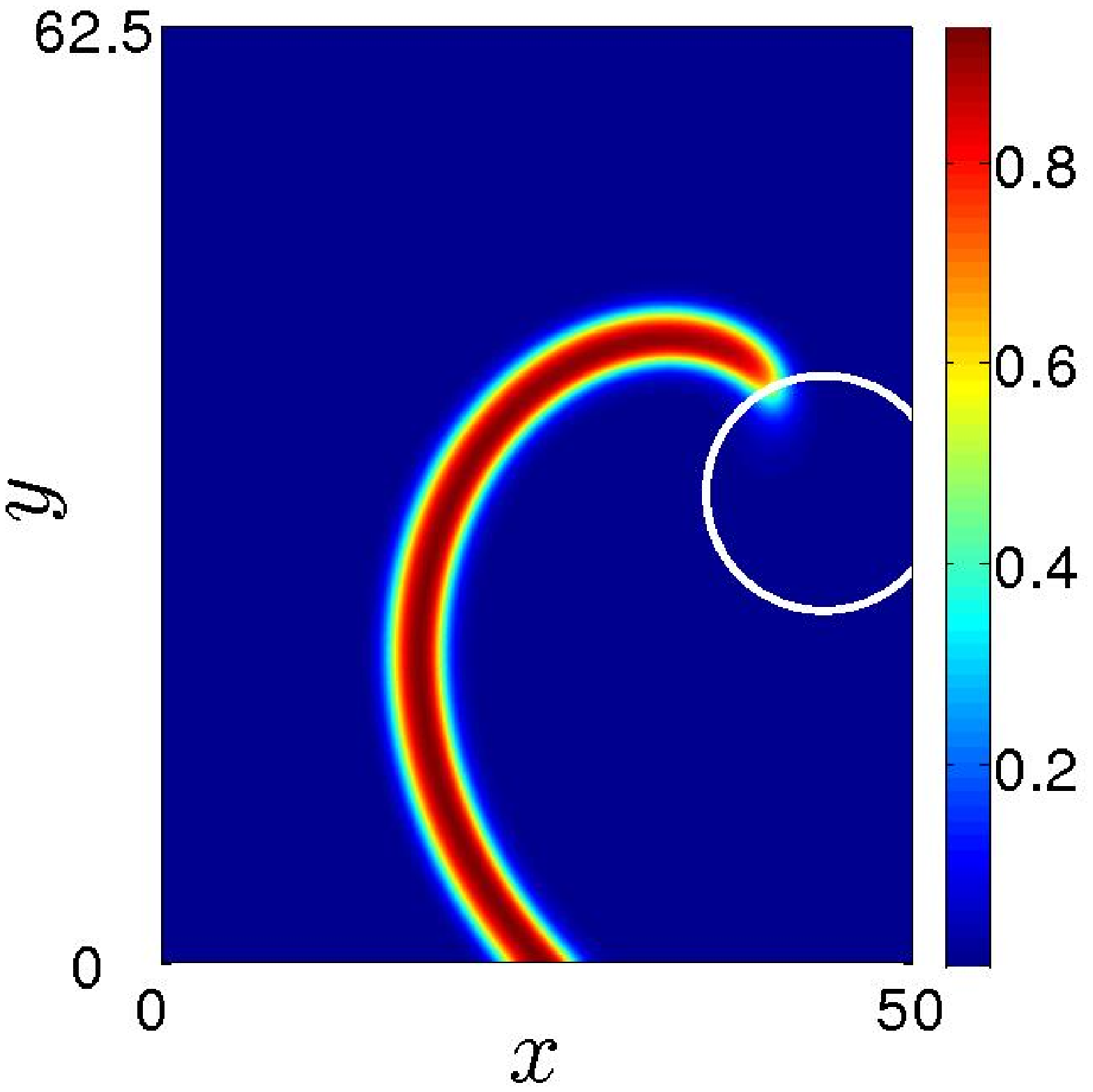}
   }    
   \vspace{0.5cm}
   \subfigure[$\epsilon=0.075$] 
   {
       \label{fig:ContourEps_0_075Cart}
       \includegraphics[width=5.9cm]{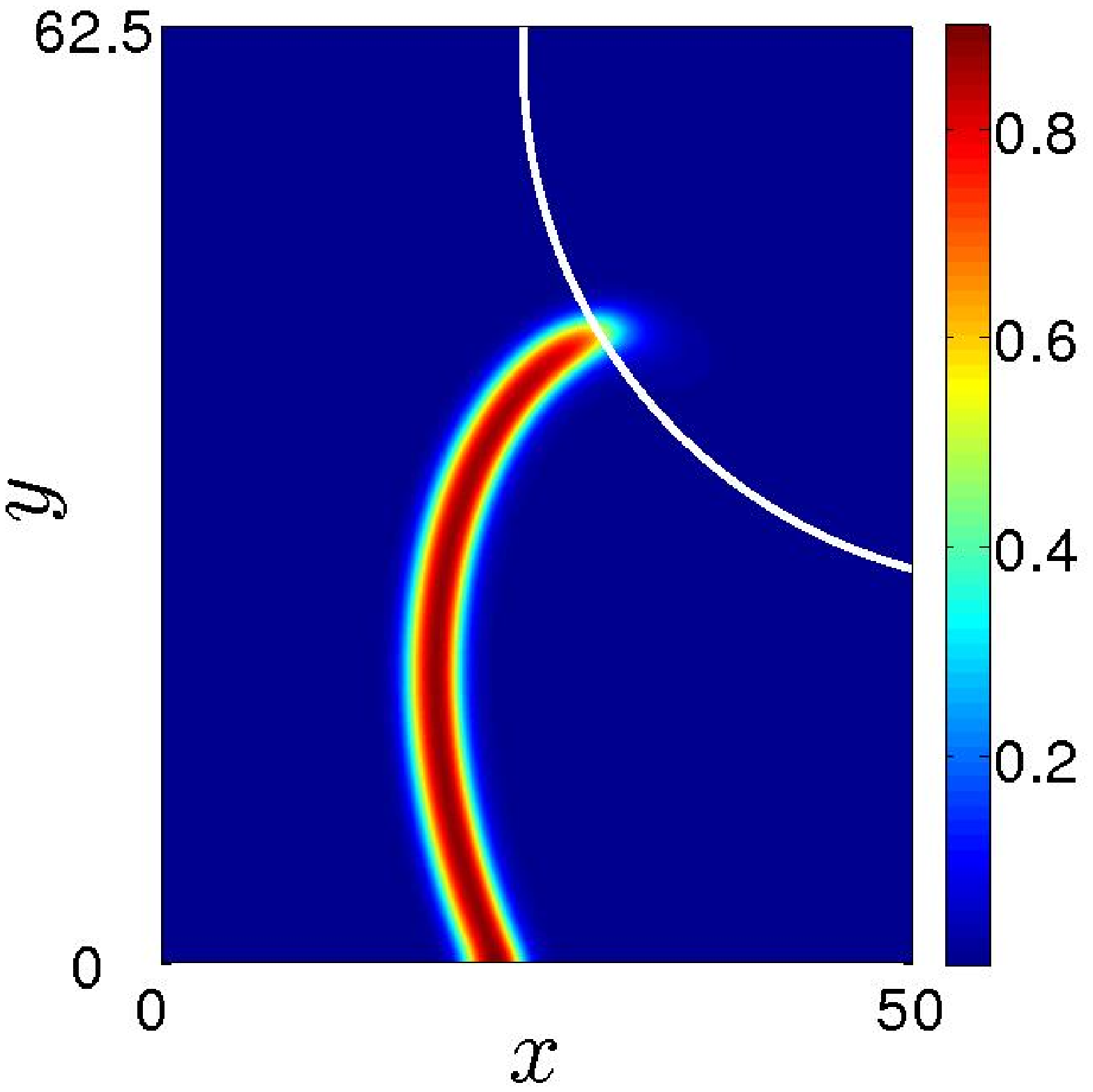}
   }
   \vspace{0.5cm}
   \subfigure[$\epsilon=0.079$] 
   {
       \label{fig:ContourEps_0_079}
       \includegraphics[width=5.9cm]{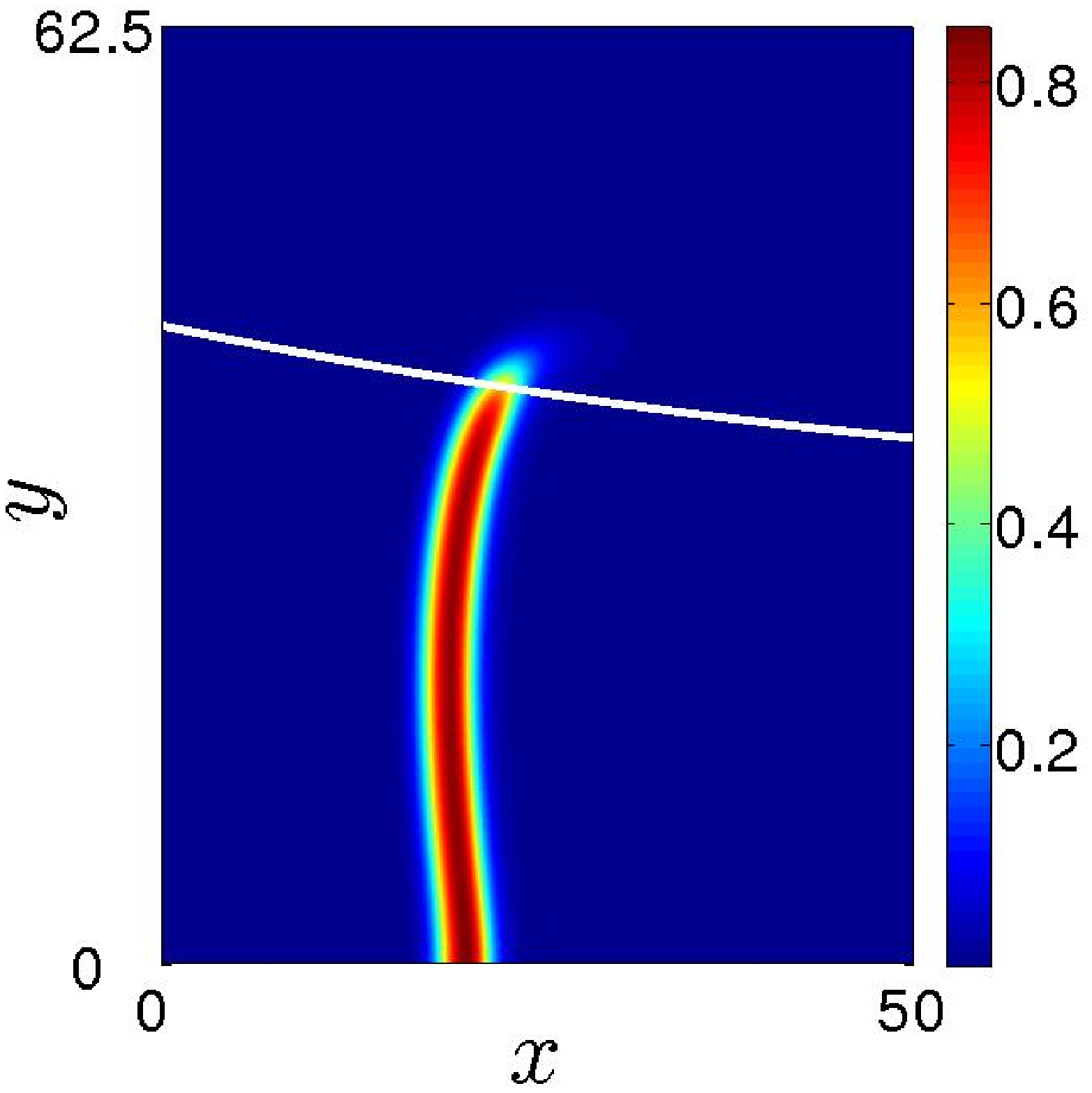}
   }
   \end{minipage}
\caption[Spiral solutions in polar coordinates (left) and Cartesian coordinates (right) for increasing $\epsilon$.]{Contour plots of the activator solution of the stationary frozen system in polar coordinates with SBCs based on approximations by involutes (left) and in Cartesian coordinates with NBCs (right) for increasing values of excitability $\epsilon$. The white circular lines indicate the trace of the tip.}
\label{fig:ContourPlotsPolarCartes} 
\end{figure}

\begin{figure}[htb]
\centering
\subfigure[SBC] 
{
    \includegraphics[width=5.9cm]{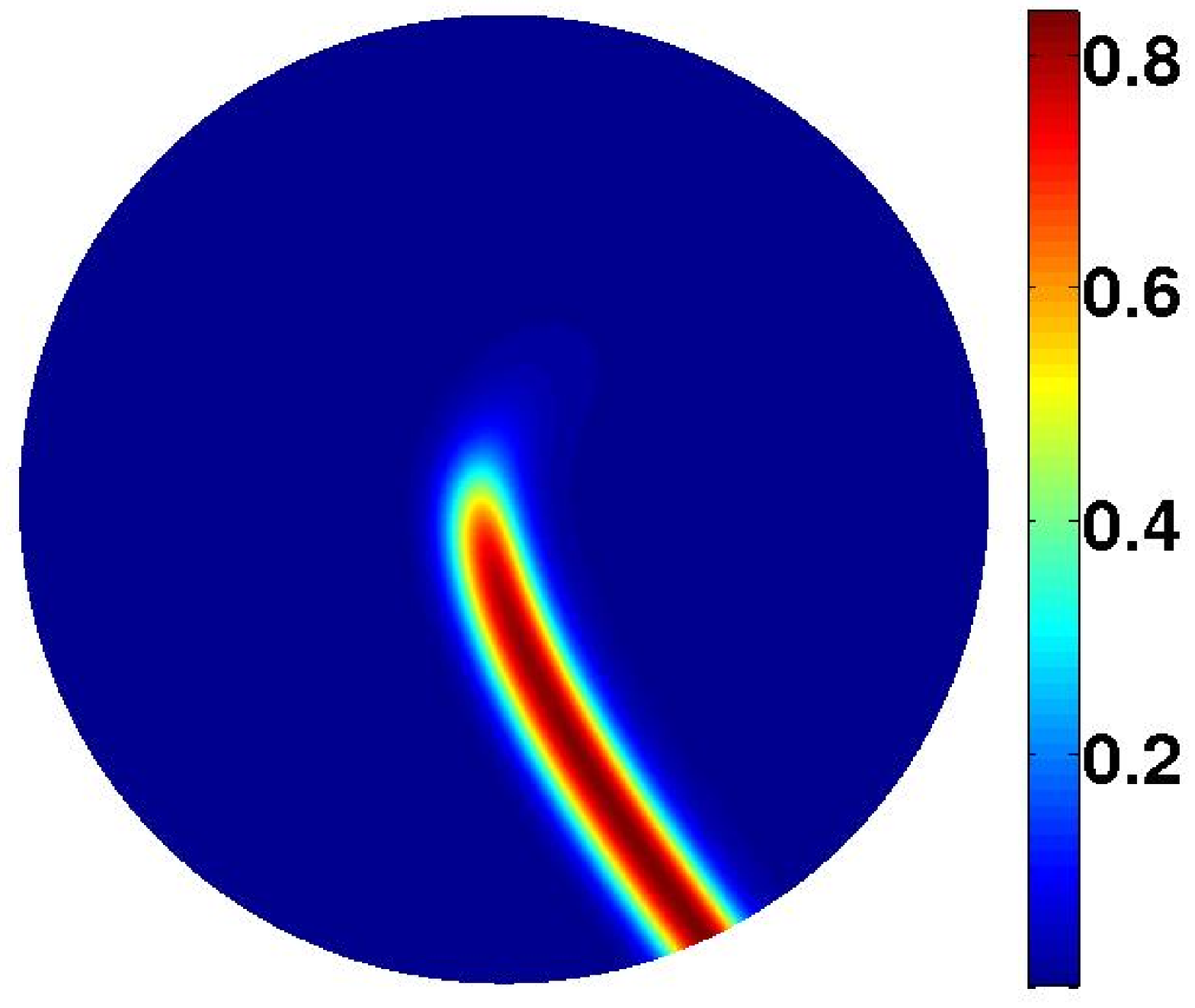}
}
\hspace{0.5cm}
\subfigure[NBC] 
{
    \includegraphics[width=5.9cm]{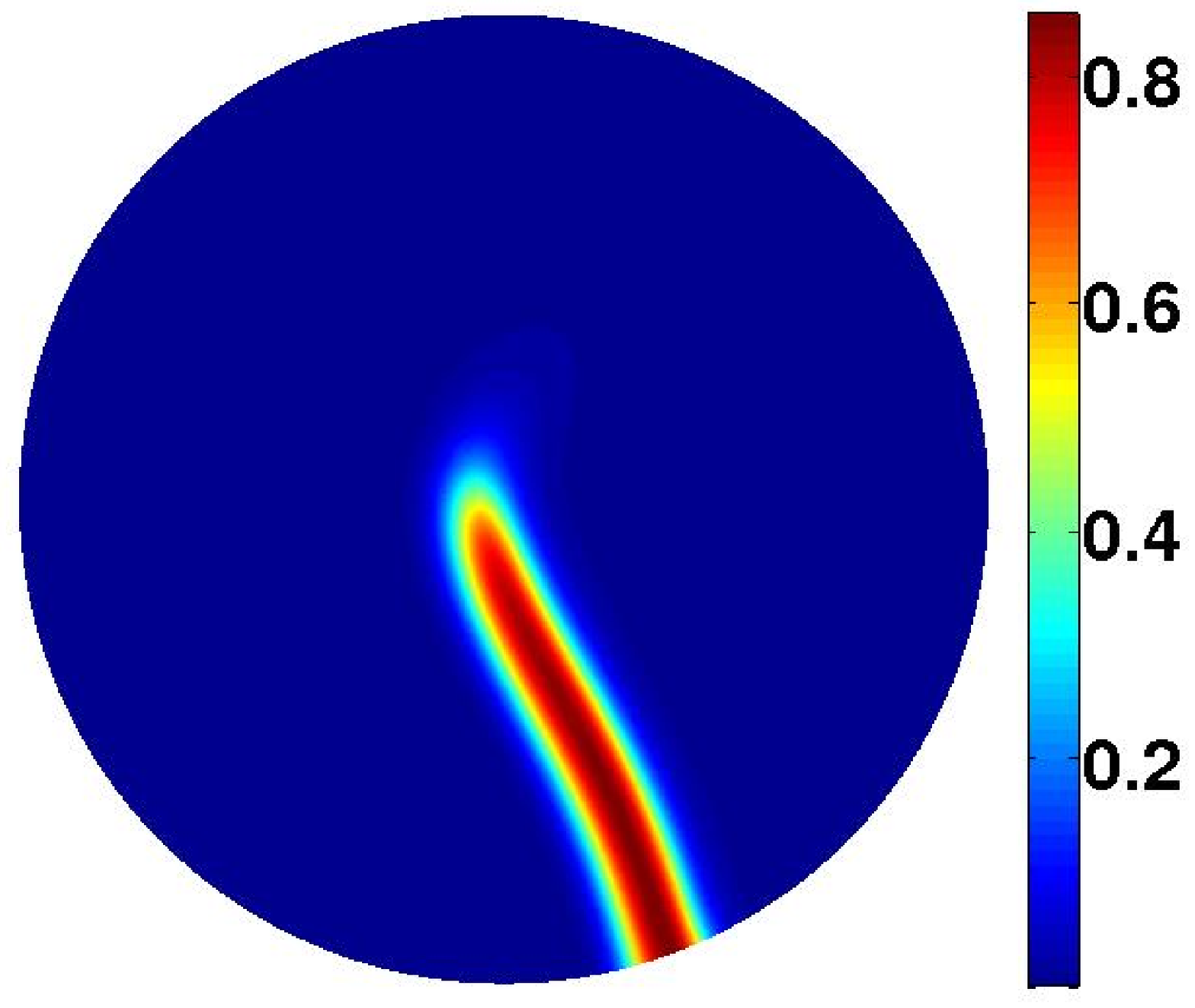}
}
\caption{Contour plot of the activator $u$ found as the solution of the stationary frozen system (\ref{eq:FrozenBarkleyStationaryPolar}) at $\epsilon=0.079$ with $R=21.74$ with SBCs (left) and NBCs (right), respectively.}
\label{fig:SBCNBCpolar} 
\end{figure}

\noindent
In Figure~\ref{fig:r-omega_plots_polar} we show results for solving the stationary frozen system (\ref{eq:FrozenBarkleyStationaryPolar}) in polar coordinates. We used two types of boundary conditions, i.e. NBCs (red $\circ$) and SBCs based on the approximation by involutes (blue $\bullet$). We further included results from direct simulations of the full Barkley model (\ref{eq:Barkley}) (black $\times$). For a computational domain of size $R=21.74$ direct simulations are limited to excitabilities of $\epsilon \lesssim 0.065$ with corresponding radii $r_c \lesssim 5.3$. One can, in principle, determine the core radius for parameter values larger than $\epsilon=0.065$, however, only with the additional computational cost of increasing the computational domain. There will be a value of $\epsilon$ where this is not possible anymore with given computational power. We stress that this value of $\epsilon$ is way below what we call the large core limit. This can be seen in Figure \ref{fig:ContourPlotsPolarCartes} (left panel) where we show contour plots of the activator at different values of $\epsilon$ corresponding to data points in Figure~\ref{fig:r-omega_plots_polar} for SBCs. The white circular lines represent the trace of the tip as defined in (\ref{eq:DefinitionTip}) and are calculated by solving equations (\ref{eq:GroupDynamics}) for the group variables. Figure \ref{fig:ContourEps_0_075} indicates that it becomes inherently difficult to calculate spirals and their characteristic parameters $r_c$ and $\omega$ for large cores by direct simulations of the full Barkley system (\ref{eq:Barkley}). It is this observation which makes the freezing method so attractive for large core spirals.\\
\indent
The insets in Figure~\ref{fig:r-omega_plots_polar} clearly show that SBCs outperform NBCs in the highly excitable regime, and allow for the determination of frozen solutions and the values of the core radius $r_c$ and the rotation frequency $\omega$ for a greater range of parameter values. The breakdown of NBCs is linked to the finite size of the computational domain as illustrated in Figure~\ref{fig:SBCNBCpolar}. We can see that the finger is not properly oriented in the finite domain and that therefore NBCs are not a suitable choice, bending the solution into an unnatural direction. This does not happen for the shape-preserving SBCs. The negative effect of the (inaccurate) NBCs to deform contour lines of the solution is proportionally larger for moderate computational domains than for larger ones (which makes polar coordinates more sensitive to these effects then Cartesian coordinates). In principle, one may extend the range of validity in $\epsilon$-parameter space for NBCs by considering larger and larger computational domains; however, this would quickly become computationally unfeasible.\\

\begin{figure}[htb]
\centering
{
    \label{fig:r-eps_mit_inset}
    \includegraphics[width=5.9cm]{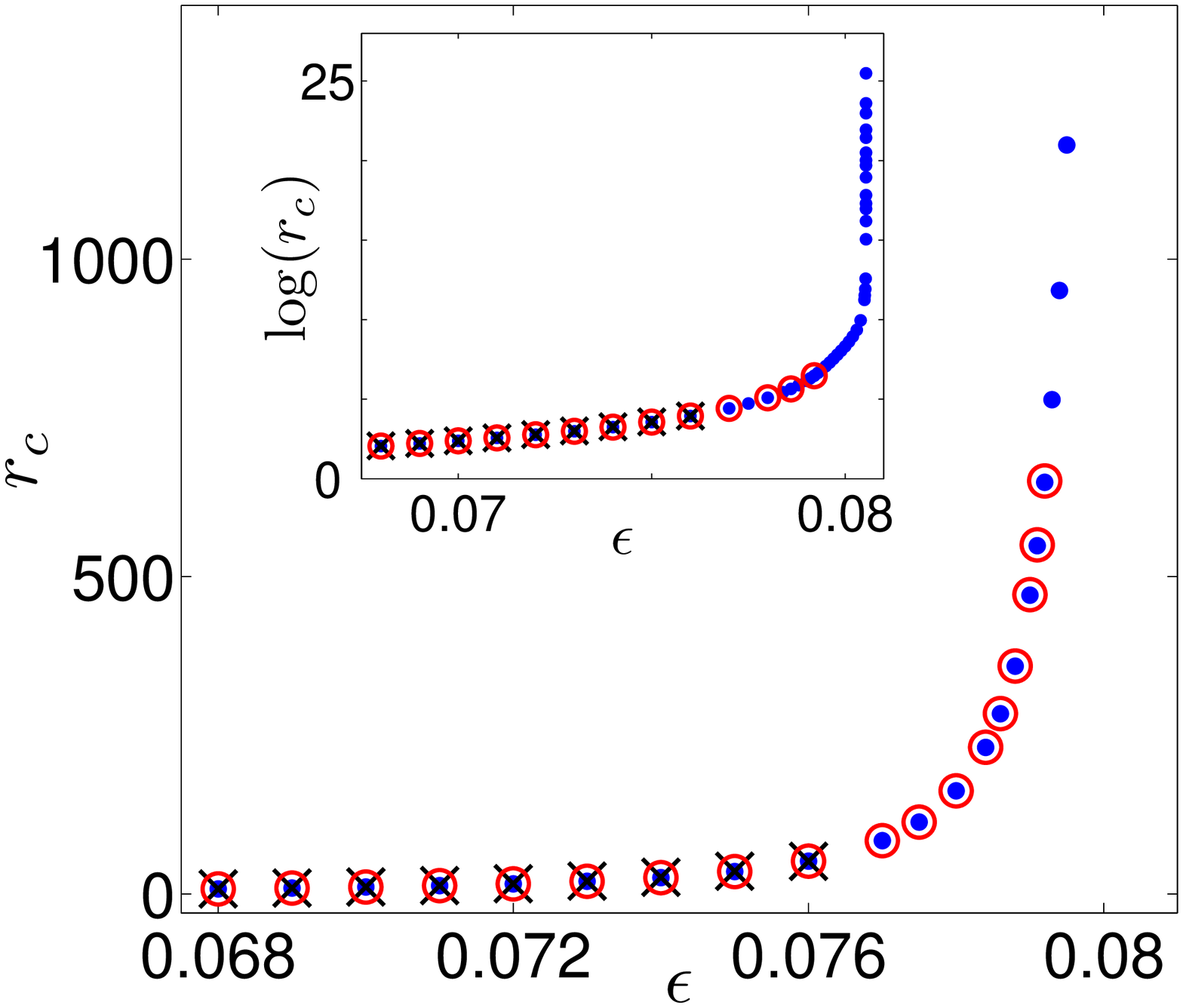}
}
\hspace{0.5cm}
{
    \label{fig:omega-eps_mit_inset}
    \includegraphics[width=5.9cm]{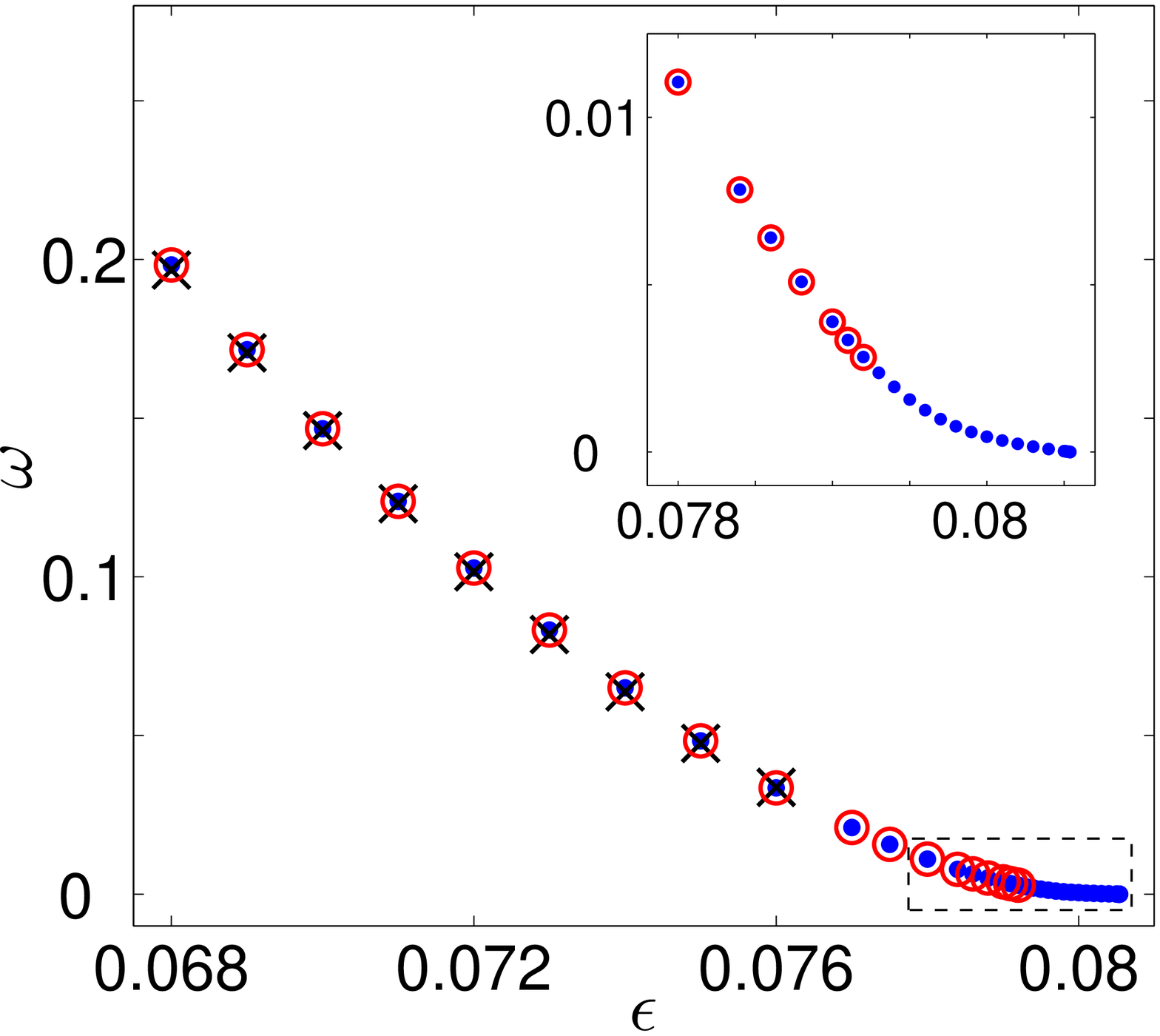}
}
\caption{Core radius $r_c$ and rotation frequency $\omega$ of spiral waves as functions of $\epsilon$ with $\epsilon<\epsilon_c$ computed in Cartesian coordinates. Here `$\times$' denotes values obtained by direct simulation of the Barkley model (\ref{eq:Barkley}), and `$\bullet$' and `$\circ$' represent results from solving the stationary frozen system (\ref{eq:FrozenBarkleyStationary}) with NBCs and SBCs respectively. A computational domain with $L_x=50$ and $L_y=62.5$ was used.}
\label{fig:r-omega_plots_cart} 
\end{figure}

\begin{figure}[htb]
\centering
\includegraphics[width=8.2cm]{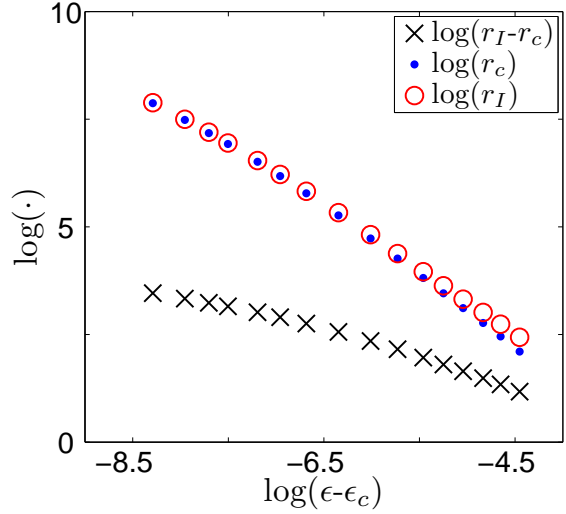}
\caption{Core radius $r_c$ and radius of the circle of the involute $r_I=c_\infty/\omega$ as a function of $\epsilon$. The critical value of the excitability parameter is $\epsilon_c=0.08054091$.}
\label{fig:rcrU} 
\end{figure}

\noindent
To approach the large core limit, we employ a Cartesian coordinate system for values of $\epsilon\geq 0.068$. In Figure~\ref{fig:r-omega_plots_cart} we show the rotation frequency and the core radius of simulations where we approach the critical point at $\epsilon_c\approx 0.08054091$. Here we use $L_x=50$ and $L_y=62.5$. In Figure~\ref{fig:ContourPlotsPolarCartes} (right panel) we show contour plots of the activator solution of the frozen stationary system (\ref{eq:FrozenBarkleyStationary}) at different values of $\epsilon$ corresponding to data points in Figure~\ref{fig:r-omega_plots_polar} for NBCs. The exact value of $\epsilon_c$ depends on the discretization and also on the size of the computational domain. See Section~\ref{sec:scaling} for a discussion on this issue. Compare the range of $r_c$ attainable in polar coordinates and Cartesian coordinates (cf. Figure~\ref{fig:r-omega_plots_polar} and inset of Figure~\ref{fig:r-omega_plots_cart}). Contrary to the results with polar coordinates, in Cartesian coordinates NBCs are applicable for a larger range of parameter values than SBCs. As discussed in Section~\ref{sec:SBCappl}, SBCs are not applicable in the large core limit, when the spiral wave appears as a finger in computational domains of finite size. The part of the spiral wave solution, starting at the tip, which cannot be approximated by an Archimedean spiral or an involute of a circle, increases the closer one is to criticality. As a proxy for the extent of this region we depict in Figure~\ref{fig:rcrU}, how $r_I-r_c$ grows with $\epsilon \to \epsilon_c$. For values $\epsilon>0.0797$, corresponding to core radii $r_c>2523$, we have $r_I-r_c>L_y$, and the freezing method using SBCs is not applicable anymore. NBCs, on the other hand,  become a better approximation the closer we are to the critical point, where the spiral wave solution approaches zero curvature -- provided that the finger is appropriately oriented within the computational domain to assure a perpendicular intersection with the boundary.

\begin{figure}[htbp]
\centering
\subfigure[Activator $u$] 
{
    \label{fig:contour_plots_at_eps_crit_u}
    \includegraphics[width=5.9cm]{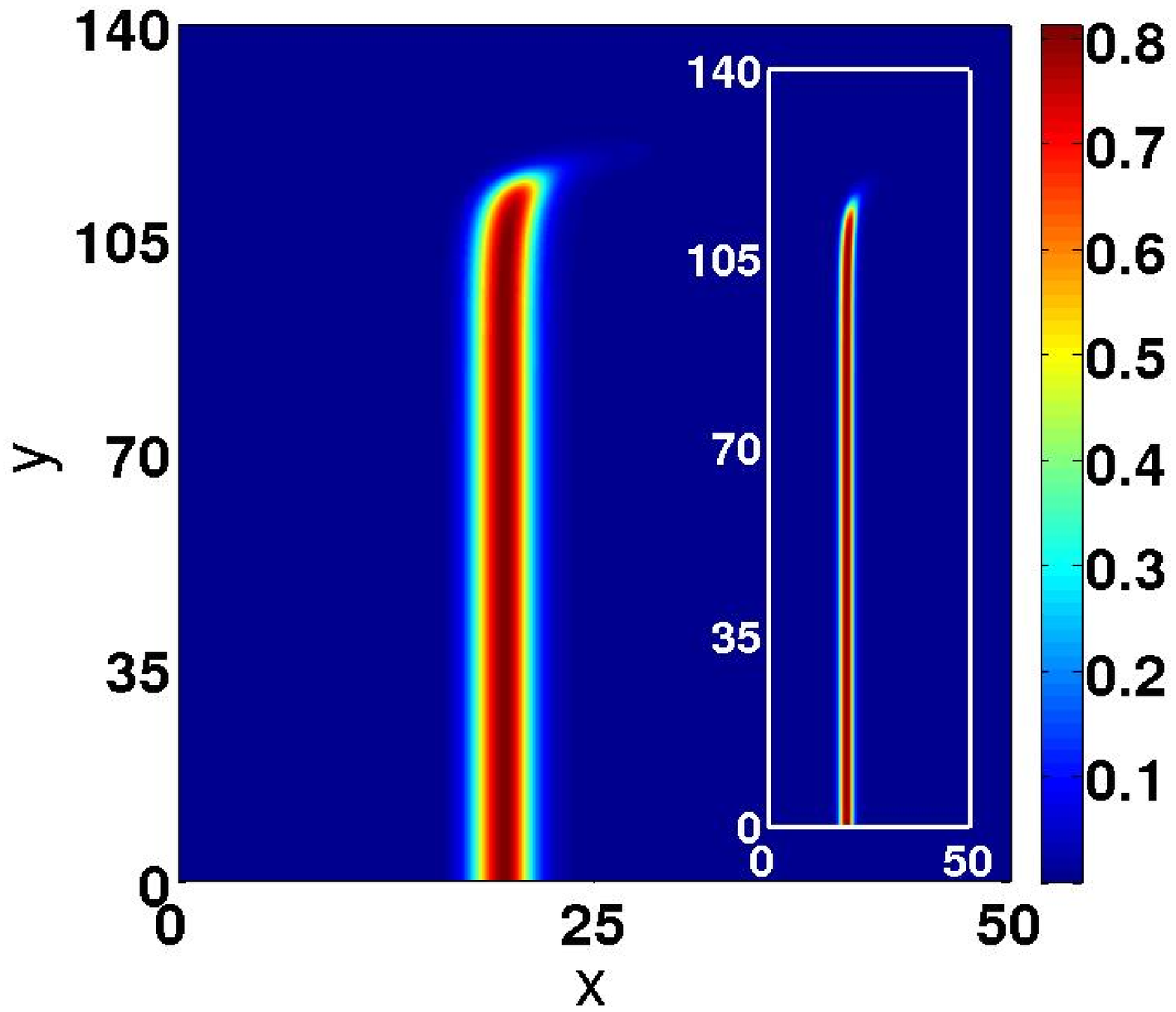}
}
\hspace{0.5cm}
\subfigure[Inhibitor $v$] 
{
    \label{fig:contour_plots_at_eps_crit_v}
    \includegraphics[width=5.9cm]{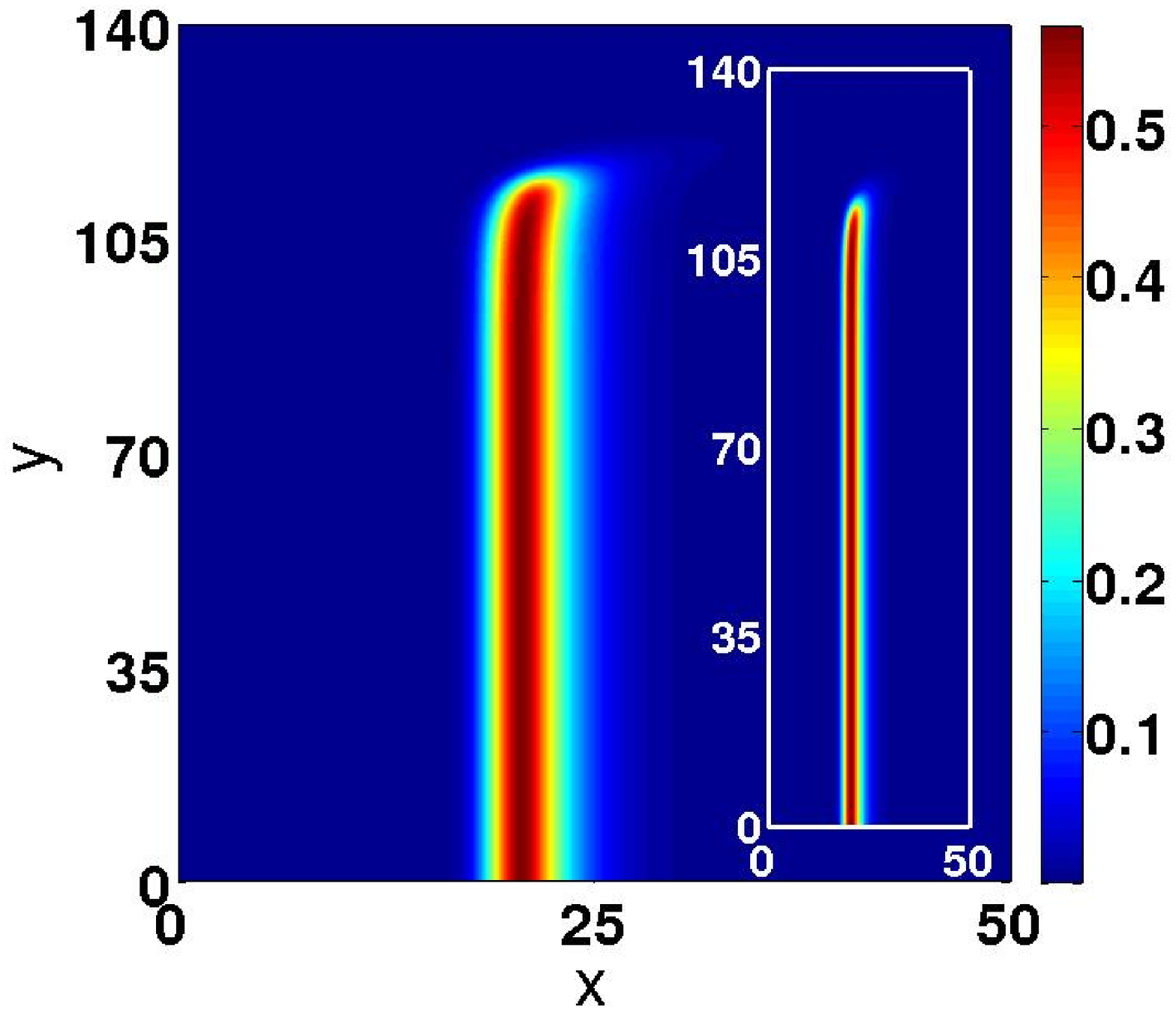}
}
\caption{Contour plot of a spiral wave solution of system (\ref{eq:FrozenBarkleyStationary}) at $\epsilon=0.08052$, close to the critical excitability $\epsilon_c$. The (a) activator and (b) inhibitor are shown with the inset depicting them with the correct aspect ratio. A computational domain with $L_x=50$ and $L_y=140$ was used.}
\label{fig:contour_plots_at_eps_crit} 
\end{figure}

To study the behaviour of the wavelength, the rotation frequency and the core radius of spiral waves in the large core limit, we therefore use from now on Cartesian coordinates and Neumann boundary conditions. In Figure~\ref{fig:contour_plots_at_eps_crit} we show contour plots of the activator and the inhibitor close to criticality, illustrating the appropriateness of Neumann boundary conditions in Cartesian coordinates in this limit. 
%


\subsection{Wavelength}
\label{sec:lambda}

\begin{figure}[htb]
\centering
\subfigure[]
{
    \label{fig:log(r)-log(2pi_cInfty_o_omega)_close-up}
    \includegraphics[width=5.9cm]{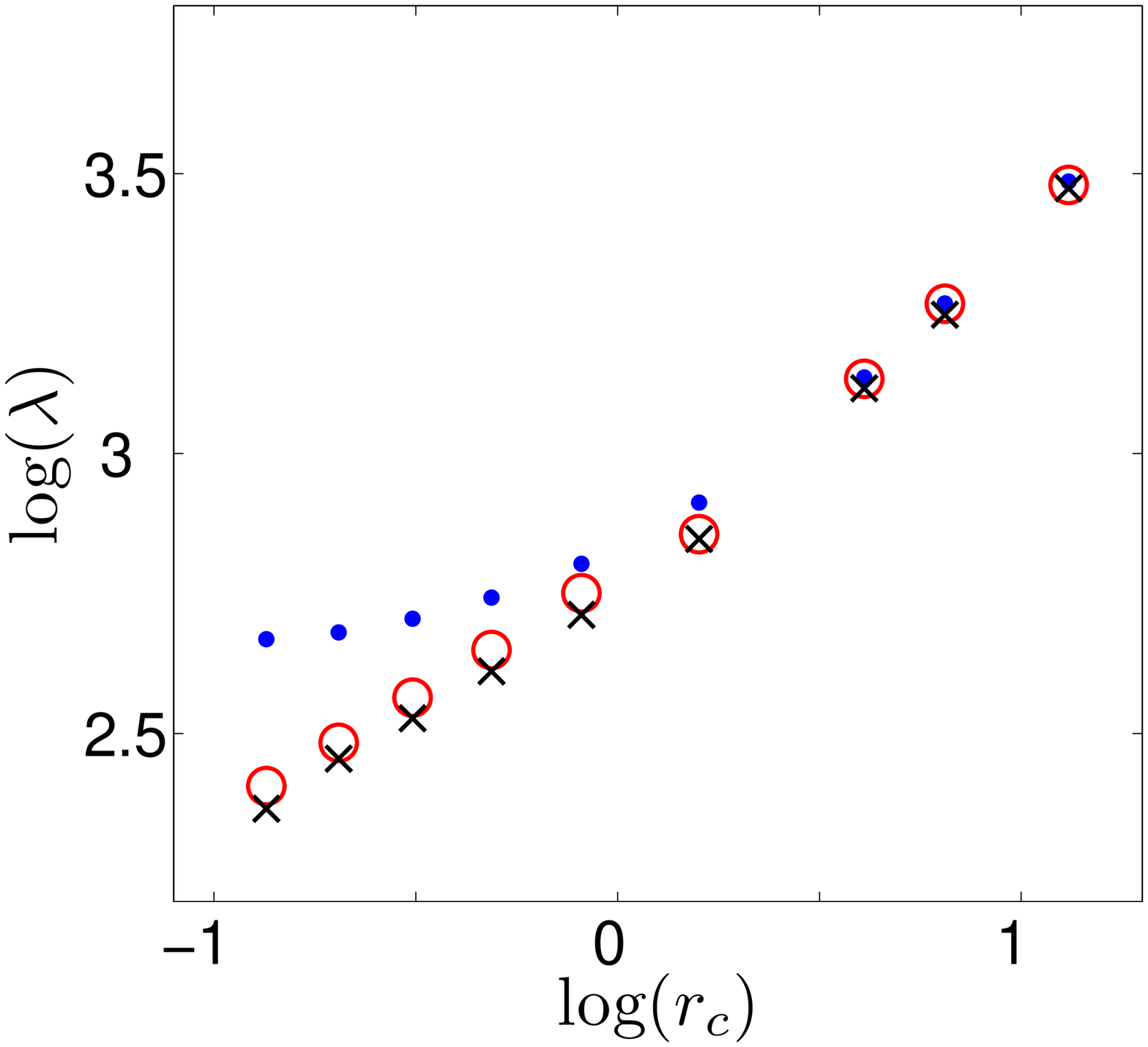}
}
\hspace{0.5cm}
\subfigure[]
{
    \label{fig:log(r)-log(2pi_cInfty_o_omega)}
    \includegraphics[width=5.9cm]{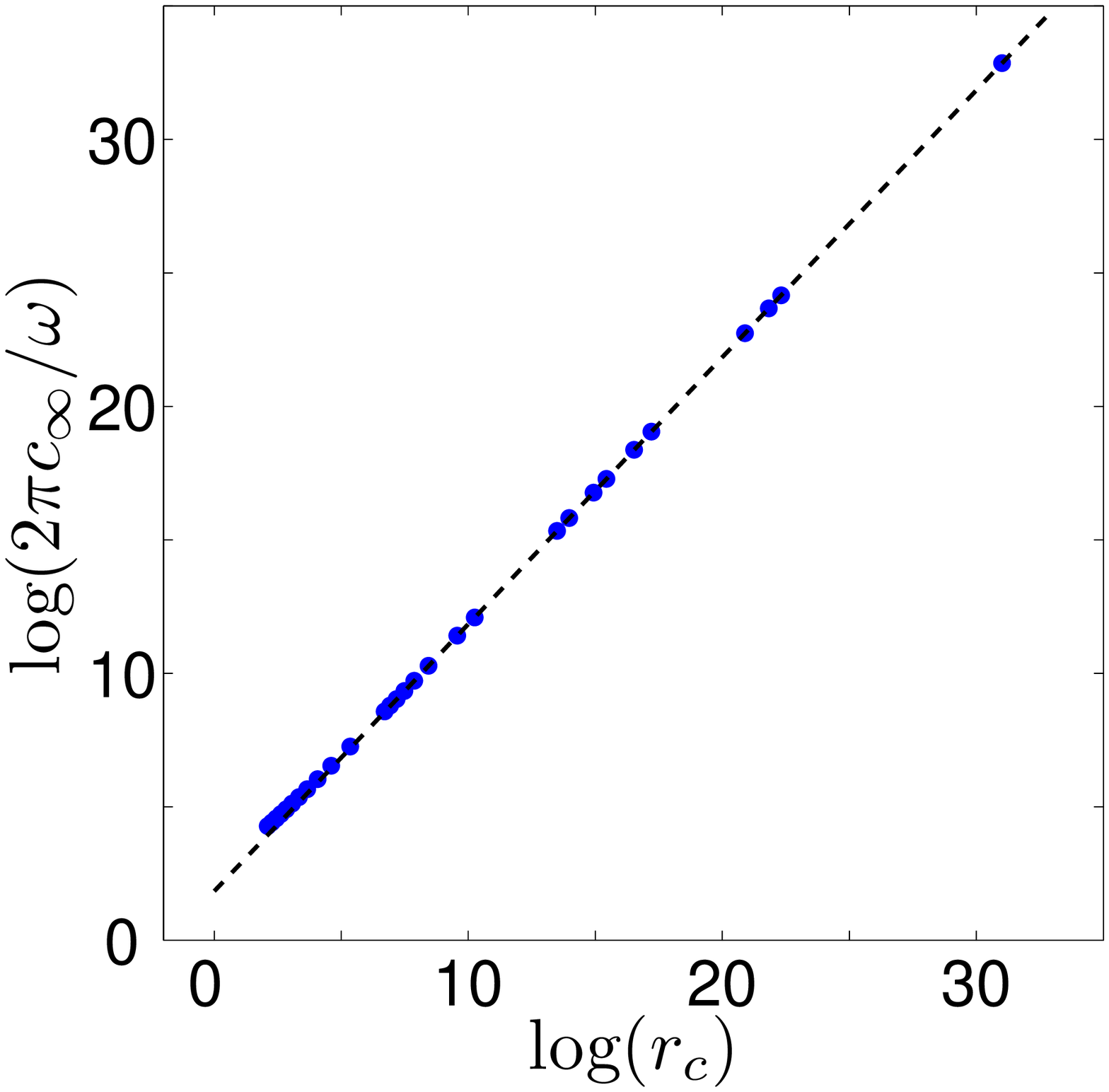}
}
\caption{Different estimates of the wavelength $\lambda$. Red `$\circ$': $\lambda$ obtained by using the nonlinear dispersion relation of $1$D wave trains $c_w(\lambda)=(\omega/2\pi)\lambda$. Blue `$\bullet$': $\lambda = 2\pi c_\infty/\omega$. (a) Small core limit. Black `$\times$': Values obtained from direct simulations of the full Barkley model (\ref{eq:Barkley}). 
(b) Large core limit. The dashed reference line corresponds to $\lambda=2\pi r_c$. Here $\omega$ and $r_c$ are obtained by solving the stationary frozen system (\ref{eq:FrozenBarkleyStationary}).}
\label{fig:log(r)-log(2pi_cInfty_o_omega_plots} 
\end{figure}

\noindent
The application of the spiral boundary conditions using Archi\-medean spirals requires the knowledge of the wavelength $\lambda$. In the small core limit the wavelength can be determined as the solution of the implicit equation 
\begin{equation}
\label{eq:1dc}
c_w(\lambda)=\frac{\omega}{2 \pi}\lambda\,,
\end{equation}
where $c_w(\lambda)$ is the velocity of a 1D wave train with wavelength $\lambda$ (see Section~\ref{sec:detpara}). In the large core limit, where the inhibitor decays sufficiently quickly and spiral wave coils do not interact, we may further simplify to $\lambda=2 \pi c_\infty/\omega$, where $c_\infty$ is the velocity of an isolated 1D pulse. In Figure~\ref{fig:log(r)-log(2pi_cInfty_o_omega)_close-up} we show a comparison of these expressions with numerical results from a direct simulation of the full Barkley model (\ref{eq:Barkley}). The velocities $c_w(\lambda)$ and $c_\infty$ are determined by freezing pulses in the corresponding 1D-model with box length $\lambda$ using the same discretization $\Delta x=0.125$ as in two dimensions. We see that the wavelength determined by (\ref{eq:1dc}) is a reasonably good approximation of the true wavelength even for small core radii. For larger radii the two methods to determine $\lambda$ converge.\\
\indent
In the large core limit we can deduce a simpler approximation for the wavelength which does not require the independent determination of the 1D velocities. One can define two approximate temporal periods for rigidly rotating spiral waves with wavelength $\lambda$ and curvature $\kappa$. First, the temporal period $T_p=\lambda/c(\lambda,\kappa)$ of a spiral wave with velocity given to first approximation by $c(\lambda,\kappa)$, and second, the time $T_r=2\pi(r_c+\delta)/c_n(\lambda,\kappa)$ which measures the time of one revolution of a spiral wave tip around a circle with radius $r_c+\delta$ chosen such that the normal velocity $c_n(\lambda,\kappa)$ of the spiral tip is tangential to that circle. Equating these two temporal periods leads to the kinematic relation \cite{Kheowan01}
\begin{equation}
\label{eq:lambdafull}
\lambda = 2\pi(r_c+\delta)\frac{c(\lambda,\kappa)}{c_n(\lambda,\kappa)}\, .
\end{equation}
In the large core limit, we expect $c_n(\lambda,\kappa)=c(\lambda,\kappa)=c_\infty$, and $\delta\ll r_c$ (see Figure~\ref{fig:rcrU}). In this case (\ref{eq:lambdafull}) reduces to the simple relationship
\begin{equation}
\label{eq:lambdarc}
\lambda = 2\pi r_c\; .
\end{equation}
In  Figure~\ref{fig:log(r)-log(2pi_cInfty_o_omega)} we show that (\ref{eq:lambdarc}) is a good approximation of the wavelength in the large core limit and matches well with $\lambda=2 \pi c_\infty/\omega$. Figure~\ref{fig:log(r)-log(2pi_cInfty_o_omega)} shows that $\omega \sim r_c^{-1}$ as already indicated by (\ref{eq:Centre0M}) implying $\mu_2^2+\mu_3^2\approx c_\infty^2$, resembling the motion of travelling waves.\\

%


\subsection{Scaling behaviour of the large core limit}
\label{sec:scaling}
\begin{figure}[htpb]
\centering
{
    \label{fig:log(r)-log(eps-eps_c)}
    \includegraphics[width=5.9cm]{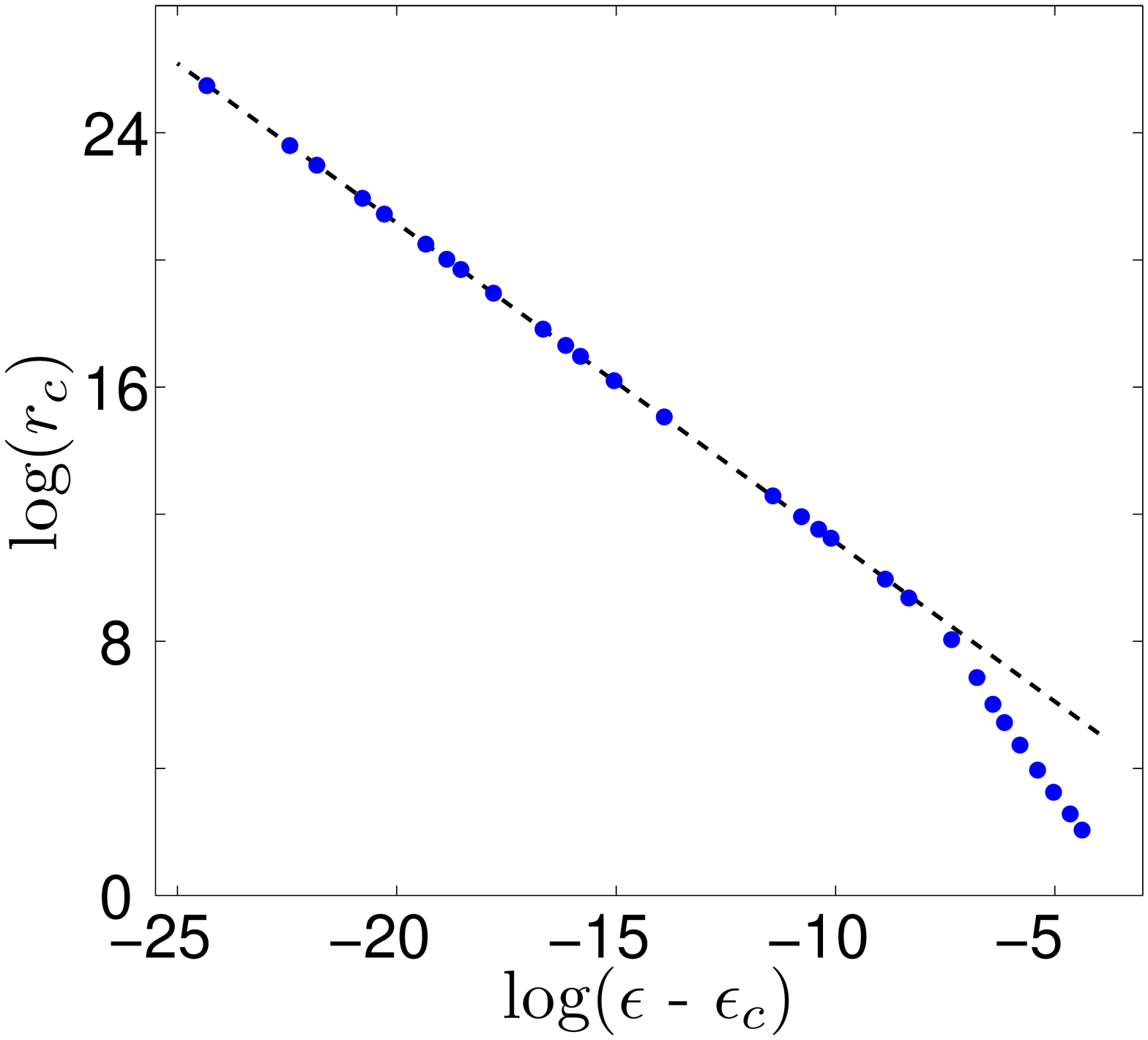}
}
\hspace{0.5cm}
{
    \label{fig:log(omega)-log(eps-eps_c)}
    \includegraphics[width=5.9cm]{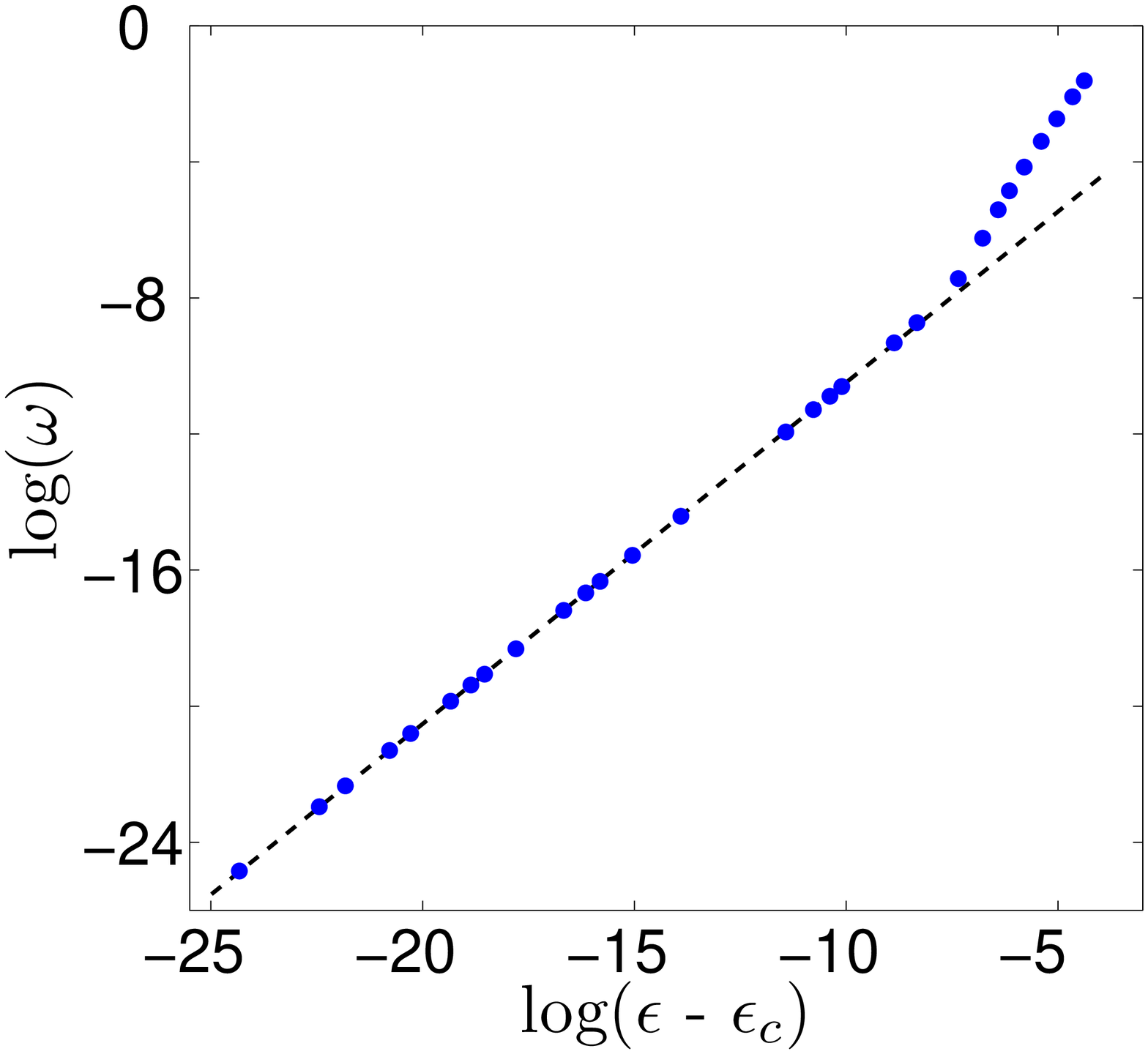}
}
\caption{Scaling behaviour of core radius $r_c$ (left) and rotation frequency $\omega$ (right). We show results for frozen solutions of the stationary system (\ref{eq:FrozenBarkleyStationary}) using Cartesian coordinates and NBCs, in computational domains with $L_x=50$ and  $L_y=62.5$. The dashed reference lines have slope $-1.004$ for $r_c$ and $1.0038$ for $\omega$, respectively. The critical value of the excitability parameter is $\epsilon_c=0.08054091$. }
\label{fig:RadiiPlots} 
\end{figure}

In Figure~\ref{fig:RadiiPlots} we show results on the scaling behaviour of the rotation frequency $\omega$ and the core radius $r_c$ as a function of the distance to criticality $(\epsilon-\epsilon_c)$. We can clearly identify a linear scaling regime 
\begin{equation}
\omega \sim (\epsilon-\epsilon_c)^{1} \qquad {\rm and} \qquad r_c \sim  (\epsilon-\epsilon_c)^{-1}
\end{equation}
at the bifurcation.\\

\noindent
This scaling behaviour was predicted in \cite{Elkin98} using kinematic theory, and in \cite{Ashwin99} using equivariant bifurcation theory. The change from a rigidly rotating spiral to a travelling wave finger was described as a so called drift bifurcation\footnote{Note that the term {\emph {drift bifurcation}} is also used in a different context for a pitchfork bifurcation of stationary patterns with reflection symmetry in an ${\rm O}(2)$-system \cite{Krupa90,Kness92}.} which occurs in the group dynamics rather than in the shape dynamics. To understand the bifurcation from rigidly rotating spirals to retracting fingers using the symmetry reduction method, one needs to look at the assumptions necessary for the orthogonal splitting of the full dynamics into the shape dynamics and the group dynamics -- which underlies the freezing method as well as the theory in \cite{Ashwin99}. Symmetry reduction relies on the existence of a centre manifold. Finite-dimensional centre manifold reductions can be proven for spiral waves in unbounded domains assuming the existence of a spectral gap \cite{Sandstede97b}. The spectral gap corresponds to a non-zero distance of the essential spectrum, which consists of the complement of the spectrum of the set of isolated eigenvalues of finite multiplicity, and the imaginary axis, thereby assuring normal hyperbolicity of the solutions. 

\begin{figure}[htpb]
\centering
  \includegraphics[width=8.2cm]{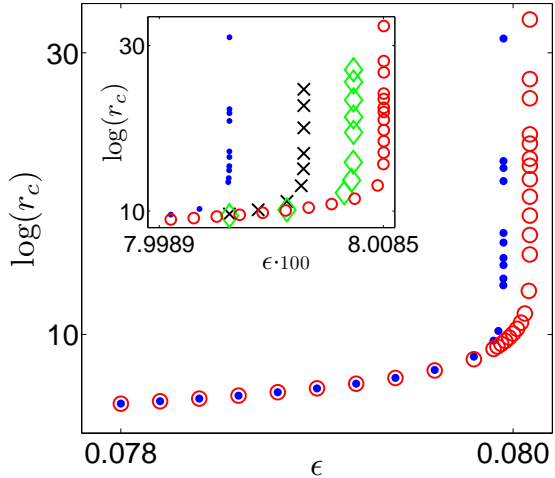}
\caption[Sensitivity of the critical limit for different computational domains.]{Demonstration of sensitivity of the critical excitability on the length of the resolved finger solution. Results are obtained by freezing the same spiral wave on rectangular domains with $L_x=50$ and $L_y=62.5$ ($\bullet$), $L_y=80$ ($\times$), $L_y=120$ ($\diamond$) and $L_y=140$ ($\circ$). The resolved finger lengths are approximately $L_{62.5}\approx 39.5$ ($\bullet$), $L_{80}\approx 59$ ($\times$), $L_{120}\approx 97$ ($\diamond$) and $L_{140}\approx 115$ ($\circ$). }
\label{fig:ComparePolarCartesianCoordinates} 
\end{figure}

In \cite{Sandstede99} it was shown that this gap, in fact, does not exist for rigidly rotating spirals in unbounded domains. It was shown, however, that small perturbations to the unboundedness of the domain, i.e. spiral wave solutions in `very large' domains with imposed boundary conditions, open up a large spectral gap. This is in contrast to the situation for roll solutions say, where the spectral gap obtained in this way is negligible. Hence for many purposes it is reasonable to proceed as if a spectral gap is present so that a symmetry reduction can be performed.  Such an approach has proved useful in understanding the transition to meandering and linearly drifting spirals \cite{Barkley94,Wulff96,Fiedler96,Golubitsky97,Sandstede97a,Sandstede97b,Sandstede99}. This gap, however, becomes smaller the closer one is to criticality, at which point the spectral gap becomes zero. Close to criticality, when the overall curvature of the spiral wave becomes zero, and the solution has morphed into a semi-infinite travelling finger, finite-dimensional centre manifold theory is not applicable anymore and ought to be replaced by an infinite-dimensional Ginzburg-Landau type description. A rigidly rotating spiral wave may become unstable to an infinite number of modes of the continuous spectrum which cannot be captured by the freezing method or the bifurcation theory of \cite{Ashwin99}. However, we argue that the (possibly unstable) solution obtained by the finite dimensional reduction will, at least, function as an organizing centre for the full dynamics which takes into account the interactions with the continuous spectrum.\\ \indent Close to criticality, the break-down of the finite-dimensional description manifests itself in the requirement for an ever increasing resolution and accuracy in numerical simulations. Phenomenologically, one needs to resolve greater and greater parts of the spiral wave close to criticality, to resolve the behaviour of the far field spiral wave. A too small segment of the solution appears like a travelling wave without curvature. The tip region (i.e. the region which is not described by simple geometric constructs such as Archimedean spirals or involutes of circles) grows when criticality is approached, as was already encountered for small core spirals (cf. Figure~\ref{fig:FittedSpirals} and Figure~\ref{fig:rcrU}).\\ \indent We found experimentally, that the actual values of $r_c$ and $\omega$ at criticality, as well as the actual critical value of the excitability $\epsilon_c$ depend strongly on the numerical resolution. For example, the orientation of the finger within the computational domain has a strong influence, especially for moderate sizes of the computational domain. For a sufficiently small length of a spiral wave solution -- which then appears finger-like, see Figure~\ref{fig:LargeCoreSpiral} -- the proportion of the region dominated by the inaccurate Neumann boundary conditions becomes unproportionally large. We therefore rotate the finger in a pre-processing procedure to maximize the validity of the NBC. Spiral waves which do not leave the computational domain perpendicularly, are rotated around the centre of the rectangular computational domain, and then subsequently mapped back onto the computational grid using bilinear interpolation before the application of the freezing procedure. This procedure inevitably leaves triangles of the computational domain which have not been assigned values for the fields $u$ and $v$ after the rotation. We manually set $u$ and $v$ to zero on grid points falling into those triangles which is justified in the large core limit for small angels of rotation. Close to criticality it proved useful to use an iterative procedure, whereby reorienting and the freezing procedure are alternated at fixed $\epsilon$. However, this method of reorientation of spiral waves is inaccurate and not methodological, and small changes will have measurable effects in the values of the group parameters \cite{HermannThesis}.\\ \indent Similarly, the numerical results become more sensitive to the actual length of the finger which is resolved within the computational domain. At criticality, infinite resolution is required. This is illustrated in Figure~\ref{fig:ComparePolarCartesianCoordinates} where we show results of the core radius $r_c$ as a function of $\epsilon$ for simulations differing only in the length of the resolved finger within the domain. Whereas the values of $r_c$ are independent on the resolved size away from the bifurcation (but note the already large magnitude of $r_c$ in the $\log$-scale), they differ strongly approaching criticality. The critical value $\epsilon_c$ also depends on the actual length of the resolved finger. This sensitivity of the results to the resolution is inherent and cannot be avoided due to the breakdown of the assumptions underlying the freezing method.

It is pertinent to mention that, despite the sensitivity of the actual numerical values of the rotation frequency, the core radius and the critical excitability, the linear scaling regime depicted in Figure~\ref{fig:RadiiPlots}  is robust against changes of size of the computational domain, the orientation of the spiral and the discretization.\\

\begin{figure}[htpb]
\centering
{
    \label{fig:log(r)-log(eps-eps_c)spurious}
    \includegraphics[width=5.9cm]{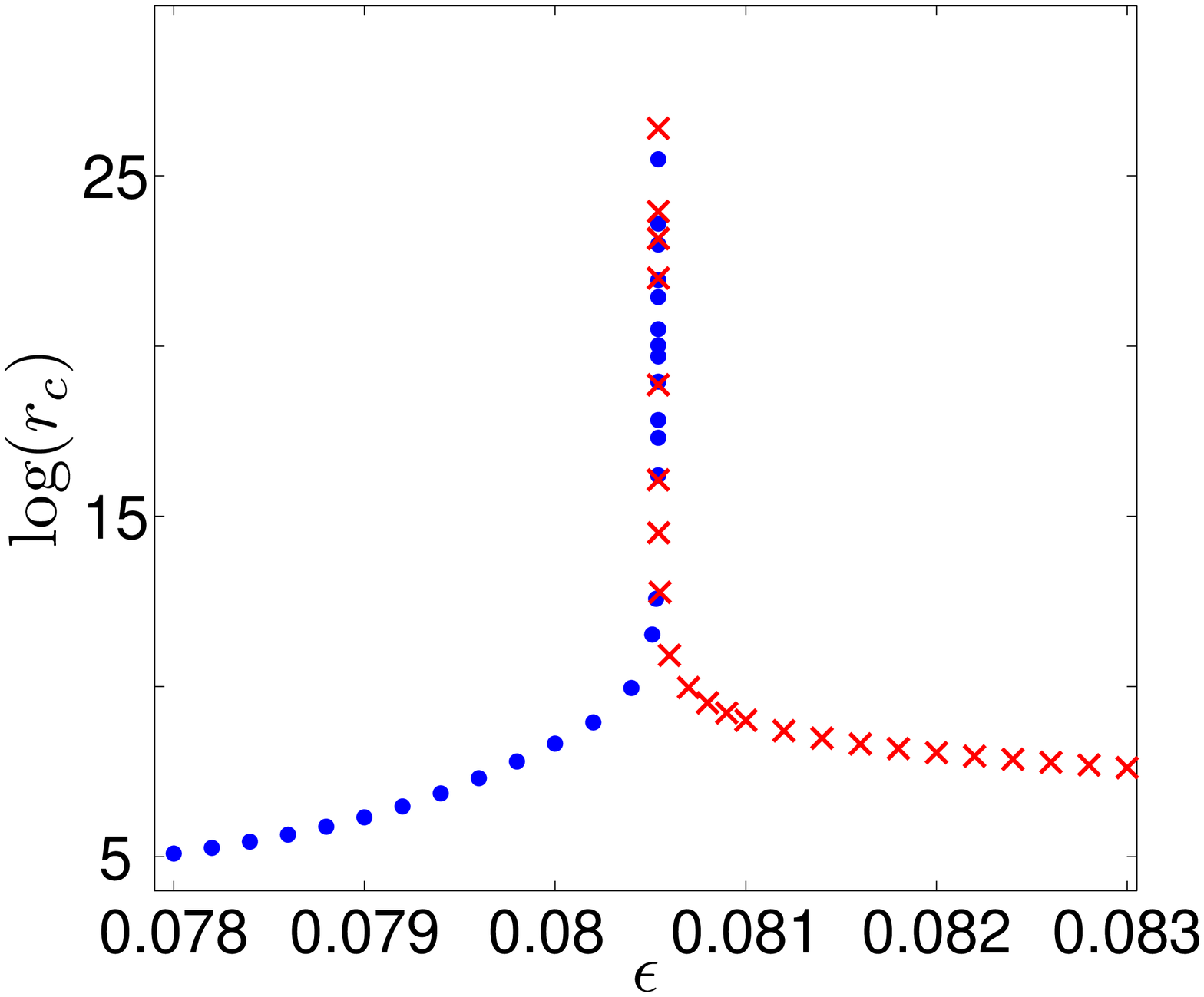}
}
\hspace{0.5cm}
{
    \label{fig:log(omega)-log(eps-eps_c)spurious}
    \includegraphics[width=5.9cm]{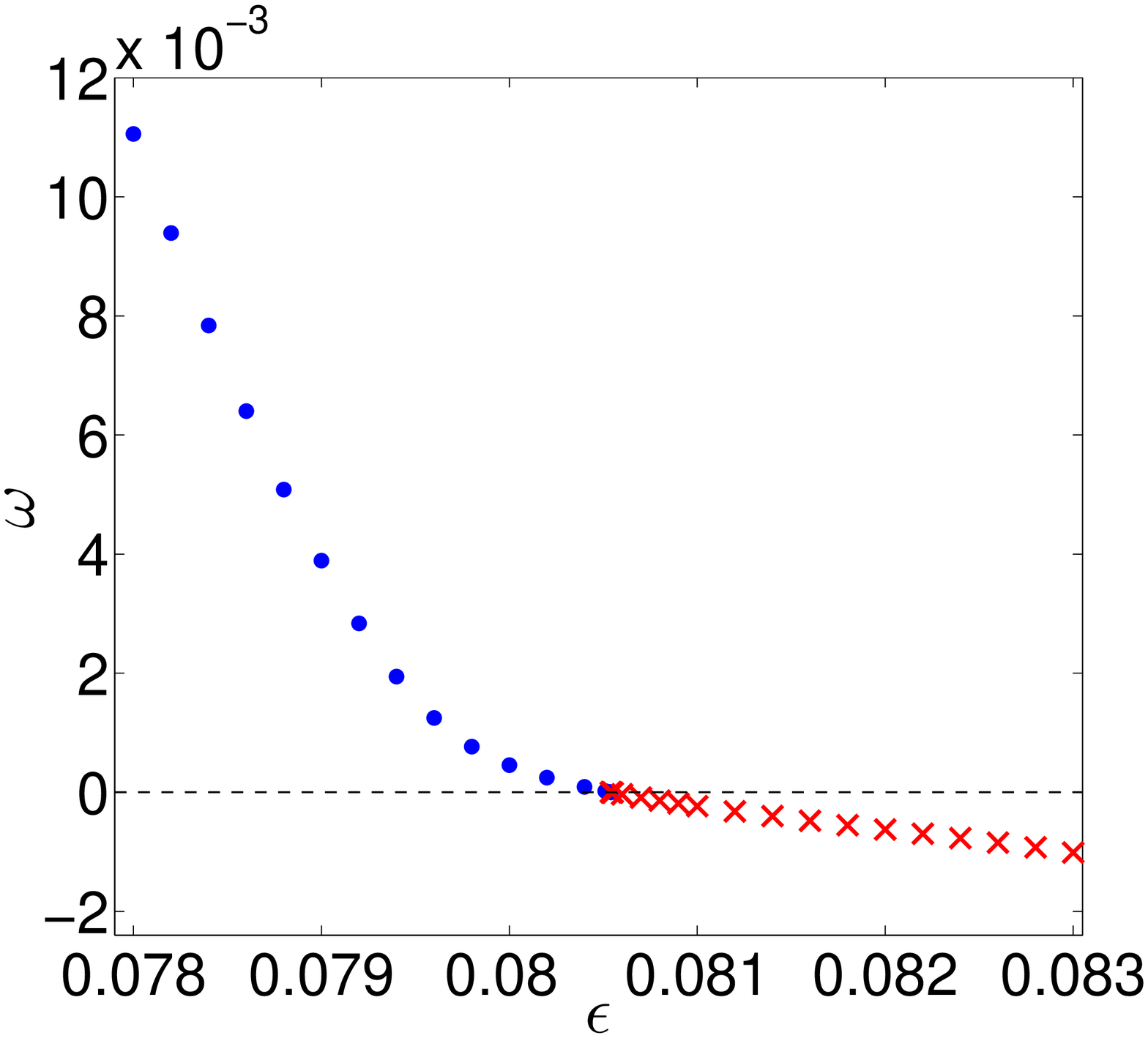}
}
\caption{Core radius $r_c$ (left) and rotation frequency $\omega$ (right) for values of the excitability $\epsilon>\epsilon_c$. At criticality the linear scaling behaviour is assumed from both sides of the bifurcation point $\epsilon_c=0.08054091$. We show results for frozen solutions of the stationary problem (\ref{eq:FrozenBarkleyStationary}) using Cartesian coordinates and NBCs, in computational domains with $L_x=50$ and  $L_y=62.5$.}
\label{fig:RadiiPlotspsurious} 
\end{figure}

\noindent
The freezing method finds frozen solutions beyond the critical $\epsilon_c$. In Figure~\ref{fig:RadiiPlotspsurious} we show results for the core radius and the rotation frequency when the excitability is varied to values $\epsilon>\epsilon_c$. We find that the linear scaling regime extends past the critical value $\epsilon_c$. Moreover, the rotation frequency becomes negative for $\epsilon>\epsilon_c$. Note that the theory of \cite{Ashwin99} does not describe the behaviour of solutions past the bifurcation point. For $\epsilon>0.0835$ we were not able to find frozen solutions of the stationary system (\ref{eq:FrozenBarkleyStationary}). We have checked that the value of $\epsilon=0.0835$ corresponds to the saddle node of travelling waves in 1D, and exhibits the typical square root behaviour close to the saddle node bifurcation. 

At the bifurcation to retracting fingers at $\epsilon_c$ there is no guarantee that the solutions of the freezing method correspond to actual solutions of the original Barkley model. However, we have verified that the frozen solutions obtained for $\epsilon>\epsilon_c$ correspond to the retracting fingers by using them as initial conditions in the full Barkley model (\ref{eq:Barkley}).

%


\section{Summary}
\label{sec:discussion}
Our aim in this work was two-fold. In a first part, we have formulated a modification of the freezing method, introduced in \cite{Beyn04}. We have formulated the freezing method in polar and Cartesian coordinates for the time-dependent and the stationary formulation to freeze spiral waves in excitable media with non-diffusive inhibitors, typical for applications in cardiac dynamics. 

In particular, we have proposed a simple method to overcome oscillations near the boundary, which have so far obstructed the investigation of the large core limit. Oscillations in the interior of the computational domain in the time-dependent problem were eliminated by employing a semi-implicit Crank-Nicolson scheme. We have further introduced spiral boundary conditions by using Archimedean spirals and involutes of circles as geometric constructs to approximate contour lines of spiral wave solutions in unbounded domains and in computational domains whose size is much smaller then the actual physical domain.

We have established the regions of applicability of our method. We found that to study spiral waves in the small core limit polar coordinates with SBCs are favourable, whereas to study spiral waves in the large core limit, Cartesian coordinates and NBCs should be used.

In a second part of this work, we have numerically investigated the large core limit of spiral waves. We have determined the shape of solutions near criticality, and have determined their rotation frequency as well as their core radius. Further, we discussed solutions of the freezing method for excitabilities beyond criticality, which could be extended to the saddle node bifurcation of travelling waves. We have presented results on the scaling behaviour of the spiral wave parameters in the large core limit and confirmed the linear scaling at the drift bifurcation developed in \cite{Ashwin99}, and discussed the limitations of the freezing method. We believe that these results may be helpful in designing kinematic theories.\\

\noindent
After submission of this work we became aware of work \cite{Foulkes10} in which the large core limit is investigated using a similar numerical method. The authors also identify a linear scaling regime in the large core limit, and study further meandering spirals and electrophoresis. In this work a phase condition is used which pins the tip of the spiral wave to the centre of the domain. We believe that this phase conditions with our implementation of the boundary conditions will prove useful in further studies of spiral wave dynamics.


\section*{Acknowledgements}
We thank Dwight Barkley, Wolf-J\"urgen Beyn, Ian Melbourne and Bj\"orn Sandstede for fruitful discussions, and are deeply indebted to Vera Th\"ummler for numerous discussions and generous help with the freezing method. GAG thanks Vadim Biktashev for stimulating discussions and for sharing their work with us before publication. 


\addcontentsline{toc}{chapter}{Bibliography}
\bibliographystyle{siam}
\bibliography{bibliography}

\end{document}